%% file: main.tex
\documentclass[11pt,a4paper]{article}
\pdfoutput=1
\usepackage[applemac]{inputenc}
\usepackage{amsmath, amsthm}
\usepackage{jheppub}
\usepackage[center]{caption}
\usepackage{subcaption}
\usepackage{multirow}
\usepackage{booktabs}
\usepackage{amsmath}
\usepackage{amsfonts}
\usepackage{amssymb}
\usepackage{graphicx}
\usepackage{indentfirst}
\usepackage{slashed}
\usepackage{mathrsfs}
\usepackage{tikz}
\usepackage{bm}

\newcommand{\refE}[1]{eq.~(\ref{#1})}
\newcommand{\refF}[1]{fig.~\ref{#1}}

\newcommand{\refS}[1]{Section~\ref{#1}}

\newcommand{\Li}[0]{\text{Li}}

\newcommand{\Disc}[0]{\operatorname{Disc}}
\newcommand{\Cut}[0]{\operatorname{Cut}}
\renewcommand\Re{\operatorname{\mathfrak{Re}}}

\newcommand{\hypgeo}[4]{\,_2F_1\left(#1,#2;#3;#4\right)}
\newcommand{\appellf}[6]{\,F_1\left(#1;#2,#3;#4;#5;#6\right)}

\newcommand{\zbar}{\bar{z}}
\newcommand{\cut}{\text{Cut}}

\newcommand{\bea}{\begin{eqnarray}}
\newcommand{\eea}{\end{eqnarray}}
\newcommand{\bean}{\begin{eqnarray*}}
\newcommand{\eean}{\end{eqnarray*}}

\def\Li{{\rm Li}}

\def\th{{\theta}}

\def\eps{\epsilon}

\def\bz{{\bar{z}}}

\newcommand{\zb}{\bar{z}}

\def\bz{\bar{z}}

\def\bw{\bar{w}}

\def\cS{{\cal S}}

\def\Label#1{\label{#1}
  \smash{\hbox to0pt{\raise1ex\hbox{\tiny[#1]}\hss}}}

\def\beq{\begin{equation}}
\def\eeq{\end{equation}}
\def\bsp#1\esp{\begin{split}#1\end{split}}

\newcommand{\cA}{\mathcal{A}}
\newcommand{\cH}{\mathcal{H}}

\renewcommand{\ln}{\log}

\preprint{IPhT-t15/030
\rightline{TCDMATH-15-02}
\rightline{Edinburgh 2015/05}
}

\title{Cuts and coproducts of massive triangle diagrams}

\author[a,b]{Samuel Abreu,}
\author[b,c,d]{Ruth Britto,}
\author[b]{and Hanna Gr\"onqvist}

\affiliation[a]{Higgs Centre for Theoretical Physics, School of Physics and Astronomy, \\
The University of Edinburgh,  
Edinburgh EH9 3JZ, Scotland, UK}
\affiliation[b]{Institut de Physique Th{\'e}orique, Universit\'e Paris Saclay, CEA, CNRS, F-91191 Gif-sur-Yvette}
\affiliation[c]{School of Mathematics, Trinity College, Dublin 2, Ireland}
\affiliation[d]{Hamilton Mathematical Institute, Trinity College, Dublin 2, Ireland}

\emailAdd{abreu.samuel@ed.ac.uk}
\emailAdd{ruth.britto@cea.fr}
\emailAdd{hanna.gronqvist@cea.fr}

\abstract{
Relations between multiple unitarity cuts and coproducts of Feynman integrals are extended to allow for internal masses.  These masses introduce new branch cuts, whose discontinuities can be derived by placing single propagators on shell and identified as particular entries of the coproduct.  First entries of the coproduct are then seen to include mass invariants alone, as well as threshold corrections for external momentum channels.
As in the massless case, the original integral can possibly be recovered from its cuts by starting with the known part of the coproduct and imposing integrability contraints.  We formulate precise rules for cuts of diagrams, and we gather evidence for the relations to coproducts through a detailed study of one-loop triangle integrals with various combinations of external and internal masses.
}


\begin{document}
\maketitle

\input{Sec_Intro.tex}

\input{Sec_DiagramsAndDefinitions.tex}

\input{Sec_FirstEntry.tex}

\input{Sec_CutRules.tex}

\input{Sec_RelationCutDelta.tex}

\input{Sec_Reconstruction.tex}

\input{Sec_Discussion.tex}

\acknowledgments

It is a pleasure to thank {\O}yvind Almelid, Claude Duhr, Einan Gardi, David Kosower, and Alexander Ochirov for helpful discussions.  Special thanks to Claude Duhr for the use of his Mathematica package {\tt PolyLogTools}, and to Claude Duhr and Einan Gardi for comments on the manuscript.  Figures were created with the help of {\tt TikZ} \cite{tantau:2013a}.
S.A. is supported by Funda\c{c}\~ao para a Ci\^encia e a Tecnologia, Portugal, through a doctoral degree fellowship (SFRH/BD/69342/2010).  S.A. and R.B. were supported by the Research Executive Agency (REA) of the European Union under the Grant Agreement number PITN-GA-2010-264564 (LHCPhenoNet).  S.A. acknowledges the hospitality of the {\em Centre National de la recherche scientifique} (CNRS) and the Institute for Theoretical Physics (IPhT) of the CEA during the early stages of this work.
H.G. thanks the Higgs Centre for Theoretical Physics at the University of Edinburgh and especially the School of Mathematics of Trinity College Dublin for their hospitality.




\appendix

\input{appendices.tex}

\bibliographystyle{JHEP}
\bibliography{bibMain.bib}

\end{document}

%% file: Sec_Intro.tex
\section{Introduction}

The evaluation of Feynman integrals is a necessary ingredient for the precise determination of physical observables in perturbative quantum field theories. The notorious difficulty of evaluating these integrals has led to the development of various integration techniques.

One family of these techniques relies on the study of the discontinuities of Feynman integrals across their branch cuts, a topic that  goes back to the early days of perturbative quantum field theory. It is based on the fact that these discontinuities can be computed directly by specific diagrammatic rules \cite{Landau:1959fi, Cutkosky:1960sp, tHooft:1973pz,Veltman:1994wz, Remiddi:1981hn}. According to these rules, partitions of Feynman diagrams into two regions are enumerated, and the particles at the boundary of the two regions are restricted to their mass shells. This operation defines the set of \emph{cut diagrams}. Collecting the cut diagrams with the same momentum flow between the regions, we construct a \emph{unitarity cut} which captures the discontinuity across the branch cut in that momentum invariant. The on-shell restrictions greatly simplify the integration and its result. The original uncut integral can then be reconstructed, traditionally through dispersion relations \cite{Landau:1959fi, Cutkosky:1960sp, tHooft:1973pz,Veltman:1994wz, Remiddi:1981hn}. More recently, modern unitarity methods use discontinuities to project an amplitude onto a basis of known master integrals \cite{Bern:1994zx, Bern:1994cg, Britto:2004nc, Britto:2005ha, Buchbinder:2005wp, Mastrolia:2006ki, Anastasiou:2006jv, Forde:2007mi, NigelGlover:2008ur, Mastrolia:2009dr, CaronHuot:2010zt,Kosower:2011ty,Johansson:2012zv,Sogaard:2013yga,Sogaard:2014ila}.  For increased efficiency, the latter techniques can incorporate {\em multiple cuts} in different channels simultaneously, and even further generalizations of cut operations.  Restricting many propagators to their mass shells separates a diagram into several on-shell elements, which are typically simpler to compute.

In order to find yet more efficient computational tools, it is useful to study the types of functions produced by Feynman integrals.  Although it is still in general an open question to determine what functions are needed, it is known that all one-loop integrals, and a large number of higher-loop integrals with a sufficiently small number of masses, can be written in terms of the transcendental functions called multiple polylogarithms. (At higher loop order, elliptic functions can appear \cite{Caffo:1998du, MullerStach:2011ru, Adams:2013nia, Bloch:2013tra, Remiddi:2013joa,Adams:2014vja}.) Recent developments on the mathematical structure of this class of functions, in particular their \emph{Hopf algebraic structure} \cite{GoncharovMixedTate, Goncharov:2005sla}, have thus had a big impact on the physics community over the last few years \cite{Goncharov:2010jf, Duhr:2012fh}.

In \cite{Abreu:2014cla}, cuts of Feynman diagrams without internal masses were studied in the light of these modern mathematical tools. One of the results of that paper was the establishment of a relation between (multiple) cuts of diagrams and the coproduct of the corresponding Feynman integral. Indeed, the coproduct, which is an operator defined in the Hopf algebra of multiple polylogarithms, is particularly useful in capturing their discontinuities. The coproduct breaks the original function down into lower-weight functions, such that discontinuities are perfectly captured by the so-called {\em first entries}, a statement known as the {\em first-entry condition}.  The result of \cite{Abreu:2014cla} was to extend this property to a correspondence of subsequent entries of the coproduct and multiple cuts.  The strategy employed was to find a relation between cuts and the coproduct by independently relating cuts to discontinuities, and coproduct entries to discontinuities. The conjectured relations were checked in several one-loop examples and a non-trivial two-loop example. 

The {\em first-entry condition} was observed in \cite{Gaiotto:2011dt} and explained mathematically in \cite{Duhr:2012fh}, but these works were restricted to integrals with massless propagators.  In these cases, the integral is a function of Mandelstam invariants constructed from the external momenta.  When we include massive propagators, there are also branch cuts in the  internal mass variables.  Thus we would expect the masses to appear plainly among the first entries of the coproduct, and in more intricate ways in subsequent entries.  
In the present paper, we find this to be true.  We generalize the relations between (multiple) cuts and the coproduct to diagrams with internal masses, gathering supporting evidence from a close study of triangle diagrams. 
The generalization requires modification of the first entry condition:  along with including the mass variables among first entries, the familiar Mandelstam invariants must now be accompanied by their thresholds.  The discontinuities in mass variables correspond to {\em single-propagator} cuts of the propagator with the same mass.   We formulate cutting rules that include these single-propagator cuts, reproducing the discontinuities precisely.\footnote{In contrast to most applications of multiple cuts in the literature, we are concerned here only with diagrammatic cuts that have clear interpretations as discontinuities.}  We then find generalized relations and check them on  scalar one-loop three-point functions with various configurations of internal and external masses. While it will be important to extend this investigation to integrals with more loops and legs, we believe that our starting point of triangle integrals already illustrates the most essential features.

It was also shown in \cite{Abreu:2014cla} that the original Feynman integral could be reconstructed from the knowledge of a single cut through purely algebraic manipulations on the coproduct tensor. In this paper, we provide evidence that the same kind of reconstruction procedure can be applied when internal masses are present.

The paper is organized as follows.  We close the introduction with a brief review of the Hopf algebra of multiple polylogarithms, as applied to Feynman integrals.
In section \ref{sec:triangles}, we present the Feynman integrals that we study as examples, define the variables used for each diagram and present their symbol alphabets.
In section \ref{sec:firstentry}, we generalize the first entry condition to diagrams with internal massive propagators. In section \ref{sec:two_types_of_cuts}, we present the rules for calculating the two different kinds of cuts considered in this paper: the familiar channel cuts, also used in \cite{Abreu:2014cla}, and the new single-propagator cuts corresponding to mass discontinuities. We explain how these cuts can be applied iteratively to reproduce multiple discontinuities. Finally, we also comment on our strategy to explicitly compute cut diagrams. Section \ref{sec:relationcutdelta} concerns the relations among discontinuities, coproducts, and cuts. We extend the relations of \cite{Abreu:2014cla} to diagrams with internal masses and comment on several subtleties that arise in the presence of massive internal propagators. In section \ref{sec:reconstruction} we discuss how, starting from a single channel cut, we can reconstruct first the symbol of a Feynman integral and then the whole integral itself. Finally, we discuss our conclusions in section \ref{sec:discussion}.

In appendix \ref{app_FeynmanRules} we give our conventions for the calculation of Feynman diagrams. In appendices \ref{app_oneMass}, \ref{app_twoMass} and \ref{app_threeMass} we present results for triangle integrals with several configurations of internal  and external masses in terms of multiple polylogarithms. We also present analytical results for their (multiple) cuts and symbols. We do not write all possible configurations of masses, because we believe the examples we give are enough to illustrate and check all our relations. We have, however, checked that the remaining cases do indeed behave in the expected way.

\subsection{The Hopf algebra for Feynman integrals}

We finish this introduction with a brief review of the Hopf algebra of multiple polylogarithms in the context of Feynman integrals. For a more detailed review, we refer the reader to the literature, e.g.~\cite{ChenSymbol, Goncharov:1998kja,Goncharov:2005sla,Goncharov:2009tja,  GoncharovMixedTate, Goncharov:2010jf,Duhr:2011zq, Duhr:2012fh, Duhr:2014woa} and references therein. Here, we focus on introducing the concepts used in this paper.

\paragraph{Multiple polylogarithms for Feynman integrals.}
As mentioned above, a large class of Feynman integrals may be expressed in terms of the transcendental functions called {\em multiple polylogarithms}.  These functions are defined by the iterated integrals
 \beq\label{eq:Mult_PolyLog_def}
 G(a_1,\ldots,a_n;z)=\,\int_0^z\,{d t\over t-a_1}\,G(a_2,\ldots,a_n;t)\,,\\
\eeq
with $a_i, z\in \mathbb{C}$ and the base case defined to be $G(z)=1$. As simple special cases, we find for example
\begin{align}\bsp
&
G(0;z)=\log z, \qquad G(a;z)=\log\left(1-\frac{z}{a}\right), \qquad G(0,a;z)=-\Li_2\left(\frac{z}{a}\right)\,,
\\ & G(a,b;z) 
= \Li_2\left(\frac{b-z}{b-a}\right) - \Li_2\left(\frac{b}{b-a}\right)
+ \ln\left(1-\frac{z}{b}\right) \ln\left(\frac{z-a}{z-b}\right)  \,,
\esp\end{align}
where the last equality holds for $a$ and $b$ different and nonzero.
The transcendental \emph{weight} of the multiple polylogarithm is defined to be $n$ for $G(a_1,\ldots,a_n;z)$.

Here, we study Feynman integrals evaluated in dimensional regularization, with dimensionality $D=4-2\eps$.  We normalize by rational functions (the leading singularities) in order to obtain {\em pure} functions: when expanded in a Laurent series around $\eps=0$, each term in the expansion has uniform weight and no factors of rational or algebraic functions of external kinematic variables.  It has been conjectured~\cite{Henn:2013pwa}, and observed in many nontrivial examples~\cite{Henn:2013tua,Henn:2013woa,Henn:2013nsa,Caron-Huot:2014lda,Gehrmann:2014bfa,Caola:2014iua,Argeri:2014qva,Grozin:2014hna,Bern:2014kca},  that such pure functions can suffice to characterize master integrals.

\paragraph{Hopf algebra, coproduct, and symbol.}
The space of multiple polylogarithms has the structure of a Hopf algebra, although one must take care with factors of $\pi$.  Specifically, if $\overline\cH_n$ is the $\mathbb{Q}$-vector space spanned by all multiple polylogarithms of weight $n$, then it is the quotient space $\cH = \overline\cH/(\pi\,\overline\cH)$ that is a Hopf algebra \cite{Goncharov:2005sla,GoncharovMixedTate}.  The quotient space $\cH$ is hence endowed with a coproduct $\Delta:\cH\to\cH\otimes\cH$.  The coproduct can be iterated and graded in a manner compatible with weight, so that we can write
\beq
\Delta_{n_1,\ldots,n_k} : \cH_{n}\to\cH_{n_1}\otimes\ldots\otimes\cH_{n_k}\,.
\eeq
The maximal iteration of the coproduct, which corresponds to the partition $(1,\ldots,1)$,
agrees\footnote{The congruence symbol is used because this relation is valid modulo factors of $\pi$ that can appear in the first entry of the coproduct tensor, see e.g.~\cite{Duhr:2011zq}. In this paper, we will often use the terms symbol and coproduct interchangeably: indeed, we will be mostly dealing with weight 2 functions, and at this weight there is little difference between these two operators.}
with the symbol of a transcendental function $F$~\cite{ChenSymbol,Goncharov:2010jf,Goncharov:2009tja,Brown:2009qja,Duhr:2011zq}:
\beq
\cS(F) \cong \Delta_{1,\ldots,1}(F)\in\cH_1\otimes\ldots\otimes\cH_1\,.
\eeq
Since every polylogarithm of weight 1 is an ordinary logarithm, the `$\log$' sign is conventionally dropped when talking about the symbol of a function.

We remark that there is an {\em integrability condition}, necessary and sufficient for an element of $\cH_1\otimes\ldots\otimes\cH_1$ to be the symbol of some function, which is that
\beq\label{eq:integrability}
\sum_{i_1,\ldots,i_n}c_{i_1,\ldots,i_n}\,d\log x_{i_k}\wedge d\log x_{i_{k+1}}\,\log x_{i_1}\otimes\ldots\otimes\log x_{i_{k-1}}\otimes\log x_{i_{k+2}}\otimes\ldots\otimes \log x_{i_n} = 0\,,
\eeq
where $\wedge$ denotes the usual wedge product on differential forms.

\paragraph{The symbol alphabet.}

We refer to the set of entries of the symbol  of a pure Feynman integral as the letters in the {\em symbol alphabet} $\cA$.\footnote{The letters of the symbol alphabet should be multiplicatively independent.  Note that the choice of symbol alphabet is not unique.} The number of independent dimensionless variables appearing in the symbol alphabet (i.e., the number of variables on which the pure function depends) is one fewer than the number of scales in the problem. These letters are in general not simple ratios of the Mandelstam invariants, and our general goal is to find a minimal set of variables in terms of which the letters are rational functions.  In the following section, we present the alphabets used for each of the triangles we study.  
We were able to obtain rational symbol alphabets in every case except for the triangle with three internal and three external masses. 
For this case, we discuss the choice of letters, and the obstacles to making them all rational, in section \ref{sec:triangles} and more fully in appendix \ref{tp1p2p3m12m23m13}.

%% file: Sec_DiagramsAndDefinitions.tex
\section{One-loop triangles}
\label{sec:triangles}

In this section we present the one-loop triangles studied in this paper. We consider the scalar one-loop three-point function with arbitrary mass configurations. In $D=4-2\epsilon$ dimensions, it is defined by 
\begin{eqnarray}
&&
T(p_1^2,p_2^2,p_3^2;m_{12}^2,m_{23}^2,m_{13}^2) \equiv
\nonumber 
\\ &&
- e^{\gamma_E \epsilon} \int \frac{d^Dk}{\pi^{D/2}} \frac{1}{\left[ k^2 - m_{13}^2+i0\right]
\left[ (k+p_1)^2 -m_{12}^2+i0 \right] \left[ (k-p_3)^2 -m_{23}^2 +i0\right] } ~,
\label{eq:def_scalar_triangle}
\end{eqnarray}
where $\gamma_E=-\Gamma^\prime (1)$ is the Euler-Mascheroni constant. We use $p_i$ with $i=1,2,3$ to denote the external four-momenta and $m_{ij}$ to denote the masses of the internal propagators. The mass  $m_{ij}$ is that of the propagator between the external legs $i$ and $j$. Throughout this paper, the masses are assumed to be generic.

Because the focus of the paper lies in the computation of cut integrals, it is important to keep track of the imaginary parts. Our conventions are the following: before any cuts are made, vertices are proportional to $i$ and propagators have a factor $i$ in the numerator, and the usual Feynman $+i0$ prescription is used. 
No factors of $i^{-1}$ are included for loops. Appendix \ref{app_FeynmanRules} contains a summary of our conventions.

\begin{figure}[h]
\centering
    \begin{subfigure}[h]{0.33\textwidth}
        \centering
\begin{tikzpicture}
\draw (0,0) -- (1,2) -- (2,0)-- (0,0);
\draw[line width=1mm] (1,2) -- (0,0);
\draw[line width=1mm] (2,0) -- (2.5,-0.5) node[anchor=north east]{1};
\draw (0,0) -- (-0.5,-0.5) node[anchor=north west]{2};
\draw (1,2) -- (1,2.71) node[anchor=south]{3};
\node (mass) at (-0.2,1) {$m_{23}^2$};
\draw[fill=black] (0,0) circle [radius=0.1];
\draw[fill=black] (1,2) circle [radius=0.1];
\draw[fill=black] (2,0) circle [radius=0.1];
\end{tikzpicture}
        \caption{$T(p_1^2,0,0;0,m_{23}^2,0)$}
         \label{fig:1mTriangle2}
    \end{subfigure}%
        ~ 
\centering
    \begin{subfigure}[h]{0.33\textwidth}
        \centering
\begin{tikzpicture}
\draw (0,0) -- (1,2) -- (2,0)-- (0,0);
\draw[line width=1mm] (2,0) -- (0,0);
\draw[line width=1mm] (2,0) -- (2.5,-0.5) node[anchor=north east]{1};
\draw (0,0) -- (-0.5,-0.5) node[anchor=north west]{2};
\draw (1,2) -- (1,2.71) node[anchor=south]{3};
\node (mass) at (1,-0.37) {$m_{12}^2$};
\draw[fill=black] (0,0) circle [radius=0.1];
\draw[fill=black] (1,2) circle [radius=0.1];
\draw[fill=black] (2,0) circle [radius=0.1];
\end{tikzpicture}
        \caption{$T(p_1^2,0,0;m_{12}^2,0,0)$}
         \label{fig:1mTriangle3}
    \end{subfigure}%
    ~
     \centering
    \begin{subfigure}[h]{0.33\textwidth}
        \centering
\begin{tikzpicture}
\draw (0,0) -- (1,2) -- (2,0)-- (0,0);
\draw[line width=1mm] (0,0) -- (1,2);
\draw[line width=1mm] (0,0) -- (2,0);
\draw[line width=1mm] (2,0) -- (2.5,-0.5) node[anchor=north east]{1};
\draw (0,0) -- (-0.5,-0.5) node[anchor=north west]{2};
\draw (1,2) -- (1,2.71) node[anchor=south]{3};
\node (mass) at (1,-0.37) {$m_{12}^2$};
\node (mass) at (-0.2,1.05) {$m_{23}^2$};
\draw[fill=black] (0,0) circle [radius=0.1];
\draw[fill=black] (1,2) circle [radius=0.1];
\draw[fill=black] (2,0) circle [radius=0.1];
\end{tikzpicture}
        \caption{$T(p_1^2,0,0;m_{12}^2,m_{23}^2,0)$}
         \label{fig:1mTriangle4}
    \end{subfigure}
        \\
     \centering
    \begin{subfigure}[h]{0.33\textwidth}
        \centering
\begin{tikzpicture}
\draw (0,0) -- (1,2) -- (2,0)-- (0,0);
\draw[line width=1mm] (2,0) -- (1,2);
\draw[line width=1mm] (0,0) -- (2,0);
\draw[line width=1mm] (2,0) -- (2.5,-0.5) node[anchor=north east]{1};
\draw (0,0) -- (-0.5,-0.5) node[anchor=north west]{2};
\draw (1,2) -- (1,2.71) node[anchor=south]{3};
\node (mass) at (1,-0.37) {$m_{12}^2$};
\node (mass) at (2,1.05) {$m_{13}^2$};
\draw[fill=black] (0,0) circle [radius=0.1];
\draw[fill=black] (1,2) circle [radius=0.1];
\draw[fill=black] (2,0) circle [radius=0.1];
\end{tikzpicture}
        \caption{$T(p_1^2,0,0;m_{12}^2,0,m_{13}^2)$}
         \label{fig:1mTriangle1}
    \end{subfigure}
        ~
     \centering
    \begin{subfigure}[h]{0.33\textwidth}
        \centering
\begin{tikzpicture}
\draw[line width=1mm] (0,0) -- (1,2) -- (2,0)-- (0,0);
\draw[line width=1mm] (2,0) -- (2.5,-0.5) node[anchor=north east]{1};
\draw (0,0) -- (-0.5,-0.5) node[anchor=north west]{2};
\draw (1,2) -- (1,2.71) node[anchor=south]{3};
\node (mass) at (1,-0.37) {$m_{12}^2$};
\node (mass) at (2,1.05) {$m_{13}^2$};
\node (mass) at (-0.2,1.05) {$m_{23}^2$};
\draw[fill=black] (0,0) circle [radius=0.1];
\draw[fill=black] (1,2) circle [radius=0.1];
\draw[fill=black] (2,0) circle [radius=0.1];
\end{tikzpicture}
        \caption{$T(p_1^2,0,0;m_{12}^2,m_{23}^2,m_{13}^2)$}
         \label{fig:1mTriangle5}
    \end{subfigure}
    \caption{Triangles with one external massive leg.}
        \label{oneMassTriangles}
\end{figure}
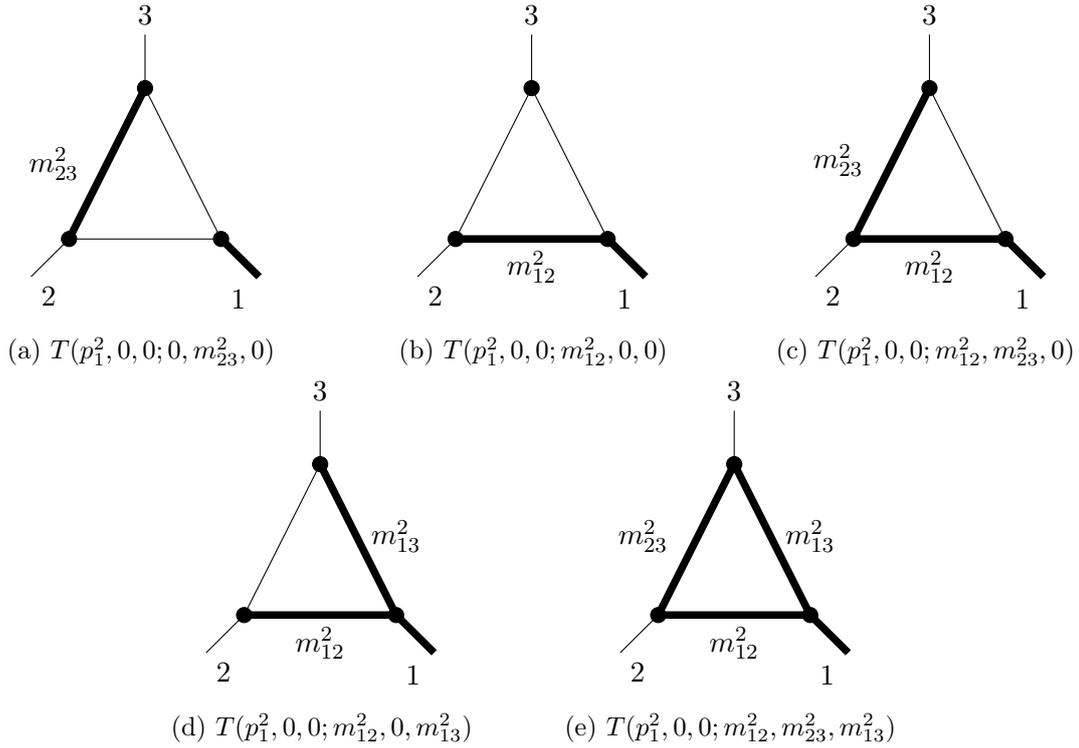

\begin{table}[h]
\begin{center}
  \begin{tabular}{ l | l }
  Triangle &  Symbol alphabet $\cA$ (up to $\mathcal{O}(\epsilon^0)$) \\
    \hline  \hline
    $T(p_1^2,0,0;0,m_{23}^2,0)$ & $ p_1^2, m_{23}^2, p_1^2+m_{23}^2 $ \\ \hline
    $T(p_1^2,0,0;m_{12}^2,0,0)$ & $ p_1^2, m_{12}^2, m_{12}^2-p_1^2 $ \\ \hline
    $T(p_1^2,0,0;m_{12}^2,m_{23}^2,0)$ & $ p_1^2, m_{12}^2, m_{23}^2, m_{12}^2-p_1^2, m_{12}^2-m_{23}^2,$\\ 
    &  $m_{12}^2-m_{23}^2-p_1^2 $ \\ \hline
 $T(p_1^2,0,0;m_{12}^2,0,m_{13}^2)$ & $w_1, \bw_1, 1-w_1, 1-\bw_1$ \\ \hline
 $T(p_1^2,0,0;m_{12}^2,m_{23}^2,m_{13}^2)$ & $w_1, \bw_1, 1-w_1, 1-\bw_1, \mu_{23}-(1-w_1)(1-\bw_1),$ \\ 
&  $     \mu_{23}, \mu_{23}-w_1 \bw_1, \mu_{23}+w_1(1-\bw_1),\mu_{23}+\bw_1(1-w_1)$ \\ \hline
    \hline
  \end{tabular}
  \end{center}
    \caption{Symbol alphabet of the triangles with one external massive leg, fig.~\ref{oneMassTriangles}.}
  \label{tab:1massTriangles}
  \end{table}

\begin{figure}[h]
        \centering
    \begin{subfigure}[h]{0.33\textwidth}
        \centering
\begin{tikzpicture}
\draw (0,0) -- (1,2) -- (2,0)-- (0,0);
\draw[line width=1mm] (0,0) -- (1,2);
\draw[line width=1mm] (0,0) -- (-0.5,-0.5) node[anchor=north east]{2};
\draw (2,0) -- (2.5,-0.5) node[anchor=north west]{1};
\draw[line width=1mm] (1,1.93) -- (1,2.71)  node[anchor=south]{3};
\node (mass) at (-0.2,1) {$m_{23}^2$};
\draw[fill=black] (0,0) circle [radius=0.1];
\draw[fill=black] (1,2) circle [radius=0.1];
\draw[fill=black] (2,0) circle [radius=0.1];
\end{tikzpicture}
        \caption{$T(0,p_2^2,p_3^2;0,m_{23}^2,0)$}
        \label{fig:2mTriangle1}
    \end{subfigure}%
    ~
 \centering
    \begin{subfigure}[h]{0.33\textwidth}
        \centering
\begin{tikzpicture}
\draw (0,0) -- (1,2) -- (2,0)-- (0,0);
\draw[line width=1mm] (0,0) -- (2,0);
\draw[line width=1mm] (0.04,0) -- (-0.5,-0.5) node[anchor=north east]{2};
\draw (2,0) -- (2.5,-0.5) node[anchor=north west]{1};
\draw[line width=1mm] (1,1.93) -- (1,2.71) node[anchor=south]{3};
\node (mass) at (1,-0.37) {$m_{12}^2$};
\draw[fill=black] (0,0) circle [radius=0.1];
\draw[fill=black] (1,2) circle [radius=0.1];
\draw[fill=black] (2,0) circle [radius=0.1];
\end{tikzpicture}
        \caption{$T(0,p_2^2,p_3^2;m_{12}^2,0,0)$}
        \label{fig:2mTriangle2}
    \end{subfigure}%
    \caption{Triangles with two external massive legs.}
    \label{twoMassTriangles}
\end{figure}
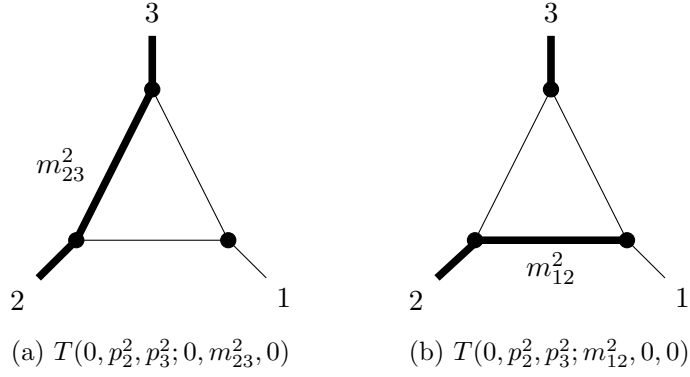

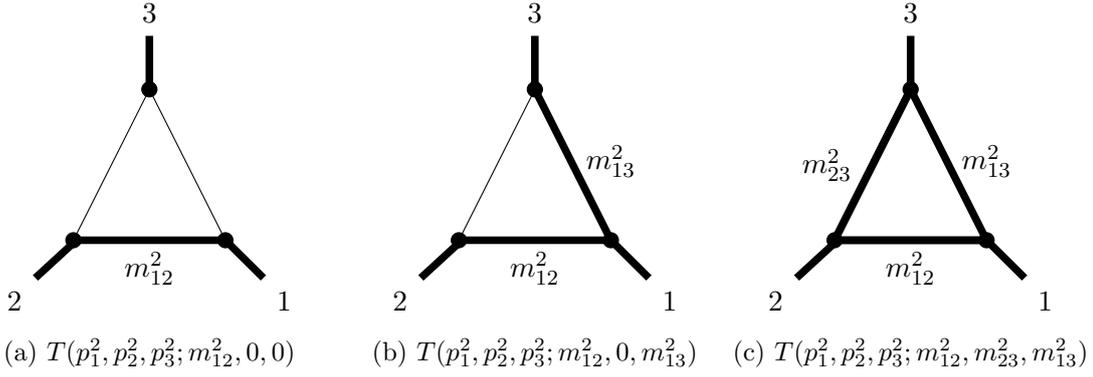
\begin{figure}[h]
    \begin{subfigure}[h]{0.31\textwidth}
        \centering
\begin{tikzpicture}        
\draw (0,0) -- (1,2) -- (2,0)-- (0,0);
\draw[line width=1mm] (0,0) -- (2,0);
\draw[line width=1mm] (0.04,0) -- (-0.5,-0.5) node[anchor=north east]{2};
\draw[line width=1mm] (2,0) -- (2.5,-0.5) node[anchor=north west]{1};
\draw[line width=1mm] (1,2) -- (1,2.71) node[anchor=south]{3};
\node (mass) at (1,-0.37) {$m_{12}^2$};
\draw[fill=black] (0,0) circle [radius=0.1];
\draw[fill=black] (1,2) circle [radius=0.1];
\draw[fill=black] (2,0) circle [radius=0.1];
\end{tikzpicture}
        \caption{$T(p_1^2,p_2^2,p_3^2;m_{12}^2,0,0)$}
        \label{fig:3mTriangle3}
    \end{subfigure}
  ~
\centering
    \begin{subfigure}[h]{0.31\textwidth}
        \centering
\begin{tikzpicture}
\draw (0,0) -- (1,2) -- (2,0)-- (0,0);
\draw[line width=1mm] (0,0) -- (2,0);
\draw[line width=1mm]  (1,2) -- (2,0);
\draw[line width=1mm] (0.04,0) -- (-0.5,-0.5) node[anchor=north east]{2};
\draw[line width=1mm] (2,0) -- (2.5,-0.5) node[anchor=north west]{1};
\draw[line width=1mm] (1,1.93) -- (1,2.71) node[anchor=south]{3};
\node (mass) at (1,-0.37) {$m_{12}^2$};
\node (mass) at (2,1.05) {$m_{13}^2$};
\draw[fill=black] (0,0) circle [radius=0.1];
\draw[fill=black] (1,2) circle [radius=0.1];
\draw[fill=black] (2,0) circle [radius=0.1];
\end{tikzpicture}
        \caption{$T(p_1^2,p_2^2,p_3^2;m_{12}^2,0,m_{13}^2)$}
        \label{fig:3mTriangle2}
    \end{subfigure}%
    ~
        \centering
    \begin{subfigure}[h]{0.31\textwidth}
        \centering
\begin{tikzpicture}
\draw[line width=1mm] (0,0) -- (1,2) -- (2,0)-- (0,0);
\draw[line width=1mm] (0,0) -- (2,0);
\draw[line width=1mm] (0,0) -- (1,2);
\draw[line width=1mm] (0.04,0) -- (-0.5,-0.5) node[anchor=north east]{2};
\draw[line width=1mm] (2,0) -- (2.5,-0.5) node[anchor=north west]{1};
\draw[line width=1mm] (1,2) -- (1,2.71)  node[anchor=south]{3};
\node (mass) at (1,-0.37) {$m_{12}^2$};
\node (mass) at (-0.1,1) {$m_{23}^2$};
\node (mass) at (2,1.05) {$m_{13}^2$};
\draw[fill=black] (0,0) circle [radius=0.1];
\draw[fill=black] (1,2) circle [radius=0.1];
\draw[fill=black] (2,0) circle [radius=0.1];
\end{tikzpicture}
        \caption{$T(p_1^2,p_2^2,p_3^2;m_{12}^2,m_{23}^2,m_{13}^2)$}
        \label{fig:3mTriangle1}
    \end{subfigure}%
        \caption{Triangles with three external massive legs.}
        \label{threeMassTriangles}
\end{figure}

We consider this triangle in cases in which  either external invariants or internal masses, or both, may become massless. See figs.~\ref{oneMassTriangles}, \ref{twoMassTriangles}, \ref{threeMassTriangles}.
Dimensional regularization is employed because of divergences in some of these cases, but in this paper we truncate the expansion at order $\epsilon^0$.
In appendices \ref{app_oneMass}, \ref{app_twoMass} and \ref{app_threeMass}, we give results for several of these functions. It is not an exhaustive list, but rather a compilation of examples that illustrates all the points we make in this paper. We have checked that the  configurations we have not listed also  satisfy the expected relations. In this section, we list the different triangles considered, define variables convenient for their analysis, and present their symbol alphabets in terms of these variables.

Our goal is to choose variables in terms of which we have a rational symbol alphabet. In cases with sufficiently few scales, we can use the invariants themselves as variables. Although the transcendental functions are only functions of ratios of the invariants, we will in general write the results explicitly in terms of the invariants. In this form, it is easy to then rewrite the result in terms of whichever dimensionless ratios are more convenient for a specific application.

\begin{table}[h]
\begin{center}
  \begin{tabular}{ l  | l }
  Triangle & Symbol alphabet $\cA$ (up to $\mathcal{O}(\epsilon^0)$) \\  \hline  \hline 
 $T(0,p_2^2,p_3^2;0,m_{23}^2,0)$  &  $ m_{23}^2, p_2^2, p_3^2, m_{23}^2-p_2^2, m_{23}^2-p_3^2$ \\ \hline
 $T(0,p_2^2,p_3^2;m_{12}^2,0,0)$  & $m_{12}^2, p_2^2, p_3^2, m_{12}^2-p_2^2, p_2^2-m_{12}^2-p_3^2 $ \\ \hline \hline
  \end{tabular}
  \caption{Symbol alphabet of the triangles with two external massive legs, fig.~\ref{twoMassTriangles}.}
  \label{tab:2massTriangles}
\end{center}
\end{table}

In more complicated cases, invariants often appear in a specific combination, the well-known K\"all\'en function, which we denote as
\begin{equation}
\lambda(a,b,c) = a^{2}+b^{2}+c^{2}-2(ab+bc+ac).
\label{eq:KallenFunction}
\end{equation}

In one-loop triangle integrals, it arises whenever three massive legs have momenta that sum to zero.  More specifically, this occurs either when all three external legs are massive, or when three massive legs meet at one of the vertices.
For these cases, the symbol letters cannot be written as rational functions of the invariants themselves, and we must thus define a different type of variable.
By convention, we will always name our external momenta so that $p_1^2$ is one of the invariants appearing as the argument of the K\"all\'en function.
Then, we define the dimensionless ratios:
\beq
\label{eq:u_i_def}
u_i = \frac{p_i^2}{p_1^2} ~,\qquad \mu_{ij} = \frac{m_{ij}^2}{p_1^2}\, .
\eeq
In terms of these dimensionless ratios, we define the following useful variables \cite{Chavez:2012kn}:
\begin{equation}\label{eq:z_def}
z=\frac{1+u_2-u_3+\sqrt{\lambda_z}}{2}~,~~
\bar{z}=\frac{1+u_2-u_3-\sqrt{\lambda_z}}{2}~,
\end{equation}
where
\begin{equation}
\lambda_z \equiv {\lambda (1,u_2,u_3)}.
\label{eq:def_lambda_z}
\end{equation}
These satisfy
\begin{equation}
z\bar{z}=u_2 ~,~~(1-z)(1-\bar{z}) =u_3~.
\end{equation}
This set of variables is useful when there are three massive external legs.

\begin{table}[h]
\begin{center}
  \begin{tabular}{ l | l  }
  Triangle & Symbol alphabet $\cA$ (up to $\mathcal{O}(\epsilon^0)$) \\  \hline  \hline 
$T(p_1^2,p_2^2,p_3^2;m_{12}^2,0,0)$&$z, \bz, 1-z, 1-\bz, \mu_{12}, 1-\mu_{12},$\\
&  $ z \bz-\mu_{12}, z-\mu_{12}, \bz-\mu_{12}$ \\ \hline
$T(p_1^2,p_2^2,p_3^2;m_{12}^2,0,m_{13}^2)$& $z, \bz, 1-z, 1-\bz, w_1, \bw_1, 1-w_1, 1-\bw_1,$   \\ 
 &$ z \bz - w_1 \bw_1,w_1-z, w_1-\bz, \bw_1-z, \bw_1-\bz,$\\
&    $(1-z)(1-\bz)-(1-w_1)(1-\bw_1) $\\ \hline
$T(p_1^2,p_2^2,p_3^2;m_{12}^2,m_{23}^2,m_{13}^2)$&$ z,  1-z, w_1, \bw_1,  1-w_1, 1-\bw_1,  \mu_{23},$ \\ 
 &   $w_1 \bar{w}_1-z \bar{z}+\mu_{23}-\sqrt{\lambda_2},$\\
& $   (1-z)(1-\zbar)-(1-w_1)(1- \bar{w}_1)-\mu_{23}+\sqrt{\lambda _3},$ \\ 
   &    $(w_1-z)(\bw_1-\zbar)-\mu_{23},
    ~ (w_1-\zbar)(\bw_1-z)-\mu_{23},$ \\
    &  $(z\zbar+w_1\bar{w}_1-\mu_{23}) (z + \bar{z})-2  z \bar{z}(w_1+\bw_1) \pm(z-\bz) \sqrt{\lambda_2},$\\ 
     & $ z^2(1-\zbar)+\zbar^2(1-z)+(w_1+\bw_1)(2 z\bz-z-\bz)$\\
     & $\quad\qquad\qquad\qquad+(\mu_{23}-w_1\bw_1)(z+\bz-2) \pm (z-\bz)\sqrt{\lambda_3} $
 \\ \hline \hline
  \end{tabular}
  \caption{Symbol alphabet of the triangles with three external massive legs, fig.~\ref{threeMassTriangles}.}
  \label{tab:3massTriangles}
  \end{center}
\end{table}

Similarly, when three massive legs meet at one vertex, we define:
\begin{align}
{w_1} &= \frac{1+\mu_{12}-\mu_{13}+\sqrt{\lambda_1}}{2} ~, ~
\bar{w}_1 = \frac{1+\mu_{12}-\mu_{13}-\sqrt{\lambda_1}}{2} ~,
\label{eq:zw1_variables_def}
\end{align}
where
\begin{equation}
\lambda_1 \equiv \lambda (1,\mu_{12},\mu_{13}).
\label{eq:def_lambda_1}
\end{equation}
These satisfy
\begin{equation}
{w_1} \bar{w}_1 = \mu_{12} ~, \qquad (1-{w_1}) (1-\bar{w}_1) = \mu_{13} ~.
\label{eq:mus_in_terms_of_w1s}
\end{equation}

In terms of these variables, we find rational symbols for all triangles except the triangle with three internal and three external masses, $T(p_1^2,p_2^2,p_3^2;m_{12}^2,m_{23}^2,m_{13}^2)$, for which we fail to find a parametrization leading to a rational symbol alphabet.\footnote{The case $T(p_1^2,p_2^2,0;m_{12}^2,m_{23}^2,m_{13}^2)$, not given as an example, requires a simple generalization of the variables presented here in order to produce a rational symbol alphabet.} Indeed, if we use a minimal set of 5 independent dimensionless variables given by $z, \bz, w_1, \bw_1, \mu_{23}$, as defined above, the symbol alphabet includes square roots of the K\"all\'en functions,
\bea
&& \lambda_2 \equiv  \lambda(z \bz , w_1 \bw_1 ,\mu_{23}), \quad \lambda_3 \equiv \lambda((1-z) (1-\bz) , (1-w_1) (1-\bw_1) ,\mu_{23}).
\label{eq:lambda3def}
\eea
The variables above are appropriate for the cut in $p_1^2$, which explains the absence of non-rational factors in the first term of \refE{eq:symbT33masses}, and the presence of square roots in the remaining terms.
Interestingly, as we show in section \ref{tp1p2p3m12m23m13}, any given cut integral computation suggests natural variables giving a rational result.  The difficulty is that each cut will suggest a different set of variables, and the variables of one cut are not rational in terms of the variables of another.

We draw the diagrams for our examples in figures \ref{oneMassTriangles}, \ref{twoMassTriangles} and \ref{threeMassTriangles}, and we give a minimal symbol alphabet of each of these examples in tables \ref{tab:1massTriangles}, \ref{tab:2massTriangles} and \ref{tab:3massTriangles}.

%% file: Sec_FirstEntry.tex
\section{The first-entry condition}
\label{sec:firstentry}

We now extend the first-entry condition \cite{Gaiotto:2011dt} to cases with internal masses. Our observations suggest that the coproduct can always be written in such a form where: {\it i)} the first entries of the coproduct component $\Delta_{1,n-1}$ are either consistent with the thresholds of Mandelstam invariants or are internal masses themselves; {\it ii)} the corresponding second entry is the discontinuity across the branch point in the first entry, as is the case for diagrams with no internal masses  \cite{Abreu:2014cla}.\\

Feynman integrals are most easily computed in the kinematic region of the invariants where the integral is well-defined independently of the $\pm i0$-prescription of the propagators, see e.g.~\cite{Landau:1959fi}. In this region, the {\it euclidean region}, we are away from any branch cut. In the most general case considered in this paper, \refE{eq:def_scalar_triangle}, the euclidean region is characterized by the following conditions:
\begin{align}\bsp
p_1^2<&\left(\sqrt{m_{12}^2}+\sqrt{m^2_{13}}\right)^2,\quad p_2^2<\left(\sqrt{m^2_{12}}+\sqrt{m^2_{23}}\right)^2,\quad p_3^2<\left(\sqrt{m^2_{13}}+\sqrt{m^2_{23}}\right)^2 ,\\
 &\qquad \qquad \qquad m_{12}^2>0\,,\quad\qquad m_{23}^2>0\,,\quad\qquad m_{13}^2>0\,.
\esp\end{align}
For all of our other examples, it can be obtained from the above by taking the appropriate limit. For instance, in the absence of internal masses, the euclidean region is the region where all external invariants are negative. As we depart from this region, we are sensitive to branch cuts of the integral. The $\pm i0$-prescription indicates which side of the branch cut we are on. Comparing results computed with different prescriptions gives the discontinuities across the branch cuts.

The coproduct structure of multiple polylogarithms provides a natural framework for analyzing their discontinuities.  Specifically, it was argued in~\cite{Duhr:2012fh} that the discontinuity acts only on the first entry of the coproduct, leaving the rest alone:
\beq\label{eq:disc_coproduct}
\Delta\,\textrm{Disc} = (\textrm{Disc}\otimes\textrm{id})\,\Delta\,.
\eeq
To be concrete, our precise definition of $\Disc$ is
\bea
\Disc_x \left[F(x\pm i 0)\right]= \lim_{\varepsilon \to 0}\left[F(x\pm i \varepsilon)-F(x \mp i \varepsilon)\right].
\label{eq:def-disc}
\eea
Here, $x$ represents either a Mandelstam invariant or an internal mass invariant (or indeed, any other kinematic variable). 
If there is no branch cut in the kinematic region being considered, or if $F$ does not depend on $x$, then the discontinuity is zero.  For instance, the discontinuity in a Mandelstam invariant exists only in the region above its threshold. The choice of the sign in $\pm i 0$ is inherited from the prescription of the propagators of the Feynman integral. Finally, for later use, we also define a sequential discontinuity operator $\Disc_{x_1,\ldots,x_k}$ as:
\bea
\Disc_{x_1,\ldots,x_k} F &\equiv& \Disc_{x_k} \left( \Disc_{x_1,\ldots,x_{k-1}} F \right),
\label{eq:disc-seq}
\eea
where the $x_i$ are associated either with internal masses or with external momentum invariants.

We note that \refE{eq:disc_coproduct} implies that the first entries of the coproduct tensor of a Feynman diagram must have the same branch cut structure as the Feynman diagram itself. In particular, when looking at the $\Delta_{1,n-1}$ component, the first entries must be simple logarithms with branch points at the boundaries of the euclidean region.

This observation, known as the  {\em first entry condition}, was first formulated in the context of integrals with massless propagators, where the first entries of their symbol can be written as logarithms of Mandelstam invariants~\cite{Gaiotto:2011dt}.

In the presence of massive propagators, we find two ways in which the first entry condition generalizes. The first is that we no longer have logarithms of Mandelstam invariants themselves, but instead logarithms with branch cuts at the mass threshold for the corresponding invariant. The second is that the squared masses of the propagators themselves appear as first entries. Both modifications are consistent with our observations above.

As predicted by \refE{eq:disc_coproduct} and already observed in the absence of internal masses \cite{Abreu:2014cla}, the second entries correspond to the discontinuities associated to the branch cut identified in the corresponding first entry. We observe the same behavior when internal masses are present, the new feature being the existence of discontinuities associated to the internal masses themselves.

In the context of the examples of this paper, we observe that when we use the Mandelstam invariants themselves (or ratios of them, if we want to work with dimensionless quantities) as variables, the above properties are very easy to check. In more complicated cases where we need to change variables to have a rational symbol alphabet, the situation is not as clear: even though the simplest expression for the symbol (or, equivalently, the coproduct) does not satisfy these properties, we can always rearrange the different terms of the symbol tensor so that they are in the form described above.

\paragraph{Example 1:}
Consider the triangle $T(p_1^2,0,0;m_{12}^2,0,0)$, whose symbol is given in \refE{eq:symbol-t-p1-m12}. The first term has $\log(m_{12}^2)$ as its first entry, and the second term has $\log(m_{12}^2-p_1^2)$ as its first entry.  The latter is written in a form in which the argument of the logarithm is positive in the euclidean region where the integral is originally evaluated.

\paragraph{Example 2:}
As another example, consider the triangle $T(p_1^2,0,0;m_{12}^2,0,m_{13}^2)$, whose symbol is given in \refE{eq:symbol-t-p1-m12-m13}.  We have changed variables according to \refE{eq:zw1_variables_def} to have a rational symbol alphabet. Because the new variables have a more complicated relation to the Mandelstam invariants, the first entry condition is not as transparent as in the previous example.  
However, as mentioned above, the symbol of ${\cal T}(p_1^2,0,0;m_{12}^2,0,m_{13}^2)$ can be rewritten as
\begin{align}\bsp
& \left(w_1 \bar{w}_1\right)\otimes\frac{{w}_1}{1-{w}_1}
+\left((1-w_1)(1-\bar{w}_1)\right)\otimes\frac{1-\bar{w}_1}{\bar{w}_1}
 +\left(w_1(1-\bar{w}_1)\right)\otimes\frac{\bar{w}_1(1-w_1)}{w_1(1-\bar{w}_1)} \, .
\label{eq:firstentry-alphabet-example}
\esp\end{align}
The first entry of the first term is $\log(m_{12}^2/p_1^2)$, and its second entry is associated with the discontinuity in the variable $m_{12}^2$.  The first entry of the second term is $\log(m_{13}^2/p_1^2)$, and its second entry is associated with the discontinuity in the variable $m_{13}^2$.  The first entry of the third term corresponds to the threshold at $p_1^2=(\sqrt{m_{12}^2}+\sqrt{m_{13}^2})^2$, and its second entry is associated with the discontinuity in the variable $p_1^2$.  The  argument of the logarithm in the first entry of this term is not a direct change of variables of $p_1^2-(\sqrt{m_{12}^2}+\sqrt{m_{13}^2})^2$, which would not be a rational function.  Nevertheless, one can verify that the condition $w_1(1-\bar{w}_1)>0$ is exactly equivalent to the condition $p_1^2>(\sqrt{m_{12}^2}+\sqrt{m_{13}^2})^2$, whenever $p_1^2>0$.  We can understand this latter condition as a weaker prerequisite for obtaining any discontinuity in the variable $p_1^2$.

In this analysis, we have neglected the denominators $p_1^2$ in each of the first entries of \refE{eq:firstentry-alphabet-example}.  We did this because the denominators are simply used to normalize the variables to be dimensionless:  as we know, the physically meaningful first entry is the one including the mass threshold, which is nonzero for the $p_1^2$ channel.  Indeed, since the three second entries sum to zero, we see that the term whose first entry would be $p_1^2$ has zero as its second entry.

We finish with a word of caution about \refE{eq:firstentry-alphabet-example}.  In this triangle, the variables $w_1$ and $\bar{w}_1$ are distinguished only by the choice of branch of the square root.  Like the function itself, the symbol has no preferred branch, and therefore it is invariant under the exchange $w_1 \leftrightarrow \bar{w}_1$.  This invariance is apparent in the form given in \refE{eq:symbol-t-p1-m12-m13} but obscure in  \refE{eq:firstentry-alphabet-example}.  Notably, the three discontinuities, as read from \refE{eq:firstentry-alphabet-example}, are not themselves invariant under this exchange.  When relating these discontinuities to cut integrals, we will be very specific about the kinematics and insist on taking the positive branch of the square root.  From that point of view, the form given in \refE{eq:firstentry-alphabet-example} will be necessarily preferred over its conjugate under $w_1 \leftrightarrow \bar{w}_1$.\footnote{Whenever we make use of the first entry condition (see section \ref{sec:reconstruction}), we will therefore always take $w_1(1-\bar{w}_1)$ for the $p_1^2$ channel, and not $\bar{w}_1(1-w_1)$.}  In the following, we will relate discontinuities to operations on individual symbol letters, and not particular combinations of them.  Thus, while we claim that one can generally write symbols of Feynman integrals in a form such as \refE{eq:firstentry-alphabet-example}, where the first entries are directly identified with kinematic invariants and thresholds, one should not immediately conclude that the corresponding second entries are unambiguously interpreted as discontinuities.

%% file: Sec_CutRules.tex
\section{Two types of cuts}
\label{sec:two_types_of_cuts}

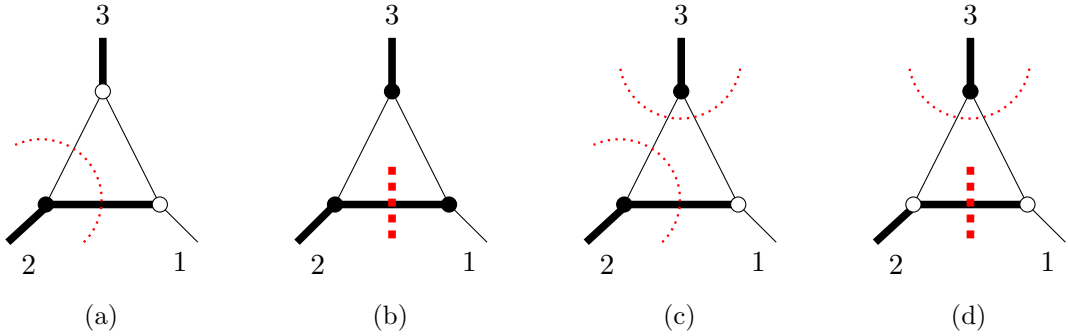
\begin{figure}
\begin{subfigure}[t]{0.245\textwidth}
\centering
\begin{tikzpicture}        
\draw (0,0) -- (0.75,1.5) -- (1.5,0)-- (0,0);
\draw[line width=1mm] (0,0) -- (1.5,0);
\draw[line width=1mm] (0.04,0) -- (-0.5,-0.5) node[anchor=north west]{2};
\draw (1.5,0) -- (2,-0.5) node[anchor=north east]{1};
\draw[line width=1mm] (.75,1.5) -- (.75,2.21) node[anchor=south]{3};
\draw[fill=black] (0,0) circle [radius=0.1];
\draw[fill=white] (1.5,0) circle [radius=0.1];
\draw[fill=white] (.75,1.5) circle [radius=0.1];
\draw[dotted, red, line width=0.3mm] (0.5,-0.5)  arc[radius = 8mm, start angle= -45, end angle= 115];
\end{tikzpicture}
\caption{ }
        \label{fig:p2cutTp2p3m12}
\end{subfigure}
\begin{subfigure}[t]{0.245\textwidth}
        \centering
\begin{tikzpicture}
\draw (0,0) -- (0.75,1.5) -- (1.5,0) -- (0,0);
\draw[line width=1mm] (1.5,0) -- (0,0);
\draw (1.5,0) -- (2,-0.5) node[anchor=north east]{1};
\draw[line width=1mm] (0,0) -- (-0.5,-0.5) node[anchor=north west]{2};
\draw[line width=1mm] (0.75,1.5) -- (0.75,2.21) node[anchor=south]{3};
\draw[fill=black] (0,0) circle [radius=0.1];
\draw[fill=black] (0.75,1.5) circle [radius=0.1];
\draw[fill=black] (1.5,0) circle [radius=0.1];
\draw[dashed, red, line width=1 mm] (.75,0.5) -- (.75,-0.5);
\end{tikzpicture}
\caption{ }
\label{fig:m12cutTp2p3m12}
\end{subfigure}
\begin{subfigure}[t]{0.245\textwidth}
\centering
\begin{tikzpicture}
\draw (0,0) -- (.75,1.5) -- (1.5,0)-- (0,0);
\draw[line width=1mm] (0,0) -- (1.5,0);
\draw[line width=1mm] (0.04,0) -- (-0.5,-0.5) node[anchor=north west]{2};
\draw (1.5,0) -- (2,-0.5) node[anchor=north east]{1};
\draw[line width=1mm] (.75,1.5) -- (.75,2.21) node[anchor=south]{3};
\draw[fill=black] (0,0) circle [radius=0.1];
\draw[fill=white] (1.5,0) circle [radius=0.1];
\draw[fill=black] (.75,1.5) circle [radius=0.1];
\draw[dotted, red, line width=0.3mm] (0.5,-0.5)  arc[radius = 8mm, start angle= -45, end angle= 115];
\draw[dotted, red, line width=0.3mm] (-0.05,1.8)  arc[radius = 8mm, start angle= -170, end angle= -5];
\end{tikzpicture}
\caption{ }
\label{fig:p2p3cutTp2p3m12}
\end{subfigure}
    \begin{subfigure}[t]{0.245\textwidth}
        \centering
\begin{tikzpicture}
\draw (0,0) -- (.75,1.5) -- (1.5,0)-- (0,0);
\draw[line width=1mm] (0,0) -- (1.5,0);
\draw[line width=1mm] (0.04,0) -- (-0.5,-0.5) node[anchor=north west]{2};
\draw (1.5,0) -- (2,-0.5) node[anchor=north east]{1};
\draw[line width=1mm] (.75,1.5) -- (.75,2.21)  node[anchor=south]{3};
\draw[fill=white] (0,0) circle [radius=0.1];
\draw[fill=white] (1.5,0) circle [radius=0.1];
\draw[fill=black] (.75,1.5) circle [radius=0.1];
\draw[dotted, red, line width=0.3mm] (-0.05,1.8)  arc[radius = 8mm, start angle= -170, end angle= -5];
\draw[dashed, red, line width=1 mm] (.75,0.5) -- (.75,-0.5);
\end{tikzpicture}
        \caption{ }
        \label{fig:p3m12cutTp2p3m12}
    \end{subfigure}
\caption{Example of cuts of $T(0,p_2^2,p_3^2;m_{12}^2,0,0)$: (a) single cut in an external channel; (b) single cut in an internal mass; (c) double cut in external channels; (d) double cut in an external channel and an internal mass. Thin dotted lines correspond to cuts on external channels and imply complex conjugation of a region of the diagram. Thick dashed lines correspond to mass cuts and do not imply any complex conjugation.}
\label{cutsTp2p3m12}
\end{figure}

Feynman diagrams with internal masses have discontinuities associated with both the external massive channels and the internal masses. These two types of discontinuities correspond to two slightly different types of diagrammatic cuts. The first type are the well-known cuts in external channels \cite{tHooft:1973pz,Veltman:1994wz,Abreu:2014cla}, and the second type are single-propagator cuts for which we give the rules in this paper. We start by reviewing the rules for cuts in external channels, then introduce single-propagator cuts, and finally explain our strategy to compute the cut diagrams for our examples. In both cases, our cutting rules are designed to reproduce the discontinuities in the corresponding variables. This will allow us to compute all types of single and double cuts considered in this paper.  See \refF{cutsTp2p3m12} for examples.

\subsection{Cut in a kinematic channel}\label{sec:cutChannel}

We start by discussing cuts in external channels $s_i$. The operator $\Cut_{s_i}$ gives a \emph{cut} Feynman integral, in which some propagators, the \emph{cut propagators}, are replaced by Dirac delta functions. The cut separates the diagram into two  parts, with the momentum flowing through the cut propagators from one part to the other corresponding to the Mandelstam invariant $s_i$.   Each cut is associated with a consistent direction of energy flow between the two parts of the diagram, in each of the cut propagators.
We briefly review the rules established in refs.~\cite{tHooft:1973pz,Veltman:1994wz} for single cuts, and generalized in ref.~\cite{Abreu:2014cla} for sequential cuts.

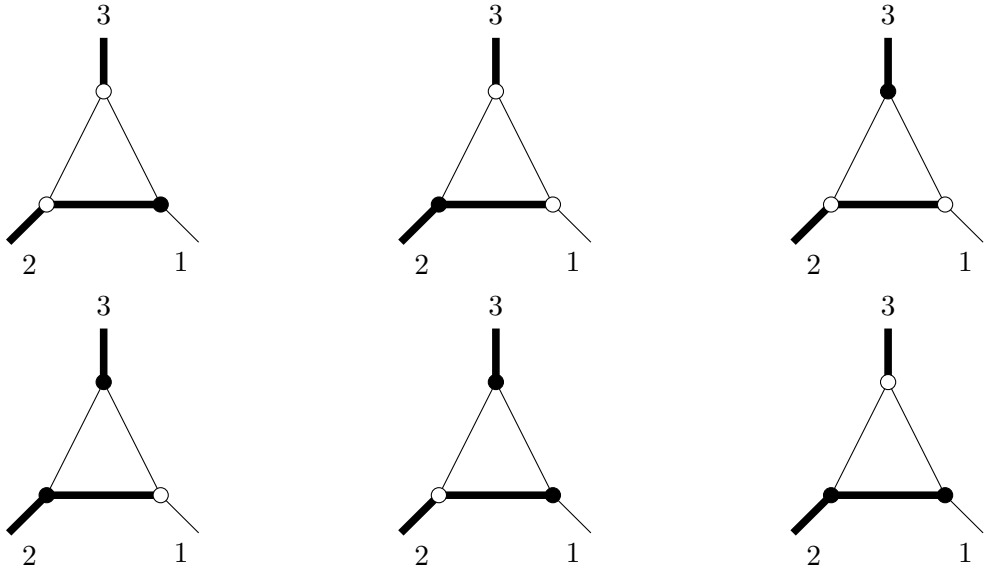
\begin{figure}
\centering
    \begin{subfigure}[t]{0.33\textwidth}
        \centering
\begin{tikzpicture}
\draw (0,0) -- (0.75,1.5) -- (1.5,0) -- (0,0);
\draw[line width=1mm] (1.5,0) -- (0,0);
\draw (1.5,0) -- (2,-0.5) node[anchor=north east]{1};
\draw[line width=1mm] (0,0) -- (-0.5,-0.5) node[anchor=north west]{2};
\draw[line width=1mm] (0.75,1.5) -- (0.75,2.21) node[anchor=south]{3};
\draw[fill=white] (0,0) circle [radius=0.1];
\draw[fill=white] (0.75,1.5) circle [radius=0.1];
\draw[fill=black] (1.5,0) circle [radius=0.1];
\end{tikzpicture}
    \end{subfigure}%
      ~
     \centering
    \begin{subfigure}[t]{0.33\textwidth}
        \centering
\begin{tikzpicture}
\draw (0,0) -- (0.75,1.5) -- (1.5,0) -- (0,0);
\draw[line width=1mm] (1.5,0) -- (0,0);
\draw (1.5,0) -- (2,-0.5) node[anchor=north east]{1};
\draw[line width=1mm] (0,0) -- (-0.5,-0.5) node[anchor=north west]{2};
\draw[line width=1mm] (0.75,1.5) -- (0.75,2.21) node[anchor=south]{3};
\draw[fill=black] (0,0) circle [radius=0.1];
\draw[fill=white] (0.75,1.5) circle [radius=0.1];
\draw[fill=white] (1.5,0) circle [radius=0.1];
\end{tikzpicture}
    \end{subfigure}%
    ~
    \begin{subfigure}[t]{0.33\textwidth}
        \centering
\begin{tikzpicture}
\draw (0,0) -- (0.75,1.5) -- (1.5,0) -- (0,0);
\draw[line width=1mm] (1.5,0) -- (0,0);
\draw (1.5,0) -- (2,-0.5) node[anchor=north east]{1};
\draw[line width=1mm] (0,0) -- (-0.5,-0.5) node[anchor=north west]{2};
\draw[line width=1mm] (0.75,1.5) -- (0.75,2.21) node[anchor=south]{3};
\draw[fill=white] (0,0) circle [radius=0.1];
\draw[fill=black] (0.75,1.5) circle [radius=0.1];
\draw[fill=white] (1.5,0) circle [radius=0.1];
\end{tikzpicture}
    \end{subfigure}%
        \\
        \centering
    \begin{subfigure}[t]{0.33\textwidth}
        \centering
\begin{tikzpicture}
\draw (0,0) -- (0.75,1.5) -- (1.5,0) -- (0,0);
\draw[line width=1mm] (1.5,0) -- (0,0);
\draw (1.5,0) -- (2,-0.5) node[anchor=north east]{1};
\draw[line width=1mm] (0,0) -- (-0.5,-0.5) node[anchor=north west]{2};
\draw[line width=1mm] (0.75,1.5) -- (0.75,2.21) node[anchor=south]{3};
\draw[fill=black] (0,0) circle [radius=0.1];
\draw[fill=black] (0.75,1.5) circle [radius=0.1];
\draw[fill=white] (1.5,0) circle [radius=0.1];
\end{tikzpicture}
    \end{subfigure}%
        ~
   \centering
    \begin{subfigure}[t]{0.33\textwidth}
        \centering
\begin{tikzpicture}
\draw (0,0) -- (0.75,1.5) -- (1.5,0) -- (0,0);
\draw[line width=1mm] (1.5,0) -- (0,0);
\draw (1.5,0) -- (2,-0.5) node[anchor=north east]{1};
\draw[line width=1mm] (0,0) -- (-0.5,-0.5) node[anchor=north west]{2};
\draw[line width=1mm] (0.75,1.5) -- (0.75,2.21) node[anchor=south]{3};
\draw[fill=white] (0,0) circle [radius=0.1];
\draw[fill=black] (0.75,1.5) circle [radius=0.1];
\draw[fill=black] (1.5,0) circle [radius=0.1];
\end{tikzpicture}
    \end{subfigure}%
    ~
     \centering
    \begin{subfigure}[t]{0.33\textwidth}
        \centering
\begin{tikzpicture}
\draw (0,0) -- (0.75,1.5) -- (1.5,0) -- (0,0);
\draw[line width=1mm] (1.5,0) -- (0,0);
\draw (1.5,0) -- (2,-0.5) node[anchor=north east]{1};
\draw[line width=1mm] (0,0) -- (-0.5,-0.5) node[anchor=north west]{2};
\draw[line width=1mm] (0.75,1.5) -- (0.75,2.21) node[anchor=south]{3};
\draw[fill=black] (0,0) circle [radius=0.1];
\draw[fill=white] (0.75,1.5) circle [radius=0.1];
\draw[fill=black] (1.5,0) circle [radius=0.1];
\end{tikzpicture}
    \end{subfigure}  %
    \caption{All possible cut diagrams of $T(0,p_2^2,p_3^2;m_{12}^2,0,0)$. A diagram in the top row and its corresponding diagram in the bottom row are associated to the same momentum channel, but opposite energy flow. }
        \label{fig:ColoringsOfTp2p3m12UnCut}
\end{figure}

\paragraph{First cut.} We enumerate all possible partitions of the vertices of a Feynman diagram into two nonempty sets, distinguished by being colored black or white; see fig.~\ref{fig:ColoringsOfTp2p3m12UnCut} for an example. Each diagram is evaluated according to the following rules for scalar theory.  (See appendix \ref{app_FeynmanRules} for a summary of our conventions.)
 
\begin{itemize}
\item Black vertices, and propagators joining two black vertices, are computed according to the usual Feynman rules.
\item White vertices, and propagators joining two white vertices, are complex-conjugated with respect to the usual Feynman rules.  
\item Propagators joining a black and a white vertex are \emph{cut}. They are replaced by an on-shell delta function, together with a factor of $2 \pi$ to capture the complex residue correctly, {\it and} a theta function restricting energy to flow in the direction 
{\em from} black {\em to} white. Stated precisely, when cut, a (complex conjugated) propagator is replaced according to
\begin{equation}
\frac{\pm i}{p^2-m^2\pm i0}\longrightarrow 2\pi\, \delta\left(p^2-m^2\right)\theta\left(p_0\right),
\end{equation}
where we assume that the momentum vector $p$ is directed from the black to the white vertex.
\end{itemize}

$\cut_{s_i}$ denotes the sum of all diagrams that isolate the channel ${s_i}$: if $p$ is the sum of the momenta flowing through cut propagators from black to white, then $p^2=s_i$. Although cut diagrams in a given momentum channel appear in pairs that are black/white color reversals (see fig.~\ref{fig:ColoringsOfTp2p3m12UnCut}) only one can be consistent with the energies of the fixed external momenta.

We stress that these rules require that one of the two regions of the diagram (the white region in our conventions) must be computed with complex conjugated Feynman rules. In particular, this implies that the $\pm i0$ of the invariants in the white region is reversed.

Consider a multiple-channel cut, $\Cut_{s_1, \ldots, s_k}F$.  It is represented by a diagram with a color-partition of vertices for each of the cut invariants $s_i=p_i^2$.  Assign a sequence of colors $(c_1(v),\ldots,c_k(v))$ to each vertex $v$  of the diagram, where each $c_i$ takes the value 0 or 1.  For a given $i$, the colors $c_i$ partition the vertices into two sets, such that the total momentum flowing from vertices labeled 0 to vertices labeled 1 is equal to $p_i$.  A vertex $v$ is finally colored according to $c(v) \equiv \sum_{i=1}^k c_i(v)$ modulo 2, with black for $c(v)=0$ and white for $c(v)=1$.  The rules for evaluating a diagram are as follows (see appendix \ref{app_FeynmanRules} for a summary of our conventions):
\begin{itemize}
\item Black vertices are computed according to the usual Feynman rules; white vertices are computed according to complex-conjugated Feynman rules.
\item A propagator joining vertices $u$ and $v$ is uncut if $c_i(u)=c_i(v)$ for all $i$.  If the vertices are black, i.e. $c(u)=c(v)=0$, then the propagator is computed according to the usual Feynman rules, and if the vertices are white, i.e. $c(u)=c(v)=1$, then the propagator is computed according to complex-conjugated Feynman rules.
\item A propagator joining vertices $u$ and $v$ is cut if $c_i(u) \neq c_i(v)$ for any $i$.  There is a theta function restricting the direction of energy flow from 0 to 1 for each $i$ for which $c_i(u) \neq c_i(v)$.  If different cuts impose conflicting energy flows, then the product of the theta functions is zero and the diagram gives no contribution.
\end{itemize}

We also restrict ourselves to real kinematics, both for the external and internal momenta. As a consequence, diagrams with massless on-shell three-point vertices vanish in dimensional regularization. Furthermore, the rules exclude crossed cuts and sequential cuts in which the channels are not all distinct. These last two restrictions are mentioned for completeness only, as they are not relevant in the examples studied in this paper. We refer the reader to ref.~\cite{Abreu:2014cla} for more details and examples of this set of rules.  We give an example of a double cut on two external channels in \refF{fig:p2p3cutTp2p3m12}.

For the examples considered in this paper, cut diagrams are computed in the region where the cut invariants are above their respective thresholds, all other consecutive Mandelstam invariants are below threshold, and the squares of all internal masses are positive.

\subsection{Cut in an internal mass}\label{sec:cutMass}

To give a complete description of the coproduct of diagrams with internal masses, we must introduce a new kind of cut, a single-propagator cut, corresponding to discontinuities across branch cuts related to the internal masses. Our discussion will be in the context of one-loop diagrams, but this is solely for the simplicity of the expressions, and all the results can be straightforwardly generalized to the multi-loop case.

Let $F$ be a one-loop planar diagram with $n$ external legs of momentum $p_i$, for $i=1,\ldots,n$, all incoming, massive or not, and with internal masses $m^2_{i,i+1}$ between legs $i$ and $i+1$, which we assume are all distinct. Furthermore, we define $q_j=\sum_{i=1}^jp_i$, for $j=1,\ldots,n$, so that $q_n=0$. Then, according to our Feynman rules,
\begin{equation}
F(q_i\cdot q_j;m_{1,2}^2,\ldots,m_{1,n}^2)=(-1)^n e^{\gamma_E \epsilon}\int \frac{d^{4-2\epsilon}k}{\pi^{2-\epsilon}}\prod_{i=1}^{n}\frac{1}{(k+q_i)^2-m^2_{i,i+1}+i0}\, .
\end{equation}

The integral $F$ is evaluated away from any branch cut in the euclidean region of the Mandelstam invariants, and for all $m^2_{i,i+1}>0$.
 In the same way that the Mandelstam invariants inherit an $+i0$ prescription from the form of the propagators, we can associate a $-i0$ prescription to the masses:
\begin{equation*}
m^2_{i,i+1} \to m^2_{i,i+1}-i0\,.
\end{equation*}

Although it does not correspond to a physical region, we can analytically continue $F$ to a region where the square of one of the masses is negative (without loss of generality, say $m_{1,n}^2<0$), while keeping all the other squared masses positive and the Mandelstam invariants in the euclidean region. In this region, we isolate the discontinuity associated with $m_{1,n}^2$:
\begin{align}\bsp\label{cutDefMassive}
\Disc_{m_{1,n}^2}F&=F(q_i\cdot q_j;m_{1,2}^2,\ldots,m_{1,n}^2-i0)-F(q_i\cdot q_j;m_{1,2}^2,\ldots,m_{1,n}^2+i0)\\
&=(-1)^ne^{\gamma_E \epsilon}\int \frac{d^{4-2\epsilon}k}{\pi^{2-\epsilon}}\left(\frac{1}{k^2-m_{1,n}^2+i0}-\frac{1}{k^2-m_{1,n}^2-i0}\right)\prod_{i=1}^{n-1}\frac{1}{(k+q_i)^2-m^2_{i,i+1}}\\
&=(-1)^{n+1}e^{\gamma_E \epsilon}\int  \frac{d^{4-2\epsilon}k}{\pi^{2-\epsilon}}(2\pi i)\delta(k^2-m_{1,n}^2)\prod_{i=1}^{n-1}\frac{1}{(k+q_i)^2-m^2_{i,i+1}}\\
&\equiv \cut_{m_{1,n}^2}F\, ,
\esp\end{align}
which shows that mass discontinuities do indeed correspond to single-particle cuts. We again stress that although we are discussing one-loop integrals, this is just for simplicity of the expressions. The same result holds for a multi-loop diagram.

Furthermore, we notice that $F$ can also be a cut Feynman diagram as long as the propagator with mass $m^2_{1,n}$ has not been cut previously. Cuts in internal masses can then be combined with cuts in external channels to compute sequential discontinuities in internal masses and external channels.

We can thus deduce the rules for single-propagator cuts, corresponding to mass discontinuities: we simply replace the cut propagator by a delta function, according to
\begin{equation}
\frac{\pm i}{p^2-m^2\pm i 0}\rightarrow 2 \pi\delta(p^2-m^2)\, ,
\end{equation}
without any condition on the energy flow or any further conjugation of other parts of the diagram. Unlike cuts in kinematic channels, the black and white colorings are unaffected by these cuts, as there is no notion of separation into two regions where one is complex-conjugated.

\subsection{Calculation of cut diagrams}\label{sec:cutCalc}

We now outline our strategy for the calculation of the cuts of the one-loop three-point functions studied in this paper. For cuts in external channels, it is a simple generalization of what is done in ref.~\cite{Abreu:2014cla}, so we will be very brief. For the single-propagator cuts, we present two alternative methods. All cuts given in appendices \ref{app_oneMass}, \ref{app_twoMass} and \ref{app_threeMass} were obtained through the methods described here.

\paragraph{Cuts in external channels.}
When computing a single cut in the channel $p_i^2$, we work in the region where $p_i^2$ is above its threshold, all other external channels are below threshold, and all masses are positive. We parametrize the external momenta as
\begin{align}\bsp
&p_i = \sqrt{p_i^2}(1,0,{\bf 0}_{D-2}), \qquad p_j = \sqrt{p_j^2}\left(\alpha,\sqrt{\alpha^2-1},{\bf 0}_{D-2}\right),
\esp\label{paramMom}
\end{align}
where $\alpha$ is trivial to determine in terms of the kinematic variables.

We route the loop momentum so that the propagators of momentum $k$ and $(p_i-k)$ are cut, and if possible the propagator of momentum $k$ is massless. We parametrize $k$ as
\begin{equation}
k =  k_0(1,\beta \cos \th,\beta\sin\th ~{\bf 1}_{D-2})\,,
\end{equation}
where $\th \in [0,\pi]$, and $k_0,\beta>0$, and ${\bf 1}_{D-2}$ ranges over unit vectors in the dimensions transverse to $p_i$ and $p_j$.
If the propagator of momentum $k$ is massless, then $\beta=1$.

Using the delta function that puts the propagator of momentum $k$ on-shell, the integration measure becomes
\begin{equation}\label{intMeasure}
\int d^{4-2\epsilon}k\,\delta (k^2-m^2) \theta (k_0)=\frac{2^{1-2\epsilon}\pi^{1-\epsilon}}{\Gamma(1-\epsilon)}\int_0^\infty dk_0 \left(\sqrt{k_0^2-m^2}\right)^{1-2\epsilon}\int_0^1 dx (1-x)^{-\epsilon}x^{-\epsilon}\, ,
\end{equation}
in the most general case we need to consider for this paper. The $k_0$ integral can be trivially performed using the delta function putting the propagator of momentum $(p_i-k)$ on-shell.

The remaining uncut propagator, of momentum $(p_j+k)$, is linear in the $x$ variable, and so the most complicated result we will get for the single cut of a one-loop three-point function can be written  to all orders in $\epsilon$ as a Gauss hypergeometric function, \refE{gaussHypDef}, as can be seen in the several examples collected in appendices \ref{app_oneMass}, \ref{app_twoMass} and \ref{app_threeMass}.

If the triangle has two or three external masses (say $p_i^2\neq0$ and $p_j^2\neq0$), we can compute its sequential cuts in the external channels $p_i^2$ and $p_j^2$, in the region where they are both above threshold, while the remaining external channel is below threshold and all internal masses are positive. The extra delta function makes the $x$ integration in \refE{intMeasure} trivial (note, however, that it might restrict the kinematic region in which the cut is nonzero). The most complicated functions we get as a result are invariants raised to powers that are linear in $\epsilon$, producing powers of logarithms upon expansion in $\epsilon$. Again, our examples are collected in appendices \ref{app_oneMass}, \ref{app_twoMass} and \ref{app_threeMass}.

\paragraph{Cuts in internal masses.}

Cuts in internal masses are harder to compute than cuts in external channels. Here, we present two ways of computing them. Either way, we compute discontinuities, which are trivially related to cuts through \refE{cutDefMassive}. The first way is a brute-force method that works in all cases considered here. The second way is more elegant, but only suitable for special configurations of the external and internal masses. Because we do not have a proof that it should work, we present it as an observation. In all cases where both can be applied we find they agree, giving evidence for the validity of the  second way. We will illustrate both in the context of the triangle $T(p^2_2,p_3^2;m_{12}^2)$.

The first method relies on getting a Feynman parameter representation for the diagram, and then computing the discontinuity of the integrand across the branch cut associated with the internal mass. It is of course valid for any configuration of the internal and external masses. For example, we have:
\begin{align}
T(p_2^2,p_3^2;m_{12}^2)=i\frac{e^{\gamma_E\epsilon}\Gamma(1+\epsilon)}{\epsilon}\int_0^1d x\frac{(1-x)^{-\epsilon}}{m_{12}^2+x(p_3^2-p_2^2)}\left((-p_3^2x)^{-\epsilon}-(m_{12}^2-p_2^2x)^{-\epsilon}\right)\,,
\end{align}
which we obtain by computing the trivial and the first non-trivial Feynman parameter integrals.
We then get
\begin{equation*}
\Disc_{m^2_{12}}\left[(m_{12}^2-p_2^2x)^{-\epsilon}\right]=\frac{2\pi i\epsilon}{\Gamma(1-\epsilon)\Gamma(1+\epsilon)}(p_2^2 x-m_{12}^2)^{-\epsilon}\theta\left(\frac{m_{12}^2}{p_2^2}-x\right)\, ,
\end{equation*}
where we used $m^2_{12}=m^2_{12}-i0$ and are in the region $p_2^2<m_{12}^2<0$, to get
\begin{align}\bsp
\cut_{m^2_{12}}T(p_2^2,p_3^2;m_{12}^2)=&\Disc_{m^2_{12}}T(p_2^2,p_3^2;m_{12}^2)\\
=&\frac{2\pi e^{\gamma_E\epsilon}}{\Gamma(1-\epsilon)}\int_0^{m_{12}^2/p_2^2}d x(1-x)^{-\epsilon}\frac{(p^2_2 x-m_{12}^2)^{-\epsilon}}{m_{12}^2+x (p_3^2-p_2^2)}\, .
\label{cutm12tp2p3m12v2}
\esp\end{align}
which is trivial to compute to any desired accuracy in $\epsilon$.

The second way only works if there is a massive external leg non-adjacent to the massive internal leg being cut. More precisely, in our notation, if we look at the cut in the internal propagator of mass $m_{ij}^2$, we need $p_k^2\neq 0$. We can then compute a three-propagator cut corresponding to $\cut_{p_k^2,m_{ij}^2}$ in the region where $m_{ij}^2<0$ and $p_k^2$ is above threshold. This is trivial to evaluate. The single-propagator cut is finally obtained through dispersive integration in the $p_k^2$-channel of the three-propagator cut. This is not guaranteed to work a priori, because we have no proof that the $m_{ij}^2$ discontinuity function has a dispersive representation. However, it does give the correct answer in all the cases we have considered. The reason why this method is not valid for any configuration of internal and external masses is because there is no sequential cut associated to an external mass and an internal mass if they are adjacent.

For our example, we have $(i,j,k)=(1,2,3)$. The three-propagator cut is computed in the region where $p_3^2>0$ and $m_{12}^2<0$ and is given by
\begin{align}
\cut_{m_{12}^2,p_3^2}T(p_2^2,p_3^2;m_{12}^2)=-\frac{4\pi^2 ie^{\gamma_E\epsilon}}{\Gamma(1-\epsilon)}\frac{(p_3^2)^{-\epsilon}(-m_{12}^2)^{-\epsilon}}{(p_3^2-p_2^2)^{1-\epsilon}}(p_3^2+m_{12}^2-p_2^2)^{-\epsilon}\theta(p_3^2+m_{12}^2-p_2^2)\, ,
\end{align}
Through a standard dispersive integral, see e.g. the brief discussion in ref.~\cite{Abreu:2014cla}, we obtain
\begin{align}\bsp
\cut_{m_{12}^2}T(p_2^2,p_3^2;m_{12}^2)&=\Disc_{m^2_{12}}T(p_2^2,p_3^2;m_{12}^2)\\
&=-2\pi\frac{e^{\gamma_E\epsilon}}{\Gamma(1-\epsilon)}(-m_{12}^2)^{-\epsilon}\int_{0}^\infty \frac{d s}{s-p_3^2}\frac{s^{-\epsilon}}{(s-p_2^2)^{1-2\epsilon}}(s+m_{12}^2-p_2^2)^{-\epsilon}\, .
\label{cutm12tp2p3m12v1}
\esp\end{align}
The integral is trivial to compute at any order in $\epsilon$, and matches the result obtained in \refE{cutm12tp2p3m12v2}.

In appendices \ref{app_oneMass}, \ref{app_twoMass} and \ref{app_threeMass}, we collect several examples of cuts in massive internal legs. Whenever possible, the cuts were computed in each of the two ways described above, and the results agreed.

%% file: Sec_RelationCutDelta.tex
\section{Relations among discontinuities}
\label{sec:relationcutdelta}

In this section, we explain how to relate cuts and coproduct entries, via their separate relations to discontinuities across branch cuts. This allows us to give a diagrammatic interpretation of coproduct entries. We generalize the relations presented in ref.~\cite{Abreu:2014cla} to diagrams with massive propagators. The generalization is straightforward, aside from two points regarding the $i0$ prescription: {\it i)} when combining channel and mass discontinuities, the $\pm i0$ associated to the masses is determined once all channel cuts have been taken; {\it ii)} the precise determination of how the $\pm i0$ prescription propagates from invariants to symbol letters is slightly more complicated than in the absence of internal masses. After establishing the general relations, we give examples to illustrate these points.

\subsection{Cut diagrams and discontinuities}
\label{sec:cut_disc}

The rules for evaluating cut diagrams are designed to compute their discontinuities.  For single cuts in internal masses, the relation is straightforward as can be seen from \refE{cutDefMassive}. For cuts in external channels, there are some subtleties which we now review.

The original relation for the first cut in an external channel follows from the largest time equation \cite{tHooft:1973pz,Veltman:1994wz,Remiddi:1981hn},
and it takes the form
\bea
\Disc_s F =-\Cut_s F.
\label{eq:oldcutting}
\eea

For sequential cuts in external channels, it was argued in ref.~\cite{Abreu:2014cla} that the relation could be generalized so that $\Cut_{s_1,\ldots,s_k} F$ captures discontinuities through the relation
\bea
 \Cut_{s_1,\ldots,s_k} F = (-1)^k \Disc_{s_1,\ldots,s_k}F,
\label{eq:cutequalsdiscInv}
\eea
where $ \Cut_{s_1,\ldots,s_k} F$ is to be computed according to the rules given above for multiple cuts.

Eq.~(\ref{eq:cutequalsdiscInv}), like \refE{eq:oldcutting}, is valid in a specific kinematic region. As mentioned in the previous section, $\Cut_{s_1,\ldots,s_k} F$ is evaluated in the region where $s_1,\ldots,\,s_k$ are above their respective thresholds, the remaining external channels are below their thresholds, and all internal masses are positive.
 On the right-hand side, we proceed step by step according to the definition in \refE{eq:disc-seq}: each $\Disc_{s_1,\ldots,\,s_i}$ is evaluated after analytic continuation to the same region in which $\Cut_{s_1,\ldots,s_i} F$ is evaluated.

The relation between cuts in internal masses and discontinuities is trivial. For a single cut, the relation is given in \refE{cutDefMassive}. It can be straightforwardly generalized as
\begin{equation}\label{eq:cutequalsdiscMass}
 \Cut_{m^2_1,\ldots,m^2_k} F = \Disc_{m^2_1,\ldots,m^2_k}F.
\end{equation}

We can now combine cuts in internal masses and external channels through
\begin{equation}\label{eq:cutequalsdisc}
 \Cut_{s_1,\ldots,s_l,m^2_1,\ldots,m^2_k} F=(-1)^l\Disc_{s_1,\ldots,s_l,m^2_1,\ldots,m^2_k}F\,.
\end{equation}
In order for \refE{eq:cutequalsdisc} to produce the correct signs, the $\pm i0$ associated to the internal masses on the right hand side are determined from the cut diagram in which all $l$ of the channel cuts have been taken. (We recall that according to our rules, channel cuts imply complex conjugation of certain regions of the diagram, which affects the $i0$-prescription of the internal propagators. Hence we make it a rule to take channel discontinuities before mass discontinuities.)
Furthermore, on the right hand side,
we take a specific order of the listed invariants. Indeed,
while sequential cuts are independent of the order in which the invariants are listed, the correspondences to $\Disc$ are derived in sequence so that the right-hand side of \refE{eq:cutequalsdisc} takes a different form when channels and masses on the left-hand side are permuted. Thus, \refE{eq:cutequalsdisc} implies relations among the different $\Disc_{s_1,\ldots,s_l,m^2_1,\ldots,m^2_k}F$. 

We note one restriction: the cut integrals reproduce sequential discontinuities through the above relations only if each additional invariant in the subscript---whether a momentum channel or a mass---introduces at least one new cut propagator in the Feynman diagrams.  For example, we would not consider $\Cut_{p_1^2,m_{12}^2}$ of a one-loop triangle, since the propagator of mass $m_{12}^2$ was already cut in the first step, $\Cut_{p_1^2}$.

In section \ref{sec:examples} we make these relations concrete in the context of specific examples.

\subsubsection{A limit on sequential mass cuts}\label{sec:seqmcut}

Suppose that two massive propagators are attached to the same external massive leg of a one-loop integral, as for example in  \refF{fig:1mTriangle1}.  Then the double discontinuity in those two internal masses will vanish.
Let us now see why this is the case.  
Without loss of generality, we consider the cut in $m_{1,2}^2$ of $\cut_{m_{1,n}^2}F$ as given in \refE{cutDefMassive}.  The integral with cuts of the two propagators of masses $m_{1,n}^2$ and $m_{1,2}^2$, which is given by 
\begin{align}\bsp\label{temp1633}
(-1)^{n}\int  \frac{d^{4-2\epsilon}k}{\pi^{2-\epsilon}}(2\pi i)^2\delta(k^2-m_{1,n}^2)\delta((k+p_1)^2-m_{1,2}^2)
\prod_{i=2}^{n-1}\frac{1}{(k+q_i)^2-m^2_{i,i+1}}\, ,
\esp\end{align}
can be used for either of the cut integrals  $\cut_{m_{1,n}^2,m_{1,2}^2}F$ or  $\cut_{p_1^2}F$, depending on the kinematic region where it is evaluated.  
If $p_1^2\neq 0$, then the uncut integral $F$ has a branch cut in $p_1^2$.  As a consequence of the largest time equation, the integral $\cut_{p_1^2}F$ is proportional to the discontinuity of $F$ across this branch cut \cite{tHooft:1973pz,Veltman:1994wz}.
In particular, the discontinuity is zero when we are below the threshold of $p_1^2$, which can be realized either for $m_{1,2}^2,m_{1,n}^2>0$ or $m_{1,2}^2,m_{1,n}^2<0$, and in this case the integral \refE{temp1633} vanishes as well.

Now, the double-cut integral $\cut_{m_{1,n}^2,m_{1,2}^2}F$ must be evaluated in the region where $m_{1,2}^2,m_{1,n}^2<0$ and all other invariants are below their thresholds.  Since $p_1^2$, in particular, is below its threshold, the integral vanishes by the argument given above.  We will see an example of this type of vanishing double cut in $T(p_1^2,0,0;m_{12}^2,0,m_{13}^2)$ in \refS{sec:ex-1-12-13}.

However, if $p_1^2=0$, then $F$ has no branch cut associated with this external channel, and the largest time equation does not give any constraint on the result of \refE{temp1633}. In this case, the double discontinuity on the masses $m_{1,2}^2$ and $m_{1,n}^2$ can indeed be nonzero.  We will see an example of this type of nonvanishing double cut in $T(p_1^2,0,0;m_{12}^2,m_{23}^2,0)$ in \refS{sec:ex-1-12-23}.

\subsubsection{Sequential cuts of triangle diagrams}
 
In the examples studied in this paper, we are restricted to $k=2$ in \refE{eq:cutequalsdiscInv}, since after a sequence of two cuts in external channels, all three propagators are cut.  
Because triangle diagrams must have at least one external massive channel, the above considerations restrict us to at most two cuts in internal masses. It follows that  the maximum value for $k$ in \refE{eq:cutequalsdiscMass} is also 2. This is consistent with the transcendental weight of the functions being two.

\subsection{Coproduct and discontinuities}
\label{sec:coprod_disc}

In \refS{sec:firstentry}, we defined the operation $\Disc_{s_1,\ldots,s_k}$. We also argued that the coproduct is a natural tool to study the discontinuity of polylogarithms. We now make the relation between the coproduct and discontinuities precise.

We start by defining the operation $\delta_{x_1,\ldots,x_k}$ on the coproduct.  Given the symbol alphabet $\cA$, we can write the $(1,1,\ldots,1,n-k)$ component of the coproduct of $F$ as
\bea
\Delta_{\underbrace{\scriptstyle 1,1,\ldots,1}_{k\textrm{ times}},n-k}F &=&
\sum_{(x_{i_1},\ldots,x_{i_k})\in\cA^k} \log x_{i_1} \otimes \cdots \otimes \log x_{i_k} \otimes 
g_{x_{i_1},\ldots,x_{i_k}}\,.
\eea 
Then, our truncation operation is defined to be
\bea\label{eq:deltaDef}
\delta_{x_{j_1},\ldots,x_{j_k}}F &\cong& 
\sum_{(x_{i_1},\ldots,x_{i_k})\in\cA^k}
\delta_{i_1j_1}\,\ldots\,\delta_{i_kj_k}\,
g_{x_{i_1},\ldots,x_{i_k}}.
\eea
The congruence symbol indicates that  $\delta_{x_{j_1},\ldots,x_{j_k}}F$ is defined only modulo $\pi $; this is an intrinsic ambiguity due to the nature of the Hopf algebra of multiple polylogarithms. If $F$ contains overall numerical factors of $\pi$, they should be factored out before performing this operation and then reinstated.

In ref.~\cite{Abreu:2014cla}, it was shown how the discontinuity of any element of the Hopf algebra is captured by the operation $\delta$ as defined in \refE{eq:deltaDef}. This relies on the relation
\bea
\Disc F \cong \mu\left[(\Disc \otimes \textrm{id}) (\Delta_{1,n-1}F)\right]\, ,
\label{temp0943}
\eea
 where $\mu$ is a linear map, $\mu: \cH\otimes\cH\to\cH$, such that $\mu\left(a\otimes b\right)=a\cdot b$, for $a,\,b\in\cH$. Eq.~(\ref{temp0943}) is a direct consequence of the relation between the coproduct and discontinuity operator presented in ref.~\cite{Duhr:2012fh}.  The whole discussion of ref.~\cite{Abreu:2014cla} in the context of massless internal propagators generalizes straightforwardly to diagrams with massive propagators, so we will simply review it briefly here.
 
The relation between sequential discontinuities and entries of the coproduct is
\bea
\Disc_{r_1,\ldots,r_k} F \cong  \Theta\, \sum_{(x_1,\ldots,x_k) \in {\cal A}^k } \left( \prod_{i=1}^k a_i(r_i,x_i) \right)  \delta_{x_1,\ldots,x_k} F,
\label{eq:discequalsdelta}
\eea
where the sum is taken over ordered sequences $(x_1,\ldots,x_k)$ of $k$ letters, and the $r_i$ can be either internal masses or external channels. The congruence symbol in \refE{eq:discequalsdelta} indicates that the right-hand side only captures terms whose coproduct is nonvanishing, and it therefore holds modulo $(2\pi i)^{k+1}$. The schematic factor $\Theta$ expresses the restriction to the kinematic region where the left-hand side will be compared with $\Cut$.
The factors $a_i(r_i,x_i)$ are the discontinuities of real-valued logarithms after analytic continuation from $R_{i-1}$, the region where the $(i-1)$-th cut is taken, to the region $R_i$, the region where the $i$-th cut is taken.  Specifically,
\bea
\label{a_def}
a_i(r_i,x_i) = \Disc_{r_i;R_i} \big[\!\big[ \log(\pm x_i) \big]\!\big]\!_{R_{i-1}},
\eea
where the double bracket means that the sign should be chosen so that the argument of the logarithm is positive in the region $R_{i-1}$. In section \ref{sec:examples} we make these relations concrete in the context of specific examples.

Eq.~(\ref{eq:discequalsdelta}) is valid independently of the order in which the discontinuities are taken. However, because for massive internal propagators the relation between $\Disc$ and $\cut$ have the correct signs only if discontinuities on channels are taken first---see the discussion below \refE{eq:cutequalsdisc}---we will in general impose the same constraint in \refE{eq:discequalsdelta}. Furthermore, we observe that it is more complicated to identify the sign of the imaginary part of the symbol letters $x_i$ inherited from the $\pm i0$ prescription of the invariants. We discuss how we overcome this difficulty in the following subsection.

\subsubsection{$\pm i0$-prescription of symbol letters}\label{sec:i0ambiguity}

In most examples considered in this paper, it is simple to determine the sign of the $i0$-prescription of a given symbol letter once we know the prescription of the invariant to which it is associated and the kinematical region in which we are working. Indeed, whenever the symbol letters are linear combinations of invariants, this is a trivial problem. However, we observe that in more complicated cases there is an ambiguity in the sign of the imaginary part of some symbol letters. We need to resolve this ambiguity, because this sign is needed to obtain the correct sign in \refE{eq:discequalsdelta}.

The simplest case where we observe this problem is the triangle with three external masses and one internal mass, $T(p_1^2,p_2^2,p_3^2;m_{12}^2,0,0)$; see table \ref{tab:3massTriangles}. For instance, when considering the double cut first in $p_2^2$ and then in $p_1^2$, we need to determine the sign of the imaginary part of $\zbar-\mu_{12}$, as inherited from the  prescription of the second cut invariant, $p_1^2-i0$. One can easily check that this sign is the same as the sign of the quantity
\begin{equation*}
\frac{\zbar(1-\zbar)-\mu_{12}(z-\zbar)}{z-\zbar}\,.
\end{equation*}
which can be either positive or negative in the region where the double cut is computed,
\begin{equation*}
z>1\,,\quad 0<\zbar<1\,,\quad0<\mu_{12}<1\, ,\quad\zbar-\mu_{12}>0\, .
\end{equation*}
If the imaginary part of $\zbar-\mu_{12}$ is negative, then we are in the subregion
\begin{equation*}
z>1\,,\qquad 0<\zbar<1\,,\qquad \frac{\zbar(1-\zbar)}{z-\zbar}<\mu_{12}<\zbar\, ,
\end{equation*}
and if it is positive, in the subregion
\begin{equation*}
z>1\,,\qquad 0<\zbar<1\,,\qquad 0<\mu_{12}<\frac{\zbar(1-\zbar)}{z-\zbar}\,.
\end{equation*}
We note that if we are in the first situation we cannot smoothly take the internal mass ($\mu_{12}$) to zero. However, if we are in the second situation, corresponding to a positive imaginary part of $\zbar-\mu_{12}$, we can take $\mu_{12}$ to zero without any problem, which is naturally a desirable property. We thus associate a positive imaginary part to the symbol letter $\zbar-\mu_{12}$. We can confirm this is indeed the correct result by considering the same double cut in the opposite order, where there are no sign ambiguities. We treat this example in detail in section \ref{sec:examples}.

All other cases where we have found sign ambiguities can be solved in the same way: {\em we always require being in a kinematic region where massless limits can be taken smoothly}. Furthermore, we have found in all of of our examples of multiple cuts that there is always an ordering of the cuts where there is no ambiguity. We have then verified that any possible ambiguities were correctly lifted through the method just described.

\subsection{Cuts and coproduct}

Having related cuts to discontinuities in section \ref{sec:cut_disc} and discontinuities to coproduct entries in section \ref{sec:coprod_disc}, it is now straightforward to relate cuts to coproduct entries. Combining the relations \refE{eq:cutequalsdisc} and \refE{eq:discequalsdelta}, we arrive at:
\bea
&& \Cut_{s_1,\ldots,s_l,m^2_1,\ldots,m^2_k} F
\cong \nonumber
\\ &&
 \Theta\, \sum_{(x_1,\ldots,x_l,y_1,\ldots,y_k) \in {\cal A}^{k+l} } 
(-1)^l\left( \prod_{i=1}^l a_i(s_i,x_i) \prod_{j=1}^k a_j(m^2_j,y_j)\right)  \delta_{x_1,\ldots,x_l,y_1,\ldots,y_k} F\, .
\label{eq:cutequalsdelta}
\eea
We recall that on the left-hand side the $s_i$ and the $m^2_j$ may be written in any order, and correspondingly permuted on the right-hand side, but we require that we act first with all the $s_i$ and then with the $m_j^2$.  It is not obvious that permutations of the sets $\{s_i\}$ and $\{m^2_j\}$ give equivalent results on the right-hand side, but this property follows from the commutativity of cuts. It implies nontrivial relations among coproduct entries.

\subsection{Examples}
\label{sec:examples}

We now illustrate eqs.~(\ref{eq:cutequalsdisc}), (\ref{eq:discequalsdelta}) and (\ref{eq:cutequalsdelta}) with different examples of triangles with internal masses, by comparing the result of the direct computation of the cuts to what is predicted by the relation between $\Disc$ and $\delta$. We have selected examples that highlight the features specific to diagrams with internal masses, and will not cover all diagrams listed in appendices \ref{app_oneMass}, \ref{app_twoMass} and \ref{app_threeMass}. All remaining cases can be treated in the same way, and we have checked that they do satisfy the expected relations.

\subsubsection{$T\left(p_1^2,0,0;m_{12}^2,m_{23}^2,0\right)$}
\label{sec:ex-1-12-23}
In this example, we illustrate iterated cuts in internal masses and iterated cuts in one external channel and one internal mass. Expressions for the integral, its symbol and cuts can be found in appendix \ref{sec:tp1m12m23}. The symbol alphabet can be found in table \ref{tab:1massTriangles}. The euclidean region, which we denote by $R_0$, is
\begin{equation}
R_0\,: \qquad m_{12}^2>0\,,\qquad m_{23}^2>0\,,\qquad p_1^2<m_{12}^2.
\end{equation}

\paragraph{Single cuts:}
For the single cut in the invariant $r$, where $r\in\{p_1^2,m_{12}^2,m_{23}^2\}$, we will move away from the euclidean region and into region $R_1^r$. These regions are, respectively,
\begin{align}\bsp
R_1^{p_1^2}\,:\qquad m_{12}^2>0\,,\qquad m_{23}^2>0\,,\qquad p_1^2>m_{12}^2\, ,\\
R_1^{m_{12}^2}\,:\qquad m_{12}^2<0\,,\qquad m_{23}^2>0\,,\qquad p_1^2<m_{12}^2\, ,\\
R_1^{m_{23}^2}\,:\qquad m_{12}^2>0\,,\qquad m_{23}^2<0\,,\qquad p_1^2<m_{12}^2\, .
\esp\end{align}
Recalling the prescriptions $p_1^2+i0$ and $m_{ij}^2-i0$, we can compute the coefficients $a_1(r,x_1)$ as defined in \refE{a_def}. They are computed respectively in $R_1^{p_1^2}$, $R_1^{m_{12}^2}$ and $R_1^{m_{23}^2}$, and turn out to be equal. We find:
\begin{align*}\bsp
a_1(p_1^2,m_{12}^2-p_1^2)=a_1(m_{12}^2,m_{12}^2)=a_1(m_{23}^2,m_{23}^2)=-2\pi i\, .
\esp\end{align*}
We then get:
\begin{align}\bsp
\cut_{p_1^2}T&=-\Disc_{p_1^2}T\cong -\frac{2\pi}{p_1^2}\Theta \delta_{m_{12}^2-p_1^2}\mathcal{T}=\frac{2\pi}{p_1^2}\ln\left(1-\frac{m_{12}^2-p_1^2}{m_{23}^2}\right),\\
\cut_{m_{12}^2}T&=\Disc_{m_{12}^2}T\cong \frac{2\pi}{p_1^2}\Theta \delta_{m_{12}^2}\mathcal{T}=-\frac{2\pi}{p_1^2}\ln\left(\frac{m_{23}^2}{m_{23}^2-m_{12}^2}\right),\\
\cut_{m_{23}^2}T&=\Disc_{m_{23}^2}T\cong \frac{2\pi}{p_1^2}\Theta \delta_{m_{23}^2}\mathcal{T}=\frac{2\pi}{p_1^2}\ln\left(1-\frac{p_1^2}{m_{12}^2-m_{23}^2}\right),
\esp\end{align}
which are consistent with the results in appendix \ref{sec:tp1m12m23}.
All relations for single cuts follow the same pattern, so we will simply list them without further details in the remaining examples.

\paragraph{Double cuts:}
According to our rules, there are two different cuts to consider: $\cut_{p_1^2,m_{23}^2}$ and $\cut_{m_{12}^2,m_{23}^2}=\cut_{m_{23}^2,m_{12}^2}$. We start with $\cut_{p_1^2,m_{23}^2}$, for which we go from the region $R_1^{p_1^2}$ to the region
\begin{equation}
R_2^{p_1^2,m_{23}^2}\,:\qquad m_{12}^2>0\,,\qquad m_{23}^2<0\,,\qquad p_1^2>m_{12}^2\, .
\end{equation}
Given our conventions for multiple cuts, we now have the prescription $m_{23}^2+i0$. Then,
\begin{equation*}
a_2\left(m_{23}^2,m_{23}^2\right)=2\pi i\, ,\qquad a_2\left(m_{23}^2,p_1^2+m_{23}^2-m_{12}^2\right)=2\pi i\theta\left(m_{12}^2-p_1^2-m_{23}^2\right)\, ,
\end{equation*}
where we have only listed the coefficients leading to nonzero contributions. We finally find
\begin{align}\bsp
\cut_{p_1^2,m_{23}^2}T&=-\Disc_{p_1^2,m_{23}^2}T\\
&\cong-\frac{4\pi^2i}{p_1^2}\Theta\left[\delta_{m_{12}^2-p_1^2,m_{23}^2}+\theta\left(m_{12}^2-p_1^2-m_{23}^2\right)\delta_{m_{12}^2-p_1^2,m_{12}^2-p_1^2-m_{23}^2}\right]\mathcal{T}\\
&=-\frac{4\pi^2i}{p_1^2}\theta(m_{23}^2+p_1^2-m_{12}^2)\, ,
\esp\end{align}
which matches the result of the direct calculation in  \ref{sec:tp1m12m23}. Interestingly, even the theta functions are correctly reproduced, which is a feature observed in all our examples. We recall that when computing multiple cuts in external channels and internal masses, we insist on taking the discontinuity first in the external invariant, and then in the mass. It can easily be checked that if we had taken the opposite order, we would have had the opposite sign in the above equation.

We now consider the double cut in the internal masses. This is an example of the behavior described in section \ref{sec:seqmcut}, where a double cut in internal masses attached to the same external massless leg is nonzero. We only give details for one order of the invariants, first $m_{12}^2$ and then $m_{23}^2$. The opposite order can be done in exactly the same way.

To compute $\cut_{m_{12}^2,m_{23}^2}$, we must go from $R_1^{m_{12}^2}$ to
\begin{equation*}
R_2^{m_{12}^2,m_{23}^2}\,:\qquad m_{12}^2<0\,,\qquad m_{23}^2<0\,,\qquad p_1^2<m_{12}^2\, .
\end{equation*}
Because mass cuts do not require complex conjugation of any region of the diagram, we still have the prescription $m_{23}^2-i0$. The coefficients $a_2(m_{23}^2,x_2)$ giving nonzero contributions are
\begin{equation*}
a_2(m_{23}^2,m_{23}^2)=-2\pi i\, ,\qquad a_2(m_{23}^2,m_{23}^2-m_{12}^2)=-2\pi i \theta(m_{12}^2-m_{23}^2).
\end{equation*}
We then find
\begin{align}\bsp
\cut_{m_{12}^2,m_{23}^2}T&=\Disc_{m_{12}^2,m_{23}^2}T\\
&=-\frac{4\pi^2i}{p_1^2}\Theta\left[\delta_{m_{12}^2,m_{23}^2}+\theta(m_{12}^2-m_{23}^2)\delta_{m_{12}^2,m_{23}^2-m_{12}^2}\right]\mathcal{T}\\
&=\frac{4\pi^2i}{p_1^2}\theta(m_{12}^2-p_1^2-p_{23}^2)\theta(m_{23}^2-m_{12}^2)\,,
\esp\end{align}
which matches the result of the direct calculation in \ref{sec:tp1m12m23}. Taking the discontinuities in the opposite order, we would have found
\begin{align}\bsp
\cut_{m_{23}^2,m_{12}^2}T&=\Disc_{m_{23}^2,m_{12}^2}T\\
&=-\frac{4\pi^2i}{p_1^2}\Theta\left[\theta(p_1^2+m_{23}^2-m_{12}^2)\delta_{m_{23}^2,m_{12}^2-p_1^2-m_{23}^2}\right.\\
&\qquad\qquad\qquad\qquad\left.+\theta(m_{23}^2-m_{12}^2)\delta_{m_{23}^2,m_{23}^2-m_{12}^2}\right]\mathcal{T}\\
&=\frac{4\pi^2i}{p_1^2}\theta(m_{12}^2-p_1^2-p_{23}^2)\theta(m_{23}^2-m_{12}^2)\,,
\esp\end{align}
which also matches the direct calculation.\\

We finish this example with a comment. As mentioned previously, for triangle integrals we cannot set up a double cut in an external momentum  and an internal mass attached to it, like $p_1^2$ and $m_{12}^2$ in this example, because there is no additional propagator to cut at the second stage.
Correspondingly, if we were to attempt to relate $\Disc$ and the coproduct for this double cut as in the above exercise, we would be stuck when taking the second discontinuity, as the $\pm i0$ prescription of the second invariant is not well-defined. Thus, even in this case, there is no conflict among $\cut$, $\Disc$ and the coproduct.

\subsubsection{$T\left(0,p_2^2,p_3^2;m_{12}^2,0,0\right)$}
In this example, we illustrate iterated cuts in external channels, and we give another example of iterated cuts in one external channel and one internal mass. Expressions for the integral, its symbol and cuts can be found in appendix \ref{sec:tp2p3m12}. The symbol alphabet can be found in table \ref{tab:2massTriangles}. The euclidean region, which we denote $R_0$ (we reuse the same notation as above for all examples, since there is no ambiguity and  to avoid having too many indices), is
\begin{equation}
R_0\,: \qquad m_{12}^2>0\,,\qquad p_2^2<m_{12}^2 \,,\qquad p_{3}^2<0.
\end{equation}
The single discontinuities are treated as above and obey the expected relations, so we will not go through the derivation. For double discontinuities, we consider two different double cuts: $\cut_{p_3^2,m_{12}^2}$ and $\cut_{p_2^2,p_3^2}=\cut_{p_3^2,p_2^2}$. The first one is very similar to what we did before so we will not address it in detail here. The second one is a new kind. In particular, we will show that both orders of taking the discontinuities give the same result.

\paragraph{$(p_2^2,p_3^2):$}
We must analytically continue the function from
\begin{equation}
R_1^{p_2^2}\,:\qquad m_{12}^2>0\,,\qquad p_2^2>m_{12}^2 \,,\qquad p_{3}^2<0\, ,
\end{equation}
to
\begin{equation}
R_{2}^{p_2^2,p_{3}^2}\,:\qquad m_{12}^2>0\,,\qquad p_2^2>m_{12}^2 \,,\qquad p_{3}^2>0.
\end{equation}
In this region, the nonvanishing coefficients $a_2(p_{3}^2,x_2)$ are
\begin{equation*}
a_2(p_{3}^2,p_{3}^2)=2\pi i \, ,\qquad a_2(p_{3}^2,p_2^2-m_{12}^2-p_3^2)=2\pi i\theta(m_{12}^2-p_2^2+p_3^2).
\end{equation*}
We then find
\begin{align}\bsp
\cut_{p_2^2,p_3^2}T&=\Disc_{p_2^2,p_{3}^2}T\\
&\cong\frac{4\pi^2i}{p_2^2-p_3^2}\Theta\left[\delta_{p_2^2-m_{12}^2,p_3^2}+\theta(m_{12}^2-p_2^2+p_3^2)\delta_{p_2^2-m_{12}^2,p_2^2-m_{12}^2-p_3^2}\right]\mathcal{T}\\
&=\frac{4\pi^2i}{p_2^2-p_3^2}\theta(p_2^2-m_{12}^2-p_3^2)\, .
\esp\end{align}

\paragraph{$(p_3^2,p_2^2):$}
We now start in the region
\begin{equation}
R_1^{p_3^2}\,:\qquad m_{12}^2>0\,,\qquad p_2^2<m_{12}^2 \,,\qquad p_{3}^2<0\,
\end{equation}
and go the same region $R_{2}^{p_3^2,p_{2}^2}$ as above. The coefficients $a_2(p_{2}^2,x_2)$ are
\begin{equation*}
a_2(p_{2}^2,m_{12}^2-p_2^2+p_3^2)=2\pi i\theta(p_2^2-m_{12}^2-p_3^2)
\end{equation*}
and thus
\begin{align}\bsp
\cut_{p_3^2,p_{2}^2}T&=\Disc_{p_3^2,p_{2}^2}T\\
&\cong\frac{4\pi^2i}{p_2^2-p_3^2}\Theta\left[\theta(p_2^2-m_{12}^2-p_3^2)\delta_{p_3^2,m_{12}^2-p_2^2+p_3^2}\right]\mathcal{T}\\
&=\frac{4\pi^2i}{p_2^2-p_3^2}\theta(p_2^2-m_{12}^2-p_3^2)\, ,
\esp\end{align}
as above.\\

As expected, the two orderings of taking discontinuities match the direct calculation of the double cut.

\subsubsection{$T\left(p_1^2,0,0;m_{12}^2,0,m_{13}^2\right)$}
\label{sec:ex-1-12-13}
In this example, we show how the relations between discontinuities and the coproduct generalize when we must use variables such as the ones defined in \refE{eq:zw1_variables_def} to get a symbol with rational letters. In particular, we hope to make clearer the discussion below \refE{eq:firstentry-alphabet-example}. We also illustrate the discussion in section \ref{sec:seqmcut}: as predicted, we show that the double cut in the two internal masses vanishes.

Expressions for the integral, its symbol, and its cuts can be found in appendix \ref{sec:tp1m12m13}. The symbol alphabet can be found in table \ref{tab:1massTriangles}. The euclidean region is
\begin{equation}
R_0\,: \qquad m_{12}^2>0\,,\qquad m_{13}^2>0\,,\qquad p_1^2<\left(\sqrt{m_{12}^2}+\sqrt{m_{13}^2}\right)^2.
\end{equation}
To simplify our discussion, we will restrict the euclidean region to the subregion $R_{0*}$, defined by
\begin{equation}
R_{0*}\,: \qquad m_{12}^2>0\,,\qquad m_{13}^2>0\,,\qquad p_1^2<0\quad\Rightarrow\quad \bar{w}_1<0\,,\qquad w_1>1.
\end{equation}
Our discussion would be similar if we had started from the other subregion of the euclidean region.

\paragraph{Single cuts:} For the single cut in the invariant $r$, with $r\in\{p_1^2,m_{12}^2,m_{23}^2\}$, we will move away from the euclidean region and into region $R_1^r$. These three regions are
\begin{align}\bsp
R_1^{p_1^2}\,:&\qquad m_{12}^2>0\,,\qquad m_{13}^2>0\,,\qquad p_1^2>\left(\sqrt{m_{12}^2}+\sqrt{m_{13}^2}\right)^2\quad\Rightarrow\quad 0<\bar{w}_1<w_1<1,\\
R_1^{m_{12}^2}\,:&\qquad m_{12}^2<0\,,\qquad m_{13}^2>0\,,\qquad p_1^2<0\quad\Rightarrow\quad 0<\bar{w}_1<1< w_1,\\
R_1^{m_{13}^2}\,:&\qquad m_{12}^2>0\,,\qquad m_{13}^2<0\,,\qquad p_1^2<0\quad\Rightarrow\quad \bar{w}_1<0< w_1<1\, .
\esp\end{align}

For the discontinuity in the $p_1^2$ channel, we first note that, in region $R_1^{p_1^2}$,  $p_1^2+i0$ implies $w_1+i0$ and $\bar{w}_1-i0$. Then, the nonzero coefficients $a_1(p_1^2,x_1)$ are
\begin{equation}
a_1(p_1^2,\bar{w}_1)=2\pi i\,,\qquad a_1(p_1^2,1-w_1)=2\pi i\, .
\end{equation}
The relation between $\cut$, $\Disc$ and the coproduct is
\begin{align}\bsp\label{cutp1ex3}
\cut_{p_1^2}T=-\Disc_{p_1^2}T\cong&\frac{2\pi}{p_1^2}\Theta\left[\delta_{\bar{w}_1}+\delta_{1-w_1}\right]\mathcal{T}\\
=&\frac{2\pi}{p_1^2}\left(\log\left(\frac{w_1}{1-w_1}\right)-\log\left(\frac{\bar{w}_1}{1-\bar{w}_1}\right)\right).
\esp\end{align}

Similarly, for the discontinuity in the mass $m_{12}^2$, we note that, in region $R_1^{m_{12}^2}$,  $m_{12}^2-i0$ implies $w_1+i0$ and $\bar{w}_1+i0$. The nonzero coefficients $a_1(m_{12}^2,x_1)$ are
\begin{equation}
a_1(m_{12}^2,\bar{w}_1)=-2\pi i\, ,
\end{equation}
and we then find
\begin{align}\bsp\label{cutm12ex3}
\cut_{m_{12}^2}T=&\Disc_{m_{12}^2}T\cong\frac{2\pi}{p_1^2}\Theta\delta_{\bar{w}_1}\mathcal{T}=\frac{2\pi}{p_1^2}\log\left(\frac{w_1}{w_1-1}\right).
\esp\end{align}

Finally, for the discontinuity in the mass $m_{13}^2$, we note that, in region $R_1^{m_{13}^2}$,  $m_{13}^2-i0$ implies $w_1-i0$ and $\bar{w}_1-i0$. The nonzero coefficients $a_1(m_{13}^2,x_1)$ are
\begin{equation}
a_1(m_{13}^2,1-w_1)=-2\pi i\, ,
\end{equation}
and we then find
\begin{align}\bsp\label{cutm13ex3}
\cut_{m_{13}^2}T=&\Disc_{m_{13}^2}T\cong\frac{2\pi}{p_1^2}\Theta\delta_{1-w_1}\mathcal{T}=-\frac{2\pi}{p_1^2}\log\left(\frac{-\bar{w}_1}{1-\bar{w}_1}\right).
\esp\end{align}

We finish the discussion of these single cuts with three comments. First, we note that eqs.~(\ref{cutp1ex3}), (\ref{cutm12ex3}) and (\ref{cutm13ex3}) reproduce the direct calculation of the cuts, as expected. Second, we have confirmed \refE{eq:firstentry-alphabet-example} as, in that form, we can indeed read the correct (symbol of the)  discontinuity across the branch cut of each of the invariants appearing in the first entry. Finally, we have shown that writing the symbol in the special form of \refE{eq:firstentry-alphabet-example} is not necessary or even natural from the point of view of the relations between $\Disc$ and $\delta$, as the relations are formulated in terms of individual symbol letters and not some particular combination of them. In other cases where similar variables are needed, we prefer to present the most compact expression of the symbol.

\paragraph{Double cuts:}
The only double cut we can consider is the double cut in the internal masses. Since the two masses are connected to an external massive leg, we claimed in section \ref{sec:seqmcut} that these double cuts should vanish. Indeed, this two-propagator cut can only be interpreted as a $p_1^2$ channel cut, which vanishes when evaluated in the region where the double mass discontinuity should be computed.

The double cut $\cut_{m_{12}^2,m_{13}^2}T=\cut_{m_{13}^2,m_{12}^2}T$ is computed in the region
\begin{equation}
R_2^{m_{12}^2,m_{13}^2}\,:\qquad m_{12}^2<0\,,\qquad m_{13}^2<0\,,\qquad p_1^2<\left(\sqrt{m_{12}^2}+\sqrt{m_{13}^2}\right)^2.
\end{equation}
In terms of the variables $w_1$ and $\bar{w}_1$, this region is split into two disconnected subregions $R_{2a}^{m_{12}^2,m_{13}^2}$ and $R_{2b}^{m_{12}^2,m_{13}^2}$,
\begin{align}\bsp
R_{2a}^{m_{12}^2,m_{13}^2}\,:\qquad \bar{w}_1<w_1<0\, ,\qquad\qquad R_{2b}^{m_{12}^2,m_{13}^2}\,:\qquad 1<\bar{w}_1<w_1.
\esp\end{align}

For $\cut_{m_{12}^2,m_{13}^2}T$, we start in region $R_1^{m_{12}^2}$. In $R_{2a}^{m_{12}^2,m_{13}^2}$, $m_{13}^2-i0$ implies $w_1+i0$, and the nonvanishing coefficients $a_2(m_{13}^2,x_2)$ are
\begin{align*}
a_2(m_{13}^2,w_1)=2\pi i\, ,\qquad a_2(m_{13}^2,1-w_1)=2\pi i\, .
\end{align*}
We then get
\begin{equation}
\cut_{m_{12}^2,m_{13}^2}T=\Disc_{m_{12}^2,m_{13}^2}T\cong-\frac{4\pi^2 i}{p_1^2}\Theta\left[\delta_{\bar{w_1},w_1}+\delta_{\bar{w_1},1-w_1}\right]\mathcal{T}=0\,.
\end{equation}
In $R_{2b}^{m_{12}^2,m_{13}^2}$, all the coefficients $a_2(m_{13}^2,x_2)$ vanish so that we again find
\begin{equation}
\cut_{m_{12}^2,m_{13}^2}T=\Disc_{m_{12}^2,m_{13}^2}T=0\,.
\end{equation}

For $\cut_{m_{13}^2,m_{23}^2}T$, we start in region $R_1^{m_{13}^2}$. In $R_{2a}^{m_{12}^2,m_{13}^2}$,  all the coefficients $a_2(m_{13}^2,x_2)$ vanish and we get
\begin{equation}
\cut_{m_{13}^2,m_{12}^2}T=\Disc_{m_{12}^2,m_{13}^2}T=0\,.
\end{equation}
In $R_{2b}^{m_{12}^2,m_{13}^2}$, $m_{12}^2-i0$ implies $\bar{w}_1+i0$. The nonvanishing coefficients $a_2(m_{13}^2,x_2)$ are
\begin{align*}
a_2(m_{12}^2,\bar{w}_1)=2\pi i\, ,\qquad a_2(m_{12}^2,1-\bar{w}_1)=2\pi i\, 
\end{align*}
and we get
\begin{equation}
\cut_{m_{13}^2,m_{12}^2}T=\Disc_{m_{12}^2,m_{13}^2}T\cong\frac{4\pi^2 i}{p_1^2}\Theta\left[\delta_{1-w_1,\bar{w_1}}+\delta_{1-w_1,1-\bar{w_1}}\right]\mathcal{T}=0\,.
\end{equation}

We thus find consistent results in all subregions and for either order of the discontinuities: for all cases, the result of the double discontinuity is zero. As already mentioned, this result illustrates the discussion in \ref{sec:seqmcut}.

\subsubsection{$T\left(p_1^2,p_2^2,p_3^2;m_{12}^2,0,0\right)$}

With our last example, we come back to the case mentioned in section \ref{sec:i0ambiguity} to show that we have lifted the ambiguity of the imaginary part of some symbol letters correctly.

The relations among cuts, discontinuities and the coproduct in this example are straightforward to obtain. Indeed, the nonzero internal mass is a simple generalization of the example studied in ref.~\cite{Abreu:2014cla}. We give the full set of relations for cuts in external channels, to verify that the procedure described in section \ref{sec:i0ambiguity} to fix this ambiguity does indeed give the correct result. We will not present cuts in the internal mass here, because we have already given several examples of this type of discontinuity, and they would not teach us anything new.

To get rational symbol letters, we use the variables defined in  \refE{eq:z_def}, and also define as usual $\mu_{12}=m^2_{12}/p_1^2$. Expressions for the integral, its symbol, and its cuts can be found in appendix \ref{sec:tp1p2p3m12}. The symbol alphabet can be found in table \ref{tab:3massTriangles}. The regions where single cuts are computed are
\begin{align}\bsp
R_1^{p_1^2}\,:&\qquad p_1^2>m^2_{12}\,,\qquad p_2^2<0\,,\qquad p_3^2<0\,,\qquad m^2_{12}>0\, ,\\
R_1^{p_2^2}\,:&\qquad p_1^2<0\,,\qquad p_2^2>m^2_{12}\,,\qquad p_3^2<0\,,\qquad m^2_{12}>0\, ,\\
R_1^{p_3^2}\,:&\qquad p_1^2<0\,,\qquad p_2^2<0\,,\qquad p_3^2>0\,,\qquad m^2_{12}>0\, .
\esp\end{align}
We note that these regions are not the complete regions in which single cuts are nonzero. For instance, in $R_1^{p_1^2}$ we could have allowed $0<p_2^2<m_{12}^2$. This complicates the discussion in terms of the $z$ and $\bz$ variables, and does not teach us anything new, so in this discussion we restrict the cut regions to the subregions defined above. In terms of $z$, $\bz$ and $\mu_{12}$, they are
\begin{align}\bsp
R_1^{p_1^2}\,:&\quad z>1\,,\quad\zbar<0\,,\quad 0<\mu_{12}<1\,,\quad z-\mu_{12}>0\,,\quad \zbar-\mu_{12}<0\,,\quad z\zbar-\mu_{12}<0\, \\
R_1^{p_2^2}\,:&\quad 0<z<1\,,\quad \zbar<0\,,\quad \mu_{12}<0\,,\quad z-\mu_{12}>0\,,\quad \zbar-\mu_{12}<0\,,\quad z\zbar-\mu_{12}<0\,\\
R_1^{p_3^2}\,:&\quad  z>1\,,\quad 0<\zbar<1\,,\quad \mu_{12}<0\,,\quad z-\mu_{12}>0\,,\quad \zbar-\mu_{12}>0\,,\quad z\zbar-\mu_{12}>0\, .
\esp\end{align}
For single cuts, knowing that $p_i^2=p_i^2+i0$, there is no ambiguity in determining the sign of the imaginary part of the relevant symbol letters in the relevant kinematic region. We then find
\begin{align}\bsp
\cut_{p_1^2}T=-\Disc_{p_1^2}T&\cong\frac{2\pi}{p_1^2(z-\bz)}\Theta\left[\delta_{1-z}+\delta_{\mu_{12}}+\delta_{z\bz-\mu_{12}}\right]\mathcal{T}\\
&=-\frac{2\pi}{p_1^2(z-\bz)}\Theta\delta_{1-\mu_{12}}\mathcal{T},\\
\cut_{p_2^2}T=-\Disc_{p_2^2}T&\cong-\frac{2\pi}{p_1^2(z-\bz)}\Theta\delta_{z\bz-\mu_{12}}\mathcal{T},\\
\cut_{p_3^2}T=-\Disc_{p_3^2}T&\cong-\frac{2\pi}{p_1^2(z-\bz)}\Theta\delta_{1-z}\mathcal{T}.
\esp\end{align}
For the $p_1^2$ channel cut, we used the fact that there is no branch point at $p_1^2=0$ to find a simpler relation.

The double cuts are computed in the regions
\begin{align}\bsp
R_2^{p_1^2,p_2^2}\,:&\qquad p_1^2>m^2_{12}\,,\qquad p_2^2>m_{12}^2\,,\qquad p_3^2<0\,,\qquad m^2_{12}>0\, ,\\ 
R_1^{p_1^2,p_3^2}\,:&\qquad p_1^2>m^2_{12}\,,\qquad p_2^2<0\,,\qquad p_3^2>0\,,\qquad m^2_{12}>0\, ,\\
R_1^{p_2^2,p_3^2}\,:&\qquad p_1^2<0\,,\qquad p_2^2>m_{12}^2\,,\qquad p_3^2>0\,,\qquad m^2_{12}>0\,  .
\esp\end{align}
We leave it as an exercise to determine the sign of the symbol letters and their imaginary parts in each of these regions. This is straightforward for all double discontinuities, except for $\cut_{p_2^2,p_1^2}T$, in which case the imaginary part of $\zbar-\mu_{12}$ does not have a definite sign in the cut region. We showed how this issue could be addressed in section  \ref{sec:i0ambiguity}, where we also mentioned we could check we had the correct result by comparing $\cut_{p_2^2,p_1^2}T$ and $\cut_{p_1^2,p_2^2}T$.

The full set of relations among cuts, discontinuities and coproducts is
\begin{align}\bsp
\cut_{p_1^2,p_2^2}T&=\Disc_{p_1^2,p_2^2}T\cong\frac{4\pi^2 i}{p_1^2(z-\bz)}\Theta\delta_{1-\mu_{12},\zbar-\mu_{12}}\mathcal{T}\, ,\\
\cut_{p_2^2,p_1^2}T&=\Disc_{p_2^2,p_1^2}T\cong-\frac{4\pi^2 i}{p_1^2(z-\bz)}\Theta\left[\delta_{z\zbar-\mu_{12},1-z}+\delta_{z\zbar-\mu_{12},\zbar}+\delta_{z\zbar-\mu_{12},\zbar-\mu_{12}}\right]\mathcal{T}\, ,\\
\cut_{p_1^2,p_3^2}T&=\Disc_{p_1^2,p_3^2}T\cong \frac{4\pi^2 i}{p_1^2(z-\bz)}\Theta\delta_{1-\mu_{12},1-z}\mathcal{T}\, ,\\
\cut_{p_3^2,p_1^2}T&=\Disc_{p_3^2,p_1^2}T\cong-\frac{4\pi^2 i}{p_1^2(z-\bz)}\Theta \delta_{1-z,\zbar-\mu_{12}}\mathcal{T}\, ,\\
\cut_{p_2^2,p_3^2}T&=\Disc_{p_2^2,p_3^2}T\cong\frac{4\pi^2 i}{p_1^2(z-\bz)}\Theta \delta_{z\zbar-\mu_{12},1-z}\mathcal{T}\, ,\\
\cut_{p_3^2,p_2^2}T&=\Disc_{p_3^2,p_2^2}T\cong\frac{4\pi^2 i}{p_1^2(z-\bz)} \Theta \delta_{1-z,\zbar-\mu_{12}}\mathcal{T}\, .
\esp\end{align}
Using these results and the expressions given in appendix \ref{sec:tp1p2p3m12}, we indeed verify that $\cut_{p_1^2,p_2^2}T=\cut_{p_2^2,p_1^2}T$. We have also checked that all the relations are satisfied for all pairs of external channels.

%% file: Sec_Reconstruction.tex
\section{Reconstruction of Feynman integrals via the coproduct}
\label{sec:reconstruction}

A major motivation for computing cut integrals is that they contain a great deal of information, sometimes sufficient to reconstruct the original uncut integral, which can be harder to compute directly.  The relations between cuts and coproducts suggest that the Hopf algebra can be used as a tool in this reconstruction.  In this section, we show how this works in the case of the massive triangles.

In \citep{Abreu:2014cla} it was shown that for the one-loop triangle and the three-point two-loop ladder diagrams with three external and no internal masses, the symbol of the uncut  pure transcendental functions could be reconstructed  through simple algebraic manipulations starting from the symbol of a single channel cut. Two strategies were presented there: in the first one, the symbol was reconstructed by adding the simplest terms  necessary to satisfy the first-entry and integrability conditions; in the second one, an ansatz with undetermined coefficients was constructed and then constrained using the symmetries obeyed by the function along with the integrability and first-entry conditions. In both cases, the symbol was completely fixed and could then be integrated to get the full uncut Feynman integrals.

We now show how similar manipulations can be used to reconstruct the symbol of triangles with internal masses. We then comment on how to recover information invisible to the symbol to get the full uncut function. As in \citep{Abreu:2014cla}, we focus on {\em finite} triangles, because the reconstruction  procedure works for the full integral but not necessarily for individual terms in the $\epsilon$ expansion.

The reconstruction procedure detailed below requires knowledge of the symbol alphabet.  
In the presence of internal masses, we observe that some symbol letters appear only in mass cuts, and not in any of the channel cuts. Compared to the massless case, then, we need to add more rules to be able to construct the full symbol starting from a channel cut.   Nevertheless, we see that channel cuts highly constrain the full function.  Indeed, we prefer to start from channel cuts rather than mass cuts, based on the idea that dispersive representations of Feynman diagrams are written in terms of channel discontinuities.  In practice, we have found reconstruction from channel cuts to be more straightforward and successful, so that is what we present here.

In this section, we exclude from our discussion the triangle with three internal masses and three external massive legs, because it does not have a rational symbol alphabet. Indeed, although in principle we see no obstacle to reconstructing the symbol through a similar procedure, the complexity of the symbol letters does not lead so directly to clean linear relations when imposing integrability of the symbol ansatz.

\subsection{Constructing and constraining an ansatz for the symbol}

The observation that a single unitarity cut suffices to reconstruct the symbol of a Feynman integral is not surprising, given its representation as a dispersion integral \cite{Remiddi:1981hn,vanNeerven:1985xr,Ball:1991bs,Abreu:2014cla}. In this representation, a Feynman integral is written as an integral over its discontinuity across a branch cut, integrated along the branch cut itself. In our reconstruction procedure, the knowledge of the discontinuity is replaced by the knowledge of the cut, and the knowledge of the integration region by that of the first-entry condition.

Our general strategy is the following. We observe that the symbol alphabets of the scalar triangles we are investigating follow a pattern. With some experience, we are able to write an ansatz for their symbol, in terms of unknown numerical coefficients. Then, by imposing the knowledge of one channel cut, the first-entry condition, the integrability condition, the absence of trivial terms (of the form $x\otimes x$) and the symmetries of the function, we are able to fix all of the unknown coefficients. We now give more specific rules for each of the steps just mentioned.

We start by explaining how to build the ansatz. First, we note that if the diagram is a function of $n$ invariants, the pure functions concerned in this section are functions of $n-1$ dimensionless variables only. For concreteness, we always choose to normalize our variables by an external invariant. The procedure starts by listing the possible first entries. These are completely fixed by the first-entry condition; see sec.~\ref{sec:firstentry}.
Listing the second entries is more difficult than listing the first entries. It can however be done based on the knowledge of a cut integral, and, for the letters that do not appear in channel cuts, by the empirical observations we list below.
\paragraph{Listing the second entries:}We always start from a single cut in an external channel. We observe the presence of the following terms in the set of second entries:
\begin{itemize}
\item All letters of the symbol alphabet of the channel cut taken as the known starting point.
\item Differences of internal masses, or their equivalents in terms of $w_1$ and $\bar{w}_1$; see eq.~\eqref{eq:mus_in_terms_of_w1s}.
\item For triangles with two external invariants, ratios of external invariants. In our examples, this is just $p_2^2/p_3^2$. For the examples with three external massive channels where we must use the variables $z$ and $\bar{z}$ (see \eqref{eq:z_def}), this condition is replaced by the presence of the letters $z$, $\zbar$, $(1-z)$ and $(1-\zbar)$. 
\end{itemize}
The terms generated through the above rules are added as cofactors of all the first entries, each multiplied by an undetermined numerical coefficient. For the first entry corresponding to the cut assumed to be known, these coefficients are of course fixed by the cut result. For the other first entries, they must be determined from additional considerations, according to the procedure we now describe.

\paragraph{Fixing the coefficients:}We fix all coefficients according to the following steps:
\begin{enumerate}
\item We discard integrable terms of the form $x \otimes x$, as they are not needed in order to construct a minimal integrable symbol.
\item
Since the first-entry condition involves the original Mandelstam invariants, the dimensionless variables appearing in the symbol should be expanded when imposing this condition.  Notably, we sometimes normalize the invariants by a variable $p_i^2$ with a nonzero mass threshold, so that $p_i^2$ should not ultimately appear as a first entry by itself, although it shows up superficially in the expansion of the dimensionless variables.
Thus, all of the second-entry cofactors of this $p_i^2$ should combine to give zero.
\item 
We use the integrability condition, eq.~\eqref{eq:integrability}, to fix the remaining parameters.
\end{enumerate}
These three rules are already highly constraining and indeed sufficient for most examples. If they are not, in particular in cases where we use the $z,~\bar{z},~w_1,\text{and }\bar{w}_1$ variables, they can be complemented by the following:
\begin{enumerate}
\setcounter{enumi}{3}
\item  Impose antisymmetry under $z\leftrightarrow \bz$ and symmetry under $w_1\leftrightarrow \bar{w}_1$. Indeed, the Feynman integrals are functions of the invariants only and must thus be symmetric under these transformations. When $z$ and $\zbar$ are necessary, there is an antisymmetric rational prefactor, and so the pure function must be antisymmetric as well.
\item If there is a symmetry under the exchange of the legs with momenta $p_2$ and $p_3$, impose symmetry under the simultaneous transformations $z \rightarrow 1-\bar{z}$, $\bar{z} \rightarrow 1-z$, $w_1 \rightarrow 1-\bar{w}_1$, $\bar{w}_1 \rightarrow 1-w_1$.\\
\end{enumerate}

We now illustrate these rules in some examples. The example in appendix \ref{sec:tp1m23} is trivial and the one in appendix \ref{sec:tp1m12} divergent, so we will not address them. The next-simplest example is $T(p_1^2,0,0;m_{12}^2,m_{23}^2,0)$---see appendix \ref{sec:tp1m12m23}---and we now show how to construct the ansatz for this case. We normalize the internal masses by $p_1^2$, giving the dimensionless variables $m^2_{12}/p_1^2 \equiv \mu_{12}$ and $m^2_{23}/p_1^2 \equiv \mu_{23}$, and we assume knowledge of the $p_1^2$ cut, \refE{eq:cutp1tp1m12m23}. Applying the rules given above for writing the ansatz, we get
\begin{align}\bsp
(\mu_{12}-1)&\otimes [(\mu_{23}-1-\mu_{12})-\mu_{23}]\\
+\mu_{12}&\otimes[a_1(\mu_{23}-1-\mu_{12})+a_2\mu_{23}+a_3(\mu_{12}-\mu_{23})] \\
+\mu_{23}&\otimes[b_1(\mu_{23}-1-\mu_{12})+b_2\mu_{23}+b_3(\mu_{12}-\mu_{23})]
\esp\end{align}
Our task is now to fix the coefficients $a_i$ and $b_i$. In this case, using rules 1), 2) and 3) above fixes all coefficients, and we reproduce the symbol in \refE{eq:symbolTp1m12m23}.

An example of similar complexity is the triangle $T(0,p_2^2,p_3^2;m_{12}^2,0,0)$, appendix \ref{sec:tp2p3m12}. We choose to normalize by $p_2^2$, and define the variables $m^2_{12}/p_2^2 \equiv \mu$ and $p_3^2/p_2^2 \equiv u$. We assume knowledge of the $p_2^2$ cut, \refE{eq:p2cutTp2p3m12}. According to the above steps, the general ansatz for the symbol reads:
\begin{align}\bsp
(\mu - 1) &\otimes \{ u+ \mu - (\mu +u-1) \} \\
u &\otimes \{a_1 u+ a_2 \mu + a_3 (\mu +u-1) \} \\
+ \mu &\otimes \{ b_1 u + b_2 \mu +b_3 (\mu +u-1) \} 
\esp\end{align}
Our task  is now to fix the coefficients $a_i$ and $b_i$. As in the previous example, rules 1), 2) and 3) are sufficient and we reproduce the symbol in \refE{eq:symbolTp2p3m12}.

As a final example of our rules to build the ansatz, we look at the most complicated case we address, $T(p_1^2,p_2^2,p_3^2;m_{12}^2,0,m_{13}^2)$, given in appendix \ref{sec:tp1p2p3m12m13}. This requires using the variables $z$, $\zbar$, $w_1$ and $\bar{w}_1$. We assume knowledge of the $p_1^2$ cut, \refE{eq:cutp1tp1p2p3m12m13}. Following our rules, the ansatz is
\begin{align}\bsp
&w_1(1-\bar{w}_1)\otimes\Big[(z-w_1)-(z-\bar{w}_1)-(\zbar-w_1)+(\zbar-\bar{w}_1)\Big]\\
+&(z\zbar-w_1\bar{w}_1)\otimes\Big[a_1(z-w_1)+a_2(z-\bar{w}_1)+a_3(\zbar-w_1)+a_4(\zbar-\bar{w}_1)\\
&\qquad\qquad+a_5 z+a_6 \zbar+a_7(1-z)+a_8(1-\zbar)+a_9(w_1\bar{w}_1-(1-w_1)(1-\bar{w}_1))\Big]\\
+&((1-z)(1-\zbar)-(1-w_1)(1-\bar{w}_1))\otimes\Big[a_i\to b_i\Big]+w_1\bar{w}_1\otimes \Big[a_i\to c_i\Big]\\
+&(1-w_1)(1-\bar{w}_1)\otimes \Big[a_i\to d_i\Big]\,,
\esp\end{align}
and we must now determine the coefficients $a_i$, $b_i$, $c_i$ and $d_i$. Interestingly, also for this case  all we need are rules 1), 2) and 3) to fix all coefficients.

For all remaining examples, building the ansatz can be done in a similar way as illustrated above. We now list the rules we must apply to fix the coefficients of the ansatz of the remaining examples (for all cases, we assume knowledge of the $p_1^2$ cut):
\begin{itemize}
\item $T(p_1^2,0,0;m_{12}^2,0,m_{13}^2)$, appendix \ref{sec:tp1m12m13}. Rules 1), 2) and 3) are sufficient.
\item $T(p_1^2,0,0;m_{12}^2,m_{23}^2,m_{13}^2)$, appendix \ref{sec:tp1m12m23m13}. Rules 1), 2), 3), 4) and 5) are needed.
\item $T(p_1^2,p_2^2,p_3^3;m_{12}^2,0,0)$, appendix \ref{sec:tp1p2p3m12}. Rules 1), 2) and 3) are sufficient.
\end{itemize}

\subsection{Reconstructing the full function from the symbol}

We now explain how we integrate the symbol to get the full function. Although integrating a symbol is in general an unsolved problem, it is a simple problem for weight two functions where a complete basis is even known to exist in terms of classical polylogarithms, see e.g.~\cite{Duhr:2011zq}. Once we have found a function that matches our symbol, all that remains to be done is fixing terms that are invisible to the symbol, in our case weight one functions multiplied by $\pi$ and terms proportional to $\zeta_2$.

Powers of $\pi$ are typically generated by analytic continuation and appear multiplied by $i$. Working in the Euclidean region where the function is real and away from any branch cut avoids this problem.

To fix the terms proportional to $\zeta_2$, we can use two strategies. The first, which always works, is to evaluate the integrated symbol numerically at a single point and compare it to a numerically integrated Feynman parametrization of the diagram. The difference must be a rational number multiplied by $\zeta_2$, which completely determines our function. Alternatively, when possible, we can use the symmetries of the diagram to check if terms proportional to $\zeta_2$ are allowed.

As examples, consider $T(p_1^2,0,0; 0,m_{23}^2,0)$ and $T(p_1^2,p_2^2,p_3^2; m_{12}^2,0,0)$. In the first case, there is no symmetry consideration to fix terms proportional to $\zeta_2$, and we must thus rely on numerical comparisons. In the second example, there is a rational prefactor antisymmetric under $z\leftrightarrow\zbar$, and thus the pure function must be antisymmetric under this transformation (the full function must be symmetric). This forbids the existence of terms proportional to $\zeta_2$.

%% file: Sec_Discussion.tex
\section{Discussion}
\label{sec:discussion}

In this paper we have studied the analytic structure of one-loop three-point Feynman integrals with different configurations of internal and external masses. More specifically, we have investigated the structure revealed by the unitarity cuts of triangle integrals with massive internal legs, by establishing a relation between cut diagrams and specific entries of the coproduct of these integrals. This generalizes the results of \cite{Abreu:2014cla}, where only diagrams with massless propagators were considered.

The main conclusions of our investigations are the following. First, the first-entry condition has to be generalized. Indeed, for diagrams without internal masses the first-entry condition simply requires that the first entries of the coproduct must be external invariants. When internal masses are present, this is modified in two ways: we can either have external channels minus their threshold or internal masses themselves as the first entry. Stated more generally, the first entries are (arguments of) logarithms with branch points at the boundaries of the Euclidean region.

Second, we have generalized our cutting rules to correctly capture all the discontinuities of diagrams with internal masses. For discontinuities in external channels, the results of \cite{Abreu:2014cla} can be used without modification. For discontinuities in internal masses, we must define  new cutting rules incorporating single-propagator cuts. We have also discussed our strategy to compute the integrals obtained after applying our cutting rules. As also argued in \cite{Abreu:2014cla}, we observe that calculating cuts is a good way to identify appropriate variables for each diagram. Even in the case of the fully massive triangle, where we were not able to find a rational symbol alphabet, computing the individual cuts still points to useful variables.

Third, we have established our relations between discontinuities and cuts, \refE{eq:cutequalsdisc}, and between discontinuities and  coproduct entries, \refE{eq:discequalsdelta}. These are then combined to relate cuts to coproduct entries, \refE{eq:cutequalsdelta}, which is the central result of the paper. The relations we have obtained are mostly a straightforward generalization of the ones in \cite{Abreu:2014cla}, aside from the restriction that when computing multiple discontinuities in external channels and in internal masses, we should first take the discontinuities in the external channels and then in the external masses (failure to do so leads to wrong overall signs). 
We have illustrated our relations in several examples (in section \ref{sec:examples}).

Finally, we have shown how channel cuts highly constrain the symbol of triangles with internal masses. Indeed, we are able to completely constrain a general ansatz for the symbol of each triangle (except the fully massive triangle) using the knowledge of a single channel cut, the integrability condition and the symmetries of the functions. However, building the ansatz is more complicated than in the absence of internal masses \cite{Abreu:2014cla}, and we have had to postulate rules that determine how to construct symbol letters not appearing in channel cuts. These rules are obtained empirically and are specific to the class of diagrams we are studying. Once the symbol is known, we explain how to reconstruct the function by fixing terms invisible to the symbol. It would be very interesting to see whether reconstruction can also be done starting from cuts in internal masses.

While some of the relations we present are justified on firmer grounds, such as the relation between coproduct entries and discontinuities, others are conjectures that generalize well established  results. For instance, the relation between multiple cuts and discontinuities is a generalization of the largest time equation, which relates single cuts and discontinuities. While we believe we have given ample evidence for the validity of all our relations, it would be good to have proofs.

We have restricted our examples to one-loop scalar triangle diagrams with different mass configurations. However, we believe that several of our conclusions generalize in a straightforward manner to more complicated diagrams, provided they can still be written in terms of multiple polylogarithms. Indeed, the discussion of sections \ref{sec:firstentry}, \ref{sec:two_types_of_cuts} and \ref{sec:relationcutdelta} is completely general. Studying diagrams with more external legs and of higher loop order would certainly be interesting: our expectation is that the complications arising in such configurations are mostly related to the increase of the number of external channels, and that the treatment of internal masses can be done along the same lines as what is presented here. Of course, the larger number of scales in the problem will lead to more complicated symbol alphabets, but we believe that computing the cuts of such diagrams would help find the most convenient choice of variables.

More generally, we believe that a better understanding of the analytic structure of Feynman diagrams is fundamental to develop more efficient computational methods. Supported by the results of this paper, we believe that the coproduct of the Hopf algebra of multiple polylogarithms is an appropriate tool to tackle this problem.

%% file: appendices.tex
\section{Feynman rules and definitions} \label{app_FeynmanRules}
Our conventions for the Feynman rules of the scalar diagrams we consider are:
\begin{itemize}
\item Vertex:
\begin{equation}
  \begin{tikzpicture}[baseline={([yshift=-.5ex]current bounding box.center)},vertex/.style={anchor=base,
    circle,fill=black!25,minimum size=18pt,inner sep=2pt}]
 \draw[fill=black] (0,0) circle [radius=0.1];
\end{tikzpicture}
  = i
\end{equation}
\item Complex conjugated vertex:
\begin{equation}
  \begin{tikzpicture}[baseline={([yshift=-.5ex]current bounding box.center)},vertex/.style={anchor=base,
    circle,fill=black!25,minimum size=18pt,inner sep=2pt}]
 \draw[fill=white] (2,0) circle [radius=0.1];
\end{tikzpicture}
  = -i
\end{equation}
\item Propagator:
\begin{equation}
  \begin{tikzpicture}[baseline={([yshift=-.5ex]current bounding box.center)},vertex/.style={anchor=base,
    circle,fill=black!25,minimum size=18pt,inner sep=2pt}]
 \draw[line width=.5mm] (0,0) -- (1.5,0);
 \draw[fill=black] (0,0) circle [radius=0.1];
 \draw[fill=black] (1.5,0) circle [radius=0.1];
\end{tikzpicture}
  = \frac{i}{p^2-m^2 + i0}
\end{equation}
Massive (massless) propagators are drawn with a thick (thin) line.
\item Complex conjugated propagator:
\begin{equation}
  \begin{tikzpicture}[baseline={([yshift=-.5ex]current bounding box.center)},vertex/.style={anchor=base,
    circle,fill=black!25,minimum size=18pt,inner sep=2pt}]
 \draw[line width=.5mm] (0,0) -- (1.5,0);
 \draw[fill=white] (0,0) circle [radius=0.1];
 \draw[fill=white] (1.5,0) circle [radius=0.1];
\end{tikzpicture}
  = \frac{-i}{p^2-m^2 - i0}
\end{equation}
Massive (massless) propagators are drawn with a thick (thin) line.
\item Cut propagator for cut in an external channel:
\begin{equation}
  \begin{tikzpicture}[baseline={([yshift=-.5ex]current bounding box.center)},vertex/.style={anchor=base,
    circle,fill=black!25,minimum size=18pt,inner sep=2pt}]
 \draw[line width=.5mm] (0,0) -- (1,0);
 \draw[fill=white] (0,0) circle [radius=0.1];
 \draw[fill=black] (1,0) circle [radius=0.1];
 \draw[dotted,line width = 0.3mm, color = red] (0.5,0.3)--(0.5,-0.3);
 \node at (1.5,0) {=};
 \draw[line width=.5mm] (2,0) -- (3,0);
 \draw[fill=black] (2,0) circle [radius=0.1];
 \draw[fill=white] (3,0) circle [radius=0.1];
 \draw[dotted,line width = 0.3mm, color = red] (2.5,0.3)--(2.5,-0.3);
 \node at (3.5,0) {=};
 \draw[line width=.5mm] (4,0) -- (5,0);
 \draw[fill=black] (4,0) circle [radius=0.1];
 \draw[fill=black] (5,0) circle [radius=0.1];
 \draw[dotted,line width = 0.3mm, color = red] (4.4,0.3)--(4.4,-0.3);
  \draw[dotted,line width = 0.3mm, color = red] (4.6,0.3)--(4.6,-0.3);
 \node at (5.5,0) {=};
  \draw[line width=.5mm] (6,0) -- (7,0);
 \draw[fill=white] (6,0) circle [radius=0.1];
 \draw[fill=white] (7,0) circle [radius=0.1];
 \draw[dotted,line width = 0.3mm, color = red] (6.4,0.3)--(6.4,-0.3);
  \draw[dotted,line width = 0.3mm, color = red] (6.6,0.3)--(6.6,-0.3);
\end{tikzpicture}
  = 2\pi\, \delta\left(p^2-m^2\right)
\end{equation}
There is a theta function restricting the direction of energy flow in a cut propagator. For single cuts, our convention is that energy flows from black to white. For multiple cuts, there are separate color labels for each cut---see section \ref{sec:two_types_of_cuts} for details.
There can be multiple thin dotted lines indicating cuts on the same propagator without changing its value. However, each thin dotted line implies complex conjugation of a region of the diagram.
\item Cut propagator for cut in an internal mass:
\begin{equation}
  \begin{tikzpicture}[baseline={([yshift=-.5ex]current bounding box.center)},vertex/.style={anchor=base,
    circle,fill=black!25,minimum size=18pt,inner sep=2pt}]
 \draw [line width=.5mm] (0,0) -- (1.5,0);
 \draw[fill=black] (0,0) circle [radius=0.1];
 \draw[fill=black] (1.5,0) circle [radius=0.1];
 \draw[dashed,line width = 1mm, color = red] (0.75,0.3)--(0.75,-0.3);
 \node at (2,0) {=};
  \draw[line width=.5mm] (2.5,0) -- (4,0);
 \draw[fill=white] (2.5,0) circle [radius=0.1];
 \draw[fill=white] (4,0) circle [radius=0.1];
 \draw[dashed,line width = 1mm, color = red] (3.25,0.3)--(3.25,-0.3);
\end{tikzpicture}
  = 2\pi\, \delta\left(p^2-m^2\right)
\end{equation}
\item Loop factor for loop momentum $k$:
\begin{equation}
\left( \frac{e^{\gamma_E \epsilon}}{\pi^{2-\epsilon}} \right) \int \mathrm{d}^{4-2\epsilon} k ~.
\end{equation}
\end{itemize}

Results for triangles and their cuts often involve the Gauss hypergeometric function $\,_2F_1$ and one of its generalizations, the $F_1$ Appell function. They have the Euler-type integral representations
\begin{align}
\label{gaussHypDef}
\hypgeo{\alpha}{\beta}{\gamma}{z}=\frac{\Gamma(\gamma)}{\Gamma(\beta)\Gamma(\gamma-\beta)}\int_0^1dt\, t^{\beta-1}(1-t)^{\gamma-\beta-1}(1-tz)^{-\alpha}
\end{align}
for $\Re\gamma>\Re\beta>0$, and
\begin{align}
\label{appellf1Def}
\appellf{\alpha}{\beta}{\beta'}{\gamma}{x}{y}=\frac{\Gamma(\gamma)}{\Gamma(\alpha)\Gamma(\gamma-\alpha)}\int_0^1d t\frac{t^{\alpha-1}(1-u)^{\gamma-\alpha-1}}{(1-t x)^{\beta}(1-t y)^{\beta'}}
\end{align}
for $\Re\gamma>\Re\alpha>0$.

\section{One-mass triangles\label{app_oneMass}}

We give explicit expressions for the triangles with one external massive channel that are used as examples in this paper. For all the examples given, we have computed the uncut triangles both through standard Feynman parametrization and through a dispersive integral, and verified agreement of the expressions. Divergent integrals were compared with the results given in ref.\cite{Ellis:2007qk}. For all triangles with one external massive channel considered in the following subsections, we separate the rational prefactor from the pure transcendental function according to the relation
\begin{equation}
T(p_1^2,0,0;m_{12}^2,m_{23}^2,m_{13}^2)=\frac{i}{p_1^2}\mathcal{T}(p_1^2,0,0;m_{12}^2,m_{23}^2,m_{13}^2)\, ,
\end{equation}
where the internal masses are generic and can be zero. Before expansion in the dimensional regularization parameter $\epsilon$, the results will often involve the functions $\,_2F_1$ and $F_1$ defined in eqs.~(\ref{gaussHypDef}) and (\ref{appellf1Def}).

\subsection{$T(p_1^2,0,0;0,m_{23}^2,0)$}\label{sec:tp1m23}
The triangle of  \refF{fig:1mTriangle2} is given by:
\begin{align}\bsp
T(p_1^2,0,0;0,m_{23}^2,0)=&\frac{i e^{\gamma_E\epsilon}}{\epsilon}\left[\frac{(-p_1^2)^{-\epsilon}}{m_{23}^2}\frac{\Gamma(1+\epsilon)\Gamma^2(1-\epsilon)}{\Gamma(2-2\epsilon)}\hypgeo{1}{1-\epsilon}{2-2\epsilon}{-\frac{p_1^2}{m_{23}^2}}\right.\\
&\left.-\left(m_{23}^2\right)^{-1-\epsilon}\frac{\Gamma(1+\epsilon)\Gamma(1-\epsilon)}{\Gamma(2-\epsilon)}\hypgeo{1}{1}{2-\epsilon}{-\frac{p_1^2}{m_{23}^2}}\right]\\
=&\frac{i}{p_1^2}\left(\text{Li}_2\left(\frac{m_{23}^2+p_1^2}{m_{23}^2}\right)-\frac{\pi ^2}{6}\right)+\mathcal{O}(\epsilon)\, .
\esp\end{align}
The symbol is
\begin{align}
\mathcal{S}\left[\mathcal{T}(p_1^2,0,0;0,m_{23}^2,0)\right]=&m_{23}^2\otimes\left(\frac{m_{23}^2+p_1^2}{m_{23}^2}\right)-p_{1}^2\otimes\left(\frac{m_{23}^2+p_1^2}{m_{23}^2}\right)+\mathcal{O}(\epsilon)\, .
\end{align}
\subsubsection{Single cuts}
The cut in the external channel $p_1^2$ is
\begin{align}\bsp
\cut_{p_1^2}T(p_1^2,0,0;0,m_{23}^2,0)=&2 \pi \frac{e^{\gamma_E\epsilon}\Gamma(1-\epsilon)}{\Gamma(2-2\epsilon)}\frac{(p_1^2)^{-\epsilon}}{p_1^2+m_{23}^2}\hypgeo{1}{1-\epsilon}{2-2\epsilon}{\frac{p_1^2}{p_1^2+m_{23}^2}}\\
=&-\frac{2\pi}{p_1^2}\ln\left(\frac{m_{23}^2}{p_1^2+m_{23}^2}\right)+\mathcal{O}(\epsilon)\, .
\esp\end{align}
The cut in the internal mass $m_{23}^2$ is
\begin{align}\bsp
\cut_{m_{23}^2}T(p_1^2,0,0;0,m_{23}^2,0)&=
\frac{2\pi e^{\gamma_E\epsilon}}{\Gamma(2-\epsilon)}\frac{(-m_{23}^2)^{-\epsilon}}{p_1^2+m_{23}^2}\hypgeo{1}{1-\epsilon}{2-\epsilon}{\frac{p_1^2}{p_1^2+m_{23}^2}}\\
&=\frac{2\pi}{p_1^2}\log\left(\frac{m_{23}^2+p_1^2}{m_{23}^2}\right)+\mathcal{O}(\epsilon)\, .
\esp\end{align}

\subsubsection{Double cuts}
The double cut in the external channel $p_1^2$ and the internal mass $m_{23}^2$ is
\begin{align}\bsp
\cut_{p_1^2,m_{23}^2}T(p_1^2,0,0;0,m_{23}^2,0)
=&-4\pi^2 i\frac{e^{\gamma_E\epsilon}}{\Gamma(1-\epsilon)}\frac{(p_1^2)^{-1+\epsilon}(-m_{23}^2)^{-\epsilon}}{(p_1^2+m_{23}^2)^{\epsilon}}\theta(p_1^2+m_{23}^2)\\
=&-\frac{4\pi^2 i}{p_1^2}\theta(p_1^2+m_{23}^2)+\mathcal{O}(\epsilon)\, .
\esp\end{align}

\subsection{$T(p_1^2,0,0;m_{12}^2,0,0)$}\label{sec:tp1m12}

The triangle of  \refF{fig:1mTriangle3} is given by:
\begin{align}\bsp
T(p_1^2,0,0;m_{12}^2,0,0)=&-i \frac{e^{\gamma_E\epsilon}\Gamma(1+\epsilon)}{\epsilon(1-\epsilon)}(m_{12}^2)^{-1-\epsilon}\hypgeo{1}{1+\epsilon}{2-\epsilon}{\frac{p_1^2}{m_{12}^2}}\\
=&\frac{i}{p_1^2}\left[\frac{1}{\epsilon}\ln\left(1-\frac{p_1^2}{m_{12}^2}\right)-\text{Li}_2\left(\frac{p_1^2}{m_{12}^2}\right)-\log ^2\left(1-\frac{p_1^2}{m_{12}^2}\right)\right.\\
&\left.-\log \left(m_{12}^2\right) \log
   \left(1-\frac{p_1^2}{m_{12}^2}\right)\right]+\mathcal{O}\left(\epsilon\right)\, .
\esp\end{align}
The symbol is
\begin{align}\bsp
\mathcal{S}\left[\mathcal{T}(p_1^2,0,0;m_{12}^2,0,0)\right]=&\frac{1}{\epsilon}\frac{m_{12}^2-p_1^2}{m_{12}^2}+m_{12}^2\otimes\frac{m_{12}^2\left(m_{12}^2-p_1^2\right)}{p_1^2}\\
&+\left(m_{12}^2-p_1^2\right)\otimes
\frac{p_1^2}{\left(m_{12}^2-p_1^2\right)^2}+\mathcal{O}\left(\epsilon\right)\,.
\label{eq:symbol-t-p1-m12}
\esp\end{align}

\subsubsection{Single cuts}
The cut in the external channel $p_1^2$ is
\begin{align}\bsp
\cut_{p_1^2}T(p_1^2,0,0;m_{12}^2,0,0)&=-2\pi \frac{e^{\gamma_E\epsilon}\Gamma(1-\epsilon)}{\epsilon\Gamma(1-2\epsilon)}\frac{(p_1^2-m_{12}^2)^{-2\epsilon}}{(p_1^2)^{1-\epsilon}}\\
&=-\frac{2\pi}{p_1^2\epsilon}-\frac{2\pi}{p_1^2}\left(\ln\left(p_1^2\right)-2\ln\left(p_1^2-m_{12}^2\right)\right)+\mathcal{O}\left(\epsilon\right)\,.
\esp\end{align}
The cut in the internal mass $m_{12}^2$ is
\begin{align}\bsp
\Cut_{m_{12}^2}T(p_1^2,0,0;m_{12}^2,0,0)=&-\frac{2\pi e^{\gamma_E\epsilon}}{\epsilon\Gamma(1-\epsilon)}\frac{(-m_{12}^2)^{-\epsilon}}{p_1^2}\hypgeo{1}{\epsilon}{1-\epsilon}{\frac{m_{12}^2}{p_1^2}}\\
=&-\frac{2\pi}{p_1^2\epsilon}+\frac{2\pi}{p_1^2}\left(\ln\left(m_{12}^2-p_1^2\right)+\ln\left(\frac{-m_{12}^2}{-p_1^2}\right)\right)+\mathcal{O}\left(\epsilon\right)\,.
\esp\end{align}

\subsubsection{Double cuts}
The double cut in the external channel $p_1^2$ and the internal mass $m_{12}^2$ is zero.

\subsection{$T(p_1^2,0,0;m_{12}^2,m_{23}^2,0)$}\label{sec:tp1m12m23}

The triangle of \refF{fig:1mTriangle4} is given by:
\begin{align}\bsp
T(p_1^2,0,0;m_{12}^2,m_{23}^2,0)
=&i \frac{e^{\gamma_E\epsilon}\Gamma(1+\epsilon)}{\epsilon(1-\epsilon)\left(m_{12}^2-m_{23}^2\right)}\left[\left(m_{23}^2\right)^{-\epsilon}\hypgeo{1}{1}{2-\epsilon}{\frac{p_1^2}{m_{12}^2-m_{23}^2}}\right.\\
&\left.-\left(m_{12}^2\right)^{-\epsilon}\,F_1\left(1;1,\epsilon;2-\epsilon;\frac{p_1^2}{m_{12}^2-m_{23}^2};\frac{p_1^2}{m_{12}^2}\right)\right]\\
=&\frac{i}{p_1^2}\left[\text{Li}_2\left(\frac{m_{12}^2}{m_{23}^2}\right)-\text{Li}_2\left(\frac{m_{12}^2-p_1^2}{m_{23}^2}\right)\right.\\
&-\log \left(1-\frac{m_{12}^2-p_1^2}{m_{23}^2}\right) \log\left(1-\frac{p_1^2}{m_{12}^2}\right)\\
&\left.+\log\left(\frac{m_{23}^2}{m_{12}^2}\right) \log \left(1-\frac{p_1^2}{m_{12}^2-m_{23}^2}\right)\right]+\mathcal{O}\left(\epsilon\right)\, .
\esp\end{align}
The symbol is
\begin{align}\bsp \label{eq:symbolTp1m12m23}
\mathcal{S}\left[\mathcal{T}(p_1^2,0,0;m_{12}^2,m_{23}^2,0)\right]=&m_{12}^2\otimes\left(\frac{m_{12}^2-m_{23}^2}{m_{23}^2}\right)
+m_{23}^2\otimes\left(1-\frac{p_1^2}{m_{12}^2-m_{23}^2}\right)\\
&-(m_{12}^2-p_1^2)\otimes\left(1-\frac{m_{12}^2-p_1^2}{m_{23}^2}\right)+\mathcal{O}\left(\epsilon\right)\, .
\esp\end{align}

\subsubsection{Single cuts}
The cut in the external channel $p_1^2$ is
\begin{align}\bsp\label{eq:cutp1tp1m12m23}
\cut_{p_1^2}&T(p_1^2,0,0;m_{12}^2,m_{23}^2,0)=\\
=&2\pi\frac{e^{\gamma_E\epsilon}\Gamma(1-\epsilon)}{\Gamma(2-2\epsilon)}\frac{(p_1^2-m_{12})^{1-2\epsilon}}{m_{23}^2(p_1^2)^{1-\epsilon}}\hypgeo{1}{1-\epsilon}{2-2\epsilon}{\frac{m_{12}^2-p_1^2}{m_{23}^2}}\\
=&\frac{2\pi}{p_1^2}\ln\left(1-\frac{m_{12}^2-p_1^2}{m_{23}^2}\right)+\mathcal{O}\left(\epsilon\right)\, .
\esp\end{align}
The cut in the internal mass $m_{12}^2$ is
\begin{align}\bsp
\cut_{m_{12}^2}&T(p_1^2,0,0;m_{12}^2,m_{23}^2,0)=\\
=&-\frac{2\pi}{p_1^2}\frac{e^{\gamma_E\epsilon}}{\Gamma(2-\epsilon)}\frac{(-m_{12}^2)^{1-\epsilon}}{m_{12}^2-m_{23}^2}\appellf{1}{1}{\epsilon}{2-\epsilon}{\frac{m_{12}^2}{m_{12}^2-m_{23}^2}}{\frac{m_{12}^2}{p_1^2}}\\
=&-\frac{2\pi}{p_1^2}\ln\left(\frac{m_{23}^2}{m_{23}^2-m_{12}^2}\right)+\mathcal{O}\left(\epsilon\right)\, .
\esp\end{align}
The cut in the internal mass $m_{23}^2$ is
\begin{align}\bsp
\cut_{m_{23}^2}&T(p_1^2,0,0;m_{12}^2,m_{23}^2,0)=\\
=&-2\pi\frac{e^{\gamma_E\epsilon}}{\Gamma(2-\epsilon)}\frac{(-m_{23}^2)^{-\epsilon}}{m_{12}^2-m_{23}^2}\hypgeo{1}{1}{2-\epsilon}{\frac{p_1^2}{m_{12}^2-m_{23}^2}}\\
=&\frac{2\pi}{p_1^2}\ln\left(1-\frac{p_1^2}{m_{12}^2-m_{23}^2}\right)+\mathcal{O}\left(\epsilon\right)\, .
\esp\end{align}

\subsubsection{Double cuts}
The double cut in the external channel $p_1^2$ and internal mass $m_{23}^2$ is
\begin{align}\bsp
\cut_{p_1^2,m_{23}^2}T(p_1^2,0,0;m_{12}^2,m_{23}^2,0)=&-4\pi^2 i\frac{e^{\gamma_E\epsilon}}{\Gamma(1-\epsilon)}\frac{(p_1^2)^{-1+\epsilon}(-m_{23}^2)^{-\epsilon}}{(p_1^2-m_{12}^2+m_{23}^2)^{\epsilon}}\theta(p_1^2-m_{12}^2+m_{23}^2)\\
=&-\frac{4\pi^2 i}{p_1^2}\theta(p_1^2-m_{12}^2+m_{23}^2)+\mathcal{O}\left(\epsilon\right)\, .
\esp\end{align}
The double cut in the two internal masses is
\begin{align}\bsp
\cut_{m_{12}^2,m_{23}^2}T(p_1^2,0,0;m_{12}^2,m_{23}^2,0)=&-4\pi^2 i\frac{e^{\gamma_E\epsilon}}{\Gamma(1-\epsilon)}\frac{(-p_1^2)^{-1+\epsilon}(-m_{23}^2)^{-\epsilon}}{(m_{12}^2-p_1^2-m_{23}^2)^{\epsilon}}\theta(m_{12}^2-p_1^2-m_{23}^2)\\
&\theta(m_{23}^2-m_{12}^2)\\
=&\frac{4\pi^2 i}{p_1^2}\theta(m_{12}^2-p_1^2-m_{23}^2)\theta(m_{23}^2-m_{12}^2)+\mathcal{O}\left(\epsilon\right)\, .
\esp\end{align}

\subsection{$T(p_1^2,0,0;m_{12}^2,0,m_{13}^2)$}
\label{sec:tp1m12m13}

The triangle of  \refF{fig:1mTriangle1} is given by:
\begin{align}\bsp\label{tp1m12m13Int}
T(p_1^2,0,0;m_{12}^2,0,m_{13}^2)=&i \frac{e^{\gamma_E\epsilon}\Gamma(1+\epsilon)}{\epsilon(1-\epsilon)}(-p_1^2)^{-1-\epsilon}\left[\frac{\left(w_1-\bar{w}_1\right)^{-\epsilon}}{\left(1-\bar{w}_1\right)}\right.\\
&\left(\frac{\left(w_1-1\right)^{1-\epsilon}}{w_1}\appellf{1-\epsilon}{1}{\epsilon}{2-\epsilon}{\frac{w_1-1}{w_1(1-\bar{w}_1)}}{\frac{w_1-1}{w_1-\bar{w}_1}}\right.\\
&\left.-w_1^{-\epsilon}\appellf{1-\epsilon}{1}{\epsilon}{2-\epsilon}{\frac{1}{1-\bar{w}_1}}{\frac{w_1}{w_1-\bar{w}_1}}\right)\\
&\left.-\frac{\left((w_1-1)(1-\bar{w}_1)\right)^{-\epsilon}}{w_1\bar{w}_1}\hypgeo{1}{1-\epsilon}{2-\epsilon}{\frac{1}{w_1\bar{w}_1}}\right]\\
=&\frac{i}{p_1^2}\log \left(\frac{w_1}{w_1-1}\right) \log \left(\frac{-\bar{w}_1}{1-\bar{w}_1}\right)+\mathcal{O}\left(\epsilon\right)\, .
\esp\end{align}
The symbol is
\begin{align}\bsp
\mathcal{S}\left[\mathcal{T}(p_1^2,0,0;m_{12}^2,0,m_{13}^2)\right]=&\left(\frac{w_1}{1-w_1}\right)\otimes\left(\frac{\bar{w}_1}{1-\bar{w}_1}\right)+\left(\frac{\bar{w}_1}{1-\bar{w}_1}\right)\otimes\left(\frac{w_1}{1-w_1}\right)+\mathcal{O}\left(\epsilon\right)\,.
\label{eq:symbol-t-p1-m12-m13}
\esp\end{align}

\subsubsection{Single cuts}
The cut in the external channel $p_1^2$ is
\begin{align}\bsp\label{eq:cutp1tp1m12m13}
\cut_{p_1^2}&T(p_1^2,0,0;m_{12}^2,0,m_{13}^2)=\\
&=-2\pi\frac{e^{\gamma_E\epsilon}\Gamma(1-\epsilon)}{\Gamma(2-2\epsilon)}(p_1^2)^{-1-\epsilon}\frac{(w_1-\bar{w}_1)^{1-2\epsilon}}{\bar{w}_1(w_1-1)}\hypgeo{1}{1-\epsilon}{2-2\epsilon}{\frac{w_1-\bar{w}_1}{\bar{w}_1(w_1-1)}}\\
&=\frac{2\pi}{p_1^2}\left(\ln\left(\frac{w_1}{1-w_1}\right)-\ln\left(\frac{\bar{w}_1}{1-\bar{w}_1}\right)\right)+\mathcal{O}\left(\epsilon\right)\,.
\esp\end{align}
The cut in the internal mass $m_{12}^2$ is
\begin{align}\bsp
\cut_{m^2_{12}}&T(p_1^2,0,0;m_{12}^2,0,m_{13}^2)=\\
=&-2\pi \frac{e^{\gamma_E\epsilon}(-p_1^2)^{-1-\epsilon}}{\Gamma(2-\epsilon)\left(w_1-1\right)}\bar{w}_1^{-\epsilon}\left(w_1-\bar{w}_1\right)^{-\epsilon}\appellf{1-\epsilon}{1}{\epsilon}{2-\epsilon}{\frac{1}{1-w_1}}{\frac{-\bar{w}_1}{w_1-\bar{w}_1}}\\
=&\frac{2\pi}{p_1^2}\ln\left(\frac{w_1}{w_1-1}\right)+\mathcal{O}\left(\epsilon\right)\, .
\esp\end{align}
The cut in the internal mass $m_{13}^2$ is
\begin{align}\bsp
\cut_{m^2_{13}}&T(p_1^2,0,0;m_{12}^2,0,m_{13}^2)=\\
=&-2\pi \frac{e^{\gamma_E\epsilon}(-p_1^2)^{-1-\epsilon}}{\Gamma(2-\epsilon)}\left[-\frac{((1-w_1)(1-\bar{w}_1))^{-\epsilon}}{w_1\bar{w}_1}\hypgeo{1}{1-\epsilon}{2-\epsilon}{\frac{1}{w_1\bar{w}_1}}\right.\\
&\left.+\frac{(1-w_1)^{1-\epsilon}(w_1-\bar{w}_1)^{-\epsilon}}{w_1(\bar{w}_1-1)}\appellf{1-\epsilon}{1}{\epsilon}{2-\epsilon}{\frac{1-w_1}{w_1(\bar{w}_1-1)}}{\frac{w_1-1}{w_1-\bar{w}_1}}\right]\\
=&-\frac{2\pi}{p_1^2}\ln\left(\frac{-\bar{w}_1}{1-\bar{w}_1}\right)+\mathcal{O}\left(\epsilon\right)\, .
\esp\end{align}

\subsubsection{Double cuts}
All double cuts are zero.

\subsection{$T(p_1^2,0,0;m_{12}^2,m_{23}^2,m_{13}^2)$}\label{sec:tp1m12m23m13}

The triangle of \refF{fig:1mTriangle5} is given by\footnote{We wrote the result in terms of harmonic polylogarithms for simplicity. It has a longer expression in terms of classical polylogarithms which can be easily obtained using
\begin{equation*}
G(1,0,x)=\Li_2(x)+\ln(1-x)\ln(x)\,.
\end{equation*}}:
\begin{align}\bsp\label{tp1m12m23m13Int}
T(p_1^2,0,0&;m_{12}^2,m_{23}^2,m_{13}^2)\\
=&i \frac{e^{\gamma_E\epsilon}\Gamma(1+\epsilon)}{\epsilon}(-p_1^2)^{-1-\epsilon}\left[\frac{(w_1-\bar{w}_1)^{-\epsilon}}{(1-\epsilon)(\mu_{23}+w_1(1-\bar{w}_1))}\right.\\
&\left((w_1-1)^{1-\epsilon}\appellf{1-\epsilon}{1}{\epsilon}{2-\epsilon}{\frac{w_1-1}{\mu_{23}+w_1(1-\bar{w}_1)}}{\frac{w_1-1}{w_1-\bar{w}_1}}\right.\\
&-\left.w_1^{1-\epsilon}\appellf{1-\epsilon}{1}{\epsilon}{2-\epsilon}{\frac{w_1}{\mu_{23}+w_1(1-\bar{w}_1)}}{\frac{w_1}{w_1-\bar{w}_1}}\right)\\
&\left.-\frac{(-\mu_{23})^{-\epsilon}}{w_1\bar{w}_1-\mu_{23}}\appellf{1}{1}{\epsilon}{2}{\frac{1}{w_1\bar{w}_1-\mu_{23}}}{\frac{\mu_{23}+(w_1-1)(1-\bar{w}_1)}{\mu_{23}}}\right]\\
=&\frac{i}{p_1^2}\left[\log \left(\frac{w_1}{w_1-1}\right) \log \left(\frac{-\bar{w}_1}{1-\bar{w}_1}\right)-G\left(1,0,\frac{\mu _{23}}{(w_1-1) (\bar{w}_1-1)}\right)\right.\\
&+G\left(1,0,\frac{\mu _{23}}{w_1(\bar{w}_1-1)}\right)+G\left(1,0,\frac{\mu _{23}}{(w_1-1) \bar{w}_1}\right)\\
&\left.-G\left(1,0,\frac{\mu_{23}}{w_1 \bar{w}_1}\right)\right]+\mathcal{O}\left(\epsilon\right)\, .
\esp\end{align}
The symbol is
\begin{align}\bsp
\mathcal{S}\left[\mathcal{T}(p_1^2,0,0;m_{12}^2,m_{23}^2,m_{13}^2)\right]=&\mu_{23}\otimes\frac{(\mu_{23}+w_1(1-\bar{w}_1))(\mu_{23}+\bar{w}_1(1-w_1))}{(\mu_{23}-w_1\bar{w}_1)(\mu_{23}-(1-w_1)(1-\bar{w}_1))}\\
&+w_1\otimes\frac{\mu_{23}-w_1\bar{w}_1}{\mu_{23}+w_1(1-\bar{w}_1)}+\bar{w}_1\otimes\frac{\mu_{23}-w_1\bar{w}_1}{\mu_{23}+\bar{w}_1(1-w_1)}\\
&+(1-w_1)\otimes\frac{\mu_{23}-(1-w_1)(1-\bar{w}_1)}{\mu_{23}+\bar{w}_1(1-w_1)}\\
&+(1-\bar{w}_1)\otimes\frac{\mu_{23}-(1-w_1)(1-\bar{w}_1)}{\mu_{23}+w_1(1-\bar{w}_1)}+\mathcal{O}\left(\epsilon\right)\,.
\esp\end{align}

\subsubsection{Single cuts}
The cut in the external channel $p_1^2$ is
\begin{align}\bsp\label{eq:cutp1tp1m12m23m13}
\cut_{p_1^2}&T(p_1^2,0,0;m_{12}^2,m_{23}^2,m_{13}^2)=\\
=&-2\pi\frac{e^{\gamma_E\epsilon}\Gamma(1-\epsilon)}{\Gamma(2-2\epsilon)}(p_1^2)^{-1-\epsilon}
\frac{(w_1-\bar{w}_1)^{1-2\epsilon}}{\bar{w}_1(w_1-1)-\mu_{23}}\hypgeo{1}{1-\epsilon}{2-2\epsilon}{\frac{w_1-\bar{w}_1}{\bar{w}_1(w_1-1)-\mu_{23}}}\\
=&\frac{2\pi}{p_1^2}\ln\left(\frac{\mu_{23}+w_1(1-\bar{w}_1)}{\mu_{23}+\bar{w}_1(1-w_1)}\right)+\mathcal{O}\left(\epsilon\right)\, .
\esp\end{align}
The cut in the internal mass $m_{12}^2$ is
\begin{align}\bsp
\cut_{m_{12}^2}&T(p_1^2,0,0;m_{12}^2,m_{23}^2,m_{13}^2)=\\
=&\frac{2\pi e^{\gamma_E\epsilon}(-p_1^2)^{-1-\epsilon}}{\Gamma(2-\epsilon)}\frac{\bar{w}_1^{1-\epsilon}(w_1-\bar{w}_1)^{-\epsilon}}{\mu_{23}-\bar{w}_1(w_1-1)}\appellf{1-\epsilon}{1}{\epsilon}{2-\epsilon}{\frac{\bar{w}_1}{\mu_{23}-\bar{w}_1(w_1-1)}}{\frac{-\bar{w}_1}{w_1-\bar{w}_1}}\\
=&\frac{2\pi}{p_1^2}\ln\left(\frac{\mu_{23}-w_1\bar{w}_1}{\mu_{23}+\bar{w}_1(1-w_1)}\right)+\mathcal{O}\left(\epsilon\right)\, .
\esp\end{align}
The cut in the internal mass $m_{23}^2$ is
\begin{align}\bsp
\cut_{m_{23}^2}&T(p_1^2,0,0;m_{12}^2,m_{23}^2,m_{13}^2)=\\
=&2\pi\frac{e^{\gamma_E\epsilon}}{\Gamma(1-\epsilon)}\frac{(-p_1^2)^{-1-\epsilon}}{1-\epsilon}\mu_{23}^{1-\epsilon}\frac{\hypgeo{1}{1}{2-\epsilon}{\frac{\mu_{23}}{(\mu_{23}-(1-w_1)(1-\bar{w}_1))(w_1\bar{w}_1-\mu_{23})}}}{(\mu_{23}-(1-w_1)(1-\bar{w}_1))(w_1\bar{w}_1-\mu_{23})}\\
=&\frac{2\pi}{p_1^2}\ln\left(\frac{(\mu_{23}+w_1(1-\bar{w}_1))(\mu_{23}+\bar{w}_1(1-w_1))}{(\mu_{23}-w_1\bar{w}_1)(\mu_{23}-(1-w_1)(1-\bar{w}_1))}\right)\\
&+\mathcal{O}\left(\epsilon\right)\, .
\esp\end{align}
The cut in the internal mass $m_{13}^2$ is
\begin{align}\bsp
\cut_{m_{13}^2}&T(p_1^2,0,0;m_{12}^2,m_{23}^2,m_{13}^2)=\\
=&-2\pi\frac{e^{\gamma_E\epsilon}}{\Gamma(2-\epsilon)}(-p_1^2)^{-1-\epsilon}\left[\frac{(1-w_1)^{1-\epsilon}(w_1-\bar{w}_1)^{-\epsilon}}{w_1(\bar{w}_1-1)-\mu_{23}}\right.\\
&\appellf{1-\epsilon}{1}{\epsilon}{2-\epsilon}{\frac{1-w_1}{w_1(\bar{w}_1-1)-\mu_{23}}}{\frac{w_1-1}{w_1-\bar{w}_1}}\\
&\left.-\frac{((1-w_1)(1-\bar{w}_1))^{1-\epsilon}\hypgeo{1}{1-\epsilon}{2-\epsilon}{\frac{(1-w_1)(1-\bar{w}_1)}{(w_1(\bar{w}_1-1)-\mu_{23})(\bar{w}_1(w_1-1)-\mu_{23})}}}{(w_1(\bar{w}_1-1)-\mu_{23})(\bar{w}_1(w_1-1)-\mu_{23})}\right]\\
=&\frac{2\pi}{p_1^2}\ln\left(\frac{\mu_{23}-(1-w_1)(1-\bar{w}_1)}{\mu_{23}+\bar{w}_1(1-w_1)}\right)+\mathcal{O}\left(\epsilon\right)\, .
\esp\end{align}

\subsubsection{Double cuts}
The double cut in the external channel $p_1^2$ and internal mass $m_{23}^2$ is
\begin{align}\bsp
\cut_{p_1^2,m_{23}^2}T(p_1^2,0,0;m_{12}^2,m_{23}^2,m_{13}^2)=&-4\pi^2 i\frac{e^{\gamma_E\epsilon}}{\Gamma(1-\epsilon)}(p_1^2)^{-1-\epsilon}\left(\bar{w}_1(w_1-1)-\mu_{23}\right)^{-\epsilon}\\
&\left(-w_1(\bar{w}_1-1)+\mu_{23}\right)^{-\epsilon}\theta(\bar{w}_1(w_1-1)-\mu_{23})\\
&\theta(-w_1(\bar{w}_1-1)+\mu_{23})\\
=&-\frac{4\pi^2 i}{p_1^2}\theta(\bar{w}_1(w_1-1)-\mu_{23})\theta(-w_1(\bar{w}_1-1)+\mu_{23})\\
&+\mathcal{O}\left(\epsilon\right)\, .
\esp\end{align}
The double cut in the internal masses $m_{12}^2$ and $m_{23}^2$ is
\begin{align}\bsp
\cut_{m_{12}^2,m_{23}^2}T(p_1^2,0,0;m_{12}^2,m_{23}^2,m_{13}^2)=&-4\pi^2 i\frac{e^{\gamma_E\epsilon}}{\Gamma(1-\epsilon)}(-p_1^2)^{-1-\epsilon}\theta(w_1\bar{w}_1-\mu_{23})\\
&\left(\mu_{23}-(w_1\bar{w}_1-\mu_{23})(\mu_{23}-(1-w_1)(1-\bar{w}_1))\right)^{-\epsilon}\\
&\theta\left(\mu_{23}-(w_1\bar{w}_1-\mu_{23})(\mu_{23}-(1-w_1)(1-\bar{w}_1))\right)\\
=&\frac{4\pi^2 i}{p_1^2}\theta(w_1\bar{w}_1-\mu_{23})\\
&\theta\left(\mu_{23}-(w_1\bar{w}_1-\mu_{23})(\mu_{23}-(1-w_1)(1-\bar{w}_1))\right)\\
&+\mathcal{O}\left(\epsilon\right)\, .
\esp\end{align}
The double cut in the internal masses $m_{13}^2$ and $m_{23}^2$ is
\begin{align}\bsp
\cut_{m_{13}^2,m_{23}^2}T(p_1^2,0,0;m_{12}^2,m_{23}^2,m_{13}^2)=&-4\pi^2 i\frac{e^{\gamma_E\epsilon}}{\Gamma(1-\epsilon)}(-p_1^2)^{-1-\epsilon}\theta((1-w_1)(1-\bar{w}_1)-\mu_{23})\\
&\left(\mu_{23}-(w_1\bar{w}_1-\mu_{23})(\mu_{23}-(1-w_1)(1-\bar{w}_1))\right)^{-\epsilon}\\
&\theta\left(\mu_{23}-(w_1\bar{w}_1-\mu_{23})(\mu_{23}-(1-w_1)(1-\bar{w}_1))\right)\\
=&\frac{4\pi^2 i}{p_1^2}\theta((1-w_1)(1-\bar{w}_1)-\mu_{23})\\
&\theta\left(\mu_{23}-(w_1\bar{w}_1-\mu_{23})(\mu_{23}-(1-w_1)(1-\bar{w}_1))\right)\\
&+\mathcal{O}\left(\epsilon\right)\, .
\esp\end{align}
All other double cuts are zero.




\section{Two-mass triangles\label{app_twoMass}}

We give explicit expressions for the triangles with two external massive channels that are used as examples in this paper. For all the examples given, we have computed the uncut triangles both through standard Feynman parametrization and through a dispersive integral, and verified agreement of the expressions. Divergent integrals were compared with the results given in ref.~\cite{Ellis:2007qk}. For all triangles with two external massive channels considered in the following subsections, we separate the rational prefactor from the pure transcendental function according to the relation
\begin{equation}
T(0,p_2^2,p_3^2;m_{12}^2,m_{23}^2,m_{13}^2)=\frac{i}{p_2^2-p_3^2}\mathcal{T}(0,p_2^2,p_3^2;m_{12}^2,m_{23}^2,m_{13}^2)\, ,
\end{equation}
where the internal masses are generic and can be zero. Before expansion in the dimensional regularization parameter $\epsilon$, the results will often involve the functions $\,_2F_1$ and $F_1$ defined in eqs.~(\ref{gaussHypDef}) and (\ref{appellf1Def}).

\subsection{$T(0,p_2^2,p_3^2;0,m_{23}^2,0)$}
The triangle of  \refF{fig:2mTriangle1} is given by:
\begin{align}\bsp
T(0,p_2^2,p_3^2;0,m_{23}^2,0)
=&i\frac{e^{\gamma_E\epsilon}\Gamma(1+\epsilon)}{\epsilon(1-\epsilon)}\frac{(m_{23}^2)^{-\epsilon}}{p_2^2-p_3^2}\left[\frac{p_2^2}{p_2^2-m^2_{23}}\hypgeo{1}{1-2\epsilon}{2-\epsilon}{\frac{p_2^2}{p_2^2-m^2_{23}}}\right.\\
&\left.-\frac{p_3^2}{p_3^2-m^2_{23}}\hypgeo{1}{1-2\epsilon}{2-\epsilon}{\frac{p_3^2}{p_3^2-m^2_{23}}}\right]\\
=&\frac{i}{p_2^2-p_3^2}\left[\frac{1}{\epsilon}\ln\left(\frac{m_{23}^2-p_2^2}{m_{23}^2-p_3^2}\right)+\text{Li}_2\left(\frac{p_2^2}{p_2^2-m_{23}^2}\right)-\text{Li}_2\left(\frac{p_3^2}{p_3^2-m_{23}^2}\right)\right.\\
&\left.-\frac{1}{2}\log ^2\left(m_{23}^2-p_2^2\right)+\frac{1}{2} \log ^2\left(m_{23}^2-p_3^2\right)\right]+\mathcal{O}\left(\epsilon\right)\, .
\esp\end{align}
The symbol is
\begin{align}\bsp
\mathcal{S}\left[\mathcal{T}(0,p_2^2,p_3^2;0,m_{23}^2,0)\right]=&\frac{1}{\epsilon}\frac{m_{23}^2-p_2^2}{m_{23}^2-p_3^2}+
m_{23}^2 \otimes
\frac{p_3^2(m_{23}^2-p_2^2)}{p_2^2(m_{23}^2-p_3^2)}
+\left(m_{23}^2-p_2^2\right)\otimes
\frac{p_2^2}{\left(m_{23}^2-p_2^2\right)^2}
 \\
&-\left(m_{23}^2-p_3^2\right)\otimes
\frac{p_3^2}{\left(m_{23}^2-p_3^2\right)^2}
+\mathcal{O}\left(\epsilon\right)
\,.
\esp\end{align}

\subsubsection{Single cuts}
The cut in the external channel $p_2^2$ is
\begin{align}\bsp
\cut_{p_2^2}T(0,p_2^2,p_3^2;0,m_{23}^2,0)=&-2\pi\frac{e^{\gamma_E\epsilon}\Gamma(1-\epsilon)}{\epsilon\Gamma(1-2\epsilon)}\frac{(p_2^2)^\epsilon(p_2^2-m_{23}^2)^{-2\epsilon}}{p_2^2-p_3^2}\\
=&-\frac{2\pi}{\epsilon(p_2^2-p_3^2)}-\frac{2\pi}{p_2^2-p_3^2}\left[\ln\left(p_2^2\right)-2\ln\left(p_2^2-m_{23}^2\right)\right]+\mathcal{O}(\epsilon)\,.
\esp\end{align}
The cut in the external channel  $p_3^2$ is trivial to obtain from the symmetry of the function.
The cut in the internal mass $m_{23}^2$ is
\begin{align}\bsp
\cut_{m_{23}^2}T(0,p_2^2,p_3^2;0,m_{23}^2,0)
=&-\frac{\pi e^{\gamma_E\epsilon}}{\epsilon\Gamma(1-\epsilon)}\frac{\Gamma(1+2\epsilon)}{p_2^2-p_3^2}\left\{\frac{\Gamma(1-\epsilon)}{\Gamma(1+\epsilon)}(-p_2^2)^{-\epsilon}\left(1-\frac{m_{23}^2}{p_2^2}\right)^{-2\epsilon}\right.\\
&\left.-\frac{(-p_2^2)^{-\epsilon}}{\Gamma(1+2\epsilon)}\hypgeo{\epsilon}{2\epsilon}{1+2\epsilon}{1-\frac{m_{23}^2}{p_2^2}}-\left(p_2^2\leftrightarrow p_3^2\right)\right\}\\
=&\frac{2\pi}{p_2^2-p_3^2}
\left[\ln\left(\frac{-p_3^2}{m_{23}^2-p_3^2}\right)-\ln\left(\frac{-p_2^2}{m_{23}^2-p_2^2}\right)\right]+\mathcal{O}(\epsilon)\,.
\esp\end{align}

\subsubsection{Double cuts}
All double cuts are zero.

\subsection{$T(0,p_2^2,p_3^2;m_{12}^2,0,0)$}
\label{sec:tp2p3m12}
The triangle of  \refF{fig:2mTriangle2} is given by:
\begin{align}\bsp
T(0,p_2^2,p_3^2;m_{12}^2,0,0)=&i\frac{e^{\gamma_E\epsilon}\Gamma(1+\epsilon)}{\epsilon}\left[\frac{(-p_3^2)^{-\epsilon}}{m_{12}^2}\frac{\Gamma^2(1-\epsilon)}{\Gamma(2-2\epsilon)}
\hypgeo{1}{1-\epsilon}{2-2\epsilon}{\frac{p_2^2-p_3^2}{m_{12}^2}}\right.\\
&\left.-\frac{(m_{12}^2)^{-1-\epsilon}}{1-\epsilon}\appellf{1}{1}{\epsilon}{2-\epsilon}{\frac{p_2^2-p_3^2}{m_{12}^2}}{\frac{p_2^2}{m_{12}^2}}\right]\\
=&\frac{i}{(p_2^2-p_3^2)}\left[\text{Li}_2\left(\frac{p_2^2}{m_{12}^2}\right)-\text{Li}_2\left(\frac{p_2^2-m_{12}^2}{p_3^2}\right)+\frac{1}{2} \log
   ^2\left(-\frac{p_3^2}{m_{12}^2}\right)\right.\\
   &\left.-\log \left(\frac{p_2^2-m_{12}^2}{p_3^2}\right) \log
   \left(\frac{m_{12}^2-p_2^2+p_3^2}{p_3^2}\right)+\frac{\pi ^2}{3}\right]+\mathcal{O}(\epsilon)\, .
\esp\end{align}
The symbol is
\begin{align}\bsp
\label{eq:symbolTp2p3m12}
\mathcal{S}\big[\mathcal{T}(0,p_2^2,p_3^2;m_{12}^2,0,0)\big]=&m^2_{12}\otimes\left(\frac{p_2^2}{p_3^2}\right)+p_3^2\otimes\left(\frac{m_{12}^2-p_2^2+p_3^2}{m_{12}^2}\right)\\
&+\left(m_{12}^2-p_2^2\right)\otimes\left(\frac{-p_3^2m_{12}^2}{p_2^2(p_2^2-m_{12}^2-p_3^2)}\right)+\mathcal{O}(\epsilon)\, .
\esp\end{align}

\subsubsection{Single cuts}
The cut in the external channel $p_2^2$ is
\begin{align}\bsp
\label{eq:p2cutTp2p3m12}
\cut_{p_2^2}&T(0,p_2^2,p_3^2;m_{12}^2,0,0)=\\
=&-2\pi\frac{e^{\gamma_E\epsilon}\Gamma(1-\epsilon)}{\Gamma(2-2\epsilon)}\frac{(p_2^2-m_{12}^2)^{1-2\epsilon}}{(p_2^2)^{-\epsilon}m_{12}^2p_3^2}\hypgeo{1}{1-\epsilon}{2-2\epsilon}{\frac{(p_2^2-p_3^2)(p_2^2-m^2_{12})}{m^2_{12}p_3^2}}\\
=&\frac{2\pi}{p_2^2-p_3^2}\left(\ln\left(\frac{p_2^2}{m_{12}^2}\right)+\ln\left(\frac{p_2^2-m_{12}^2-p_3^2}{-p_3^2}\right)\right)+\mathcal{O}(\epsilon)\, .
\esp\end{align}
The cut in the external channel $p_3^2$ is
\begin{align}\bsp \label{eq:p3cutTp2p3m12}
\cut_{p_3^2}&T(0,p_2^2,p_3^2;m_{12}^2,0,0)=\\
=&-2\pi\frac{e^{\gamma_E\epsilon}\Gamma(1-\epsilon)}{\Gamma(2-2\epsilon)}\frac{(p_3^2)^{-\epsilon}}{p_2^2-p_3^2-m_{12}^2}\hypgeo{1}{1-\epsilon}{2-2\epsilon}{\frac{p_2^2-p_3^2}{p_2^2-p_3^2-m_{12}^2}}\\
=&\frac{2\pi}{p_2^2-p_3^2}\left(\ln\left(m_{12}^2\right)-\ln\left(m_{12}^2-p_2^2+p_3^2\right)\right)+\mathcal{O}(\epsilon)\,.
\esp\end{align}
The cut in the internal mass $m_{12}^2$ is
\begin{align}\bsp
\cut_{m_{12}^2}&T(0,p_2^2,p_3^2;m_{12}^2,0,0)=\\
=&\frac{2\pi e^{\gamma_E\epsilon}}{\Gamma(2-\epsilon)}\frac{(m_{12}^2-p_2^2)^{-\epsilon}}{p_3^2}\left(\frac{m_{12}^2}{p_2^2}\right)^{-\epsilon}\appellf{1-\epsilon}{1}{\epsilon}{2-\epsilon}{\frac{p_3^2-p_2^2}{p_3^2}}{\frac{m_{12}^2}{m_{12}^2-p_2^2}}\\
=&\frac{2\pi}{p_2^2-p_3^2}\left(\ln\left(-p_2^2\right)-\ln\left(-p_3^2\right)\right)+\mathcal{O}(\epsilon)\,.
\esp\end{align}

\subsubsection{Double cuts}
The double cut in the external channels $p_2^2$ and $p_3^2$ is
\begin{align}\bsp
\cut_{p_2^2,p_3^2}T(0,p_2^2,p_3^2;m_{12}^2,0,0)=&\frac{4\pi^2 ie^{\gamma_E\epsilon}}{\Gamma(1-\epsilon)}\frac{(p_3^2)^{-\epsilon}(p_2^2-p_3^2-m_{12}^2)^{-\epsilon}}{(m_{12}^2)^{\epsilon}(p_2^2-p_3^2)^{1-2\epsilon}}
\theta(p_2^2-p_3^2-m^2_{12})\\
=&\frac{4\pi^2 i}{p_2^2-p_3^2}\theta(p_2^2-p_3^2-m^2_{12})+\mathcal{O}(\epsilon)\,.
\esp\end{align}
The double cut in the external channel $p_3^2$ and the internal mass $m_{12}^2$ is
\begin{align}\bsp
\cut_{p_3^2,m_{12}^2}T(0,p_2^2,p_3^2;m_{12}^2,0,0)=&-4\pi^2i\frac{e^{\gamma_E\epsilon}}{\Gamma(1-\epsilon)}\frac{(p_3^2)^{-\epsilon}(-m_{12}^2)^{-\epsilon}}{(p_3^2-p_2^2)^{1-\epsilon}}(p_3^2+m_{12}^2-p_2^2)^{-\epsilon}\\
&\theta(p_3^2+m_{12}^2-p_2^2)\\
=& \frac{4\pi^2 i}{p_2^2-p_3^2}\theta(p_3^2+m_{12}^2-p_2^2)\, .
\esp\end{align}
The double cut in the external channel $p_2^2$ and the internal mass $m_{12}^2$ is zero.




\section{Three-mass triangles\label{app_threeMass}}

We now present expressions for triangles with three external massive legs. We start by describing how we computed the triangles with one or two massive propagators, for which we give a very simple expression that allows us to evaluate the diagrams very easily to arbitrary order in $\epsilon$. Our method does not work for the case with three massive propagators, where we were not able to find a rational parametrization, and we rely on the result in ref.~\cite{Denner:1991kt}. We will comment further on the choice of variables for this example in section \ref{tp1p2p3m12m23m13} below. For the cases treated in this section we will not compute mass discontinuities, as they do not add anything to what we have already illustrated in the context of previous examples. We separate the rational prefactor from the pure transcendental function according to the relation
\begin{equation}
T(p_1^2,p_2^2,p_3^2;m_{12}^2,m_{23}^2,m_{13}^2)=\frac{i}{p_1^2(z-\zbar)}\mathcal{T}(p_1^2,p_2^2,p_3^2;m_{12}^2,m_{23}^2,m_{13}^2)\, ,
\end{equation}
where the internal masses are generic and can be zero.

\subsection{Computation of $T(p_1^2,p_2^2,p_3^3;m_{12}^2,0,0)$ and $T(p_1^2,p_2^2,p_3^3;m_{12}^2,0,m_{13}^2)$}

Triangles with two external masses are easily computed with standard techniques to arbitrary order in $\epsilon$. However, that is no longer the case for triangles with three external masses \cite{Chavez:2012kn}. In ref.~\cite{Abreu:2014cla}, the triangle with three external masses and massless internal propagators was easily computable to arbitrary order in $\epsilon$ through a double dispersion integral over its double cut. We now show this is also possible when there are one or two massive propagators. In the following, we will use the variables
\begin{align*}
\alpha\bar{\alpha}=x=\frac{s_2}{p_1^2}\, ,\qquad\qquad (1-\alpha)(1-\bar{\alpha})=y=\frac{s_3}{p_1^2}\,,
\end{align*}
where $s_2$ and $s_3$ are integration variables in dispersion relations.

We will use the shorthand $T(p_i^2;m^2_{jk})$ for any of the three-mass triangles.  We now proceed as in ref.~\cite{Abreu:2014cla}:
\begin{align}\bsp\label{doubleDiscGen}
T(p_i^2;m^2_{jk})=&-\frac{1}{(2\pi i)^2}\int_{c_2}\frac{ds_2}{s_2-p_2^2}\int_{c_3}\frac{ds_3}{s_3-p_3^2}\left(\cut_{p_2^2,p_3^2}T(p_i^2;m^2_{jk})\right)\bigg\vert_{p_2^2=s_2,p_3^2=s_3}\\
=&\frac{i}{4\pi^2p_1^2}\int_{c_\alpha}d\alpha\int_{c_{\bar{\alpha}}}d\bar{\alpha}\frac{\left(\cut_{p_2^2,p_3^2}\mathcal{T}(p_i^2;m^2_{jk})\right)\bigg\vert_{z=\alpha,\zbar=\bar{\alpha}}}{(\alpha\bar{\alpha}-z \zbar)((1-\alpha)(1-\bar{\alpha})-(1-z)(1- \zbar))}\,.
\esp\end{align}
The only difference between the triangles with one and two massive propagators are the integration contours $c_2$ and $c_3$, and $c_\alpha$ and $c_{\bar{\alpha}}$. For the case with one internal mass,
\begin{align*}
c_2=[m_{12}^2,\infty)\,,\quad c_3=[0,\infty) \quad\text{and}\quad c_\alpha=[1,\infty)\, ,\quad c_{\bar{\alpha}}=(-\infty,\mu_{12}]\, ,
\end{align*}
and for the case with two internal masses,
\begin{align*}
c_2=[m_{12}^2,\infty)\,,\quad c_3=[m_{13}^2,\infty) \quad\text{and}\quad c_\alpha=[w_1,\infty)\, ,\quad c_{\bar{\alpha}}=(-\infty,\bar{w}_{1}]\,.
\end{align*}

For either case, the functions $\left(\cut_{p_2^2,p_3^2}\mathcal{T}(p_i^2;m^2_{jk})\right)\bigg\vert_{z=\alpha,\zbar=\bar{\alpha}}$ are given by powers of logarithms whose arguments are linear in both $\alpha$ and $\bar{\alpha}$. The integral in \refE{doubleDiscGen} is thus trivial to solve in terms of polylogarithms to the desired order in $\epsilon$. The change of variables
\begin{equation*}
\beta=\frac{a_\beta}{\alpha}\qquad\qquad\gamma=\frac{1-a_\gamma}{1-\bar{\alpha}}\,,
\end{equation*}
where $a_\beta=1$ or $w_1$ and $a_\gamma=\mu_{12}$ or $\bar{w}_1$ respectively for the cases with one and two internal massive propagators, makes the integration particularly simple to perform. The results for the finite terms of these two triangles, given below in eqs.~(\ref{tp1p2p3m12}) and (\ref{tp1p2p3m12m13}), were computed with this method, and checked to agree with the result in ref.~\cite{Denner:1991kt}. In our method, as mentioned above, higher orders in $\epsilon$ become trivial to compute.

\subsection{$T(p_1^2,p_2^2,p_3^3;m_{12}^2,0,0)$}
\label{sec:tp1p2p3m12}
The triangle of \refF{fig:3mTriangle3} is given by
\begin{equation}
T(p_1^2,p_2^2,p_3^3;m_{12}^2,0,0) = \frac{i}{p_1^2 (z-\zb)}\mathcal{T}\left(p_1^2,z,\zbar,\mu_{12}\right) ~,
\end{equation}
where $\mathcal{T}\left(p_1^2,z,\zbar,\mu_{12}\right)$ is a pure function given by
\begin{align}\bsp\label{tp1p2p3m12}
\mathcal{T}\left(p_1^2,z,\zbar,\mu_{12}\right)=&G\left(1,z,\mu _{12}\right)+G\left(1,\frac{\mu _{12}}{z},\zbar\right)-G\left(1,\frac{\mu _{12}}{\zbar},z\right)-G\left(1,\zbar,\mu_{12}\right)\\
&-\text{Li}_2(z)+\text{Li}_2(\zbar)+\log (1-z) \log \left(1-\frac{\mu _{12}}{z}\right)\\
&+\log \left(1-\frac{1}{\mu _{12}}\right) \log\left(\frac{1-z}{1-\zbar}\right)-\log (1-\zbar) \log \left(1-\frac{\mu _{12}}{\zbar}\right)\\
&+\log \left(1-\mu _{12}\right) \log \left(\frac{z (1-\zbar) \left(\zbar-\mu _{12}\right)}{\zbar(1-z) \left(z-\mu _{12}\right)}\right)
+\mathcal{O}(\epsilon)\,.
\esp\end{align}
The symbol of its finite part is
\begin{align}\bsp
\mathcal{S}\left[\mathcal{T}\left(p_1^2,z,\zbar,\mu_{12}\right)\right]=&\mu_{12}\otimes\frac{\zbar(z-\mu_{12})}{z(\zbar-\mu_{12})}
+(1-z)(1-\zbar)\otimes\frac{(z-\mu_{12})}{(\zbar-\mu_{12})}\\
&+(z\zbar-\mu_{12})\otimes\frac{z(1-\zbar)(\zbar-\mu_{12})}{\zbar(1-z)(z-\mu_{12})}\\
&+(1-\mu_{12})\otimes\frac{(1-z)(\zbar-\mu_{12})}{(1-\zbar)(z-\mu_{12})}\, .
\esp\end{align}

\subsubsection{Single cuts}
The cut in the $p_1^2$ channel is
\begin{align}\bsp\label{eq:cutp1tp1p2p3m12}
\cut_{p_1^2}T(p_1^2,p_2^2,p_3^3;m_{12}^2,0,0)
=&-2\pi\frac{e^{\gamma_E\epsilon}\Gamma(1-\epsilon)}{\Gamma(2-2\epsilon)}\left(p_1^2\right)^{-1-\epsilon}\frac{(1-\mu_{12})^{1-2\epsilon}}{(1-z)(\mu_{12}-\zbar)}\\
&\,_2F_1\left(1,1-\epsilon,2-2\epsilon,\frac{(1-\mu_{12})(z-\zbar)}{(1-z)(\mu_{12}-\zbar)}\right)\\
=&\frac{2\pi}{p_1^2(z-\bz)}\ln\left(\frac{(1-\bz)(z-\mu_{12})}{(z-1)(\mu_{12}-\bz)}\right)+\mathcal{O}\left(\epsilon\right)\,.
\esp\end{align}
The cut in the $p_2^2$ channel is
\begin{align}\bsp
\cut_{p_2^2}T(p_1^2,p_2^2,p_3^3;m_{12}^2,0,0)
=&-2\pi\frac{e^{\gamma_E\epsilon}\Gamma(1-\epsilon)}{\Gamma(2-2\epsilon)}\left(-p_1^2\right)^{-1-\epsilon}\frac{(\mu_{12}-z\zbar)^{1-2\epsilon}(-z\zbar)^{\epsilon}}{\bz(1-z)(z-\mu_{12})}\\
&\hypgeo{1}{1-\epsilon}{2-2\epsilon}{\frac{(z-\bz)(\mu_{12}-z \bz)}{\bz(1-z)(z-\mu_{12})}}\\
=&\frac{2\pi}{p_1^2(z-\bz)}\ln\left(\frac{-\bz(1-z)(z-\mu_{12})}{z(1-\bz)(\mu_{12}-\bz)}\right)+\mathcal{O}\left(\epsilon\right)\, .
\esp\end{align}
The cut in the $p_3^2$ channel is
\begin{align}\bsp
\cut_{p_3^2}T(p_1^2,p_2^2,p_3^3;m_{12}^2,0,0)
=&-2\pi\frac{e^{\gamma_E\epsilon}\Gamma(1-\epsilon)}{\Gamma(2-2\epsilon)}\left(-p_1^2\right)^{-1-\epsilon}\frac{((z-1)(1-\bz))^{-\epsilon}}{\mu_{12}-\bz}\\
&\hypgeo{1}{1-\epsilon}{2-2\epsilon}{\frac{z-\bz}{\mu_{12}-\bz}}\\
=&\frac{2\pi}{p_1^2(z-\bz)}\ln\left(\frac{\mu_{12}-\bz}{\mu_{12}-z}\right)+\mathcal{O}\left(\epsilon\right)\, .
\esp\end{align}

\subsubsection{Double cuts}
The double cut in the $p_i^2$ and $p_j^2$ channels is
\begin{align}\bsp
\cut_{p_i^2,p_j^2}T(p_1^2,p_2^2,p_3^3;m_{12}^2,0,0)
=& 4\pi^2i\Theta_{ij}\frac{e^{\gamma_E\epsilon}}{\Gamma(1-\epsilon)}\left((-1)^a p_1^2\right)^{-1-\epsilon}(z-\zbar)^{-1+2\epsilon}\\
&\left((z-1)(1-\zbar)(\zbar-\mu_{12})(z-\mu_{12})\right)^{-\epsilon}\\
=&\frac{4\pi^2 i}{p_1^2(z-\zbar)}(-1)^a\Theta_{ij}+\mathcal{O}(\epsilon)\, ,
\esp\end{align}
where $a=0$ for $(i,j)=(1,2)$ or $(1,3)$, and $a=1$ for $(i,j)=(2,3)$. The theta functions are
\begin{align*}
\Theta_{12}&=\theta(z-1)\theta(1-\zbar)\theta(z-\mu_{12})\theta(\zbar-\mu_{12})\\
\Theta_{13}&=\theta(1-z)\theta(1-\zbar)\theta(z-\mu_{12})\theta(\mu_{12}-\zbar)\\
\Theta_{23}&=\theta(z-1)\theta(1-\zbar)\theta(z-\mu_{12})\theta(\mu_{12}-\zbar)\,.
\end{align*}

\subsection{$T(p_1^2,p_2^2,p_3^3;m_{12}^2,0,m_{13}^2)$}\label{sec:tp1p2p3m12m13}
The triangle of \refF{fig:3mTriangle2} is given by
\begin{equation}
T(p_1^2,p_2^2,p_3^3;m_{12}^2,0,m_{13}^2) = \frac{i}{p_1^2 (z-\zb)}\mathcal{T}\left(p_1^2,z,\zbar,w_1,\bar{w}_1\right),
\end{equation}
where $\mathcal{T}\left(p_1^2,z,\zbar,w_1,\bar{w}_1\right)$ is a pure function given by
\begin{align}\bsp\label{tp1p2p3m12m13}
\mathcal{T}\left(p_1^2,z,\zbar,w_1,\bar{w}_1\right)=&G\left(\frac{w_1}{z},\frac{w_1(\bar{w}_1-1)}{\bar{w}_1+(1-z)(1-\zbar)-1},1\right)+G\left(\frac{w_1}{\zbar},w_1,1\right)\\
&-G\left(\frac{w_1}{\zbar},\frac{w_1(\bar{w}_1-1)}{\bar{w}_1+(1-z)(1-\zbar)-1},1\right)-G\left(\frac{w_1}{z},w_1,1\right)\\
&-\Li_2\left(\frac{z\zbar}{w_1\bar{w}_1}\right)-G\left(\frac{w_1}{z},\frac{w_1 \bar{w}_1}{z\zbar},1\right)-G\left(\bar{w}_1,\frac{w_1\bar{w}_1}{\zbar},z\right)\\
&+\log \left(1-\frac{1}{\bar{w}_1}\right)\log \left(\frac{w_1-z}{w_1-\zbar}\right)+\log\left(1-\frac{\zbar}{w_1}\right) \log \left(1-\frac{z}{\bar{w}_1}\right)\\
&+\mathcal{O}(\epsilon)\,.
\esp\end{align}
The symbol of its finite part is
\begin{align}\bsp\label{eq:symboltp1p2p3m12m13}
\mathcal{S}&\left[\mathcal{T}\left(p_1^2,z,\zbar,w_1,\bar{w}_1\right)\right]=(z\zbar-w_1\bar{w}_1)\otimes\frac{z(w_1-\zbar)(\bar{w}_1-\zbar)}{\zbar(w_1-z)(\bar{w}_1-z)}\\
&+((1-z)(1-\zbar)-(1-w_1)(1-\bar{w}_1))\otimes\frac{(1-\zbar)(w_1-z)(\bar{w}_1-z)}{(1-z)(w_1-\zbar)(\bar{w}_1-\zbar)}\\
&+(1-w_1)\otimes\frac{(1-z)(w_1-\zbar)}{(1-\zbar)(w_1-z)}+w_1\otimes\frac{\zbar(w_1-z)}{z(w_1-\zbar)}\\
&+(1-\bar{w}_1)\otimes\frac{(1-z)(\bar{w}_1-\zbar)}{(1-\zbar)(\bar{w}_1-z)}+
\bar{w}_1\otimes\frac{\zbar(\bar{w}_1-z)}{z(\bar{w}_1-\zbar)}\, .
\esp\end{align}

\subsubsection{Single cuts}
The cut in the $p_1^2$ channel is
\begin{align}\bsp\label{eq:cutp1tp1p2p3m12m13}
\cut_{p_1^2}T(p_1^2,p_2^2,p_3^3;m_{12}^2,0,m_{13}^2)
=&-2\pi\frac{e^{\gamma_E\epsilon}\Gamma(1-\epsilon)}{\Gamma(2-2\epsilon)}\left(p_1^2\right)^{-1-\epsilon}\frac{(w_1-\bar{w}_1)^{1-2\epsilon}}{(z-w_1)(\zbar-\bar{w}_1)}\\
&\,_2F_1\left(1,1-\epsilon,2-2\epsilon, \frac{(z-\zbar)(w_1-\bar{w}_1)}{(z-w_1)(\zbar-\bar{w}_1)}\right)\\
=&\frac{2\pi}{p_1^2(z-\zbar)}\ln\left(\frac{(z-\bar{w}_1)(w_1-\bz)}{(\bar{w}_1-\bz)(z-w_1)}\right)\, +\mathcal{O}\left(\epsilon\right) .
\esp\end{align}
The cut in the $p_2^2$ channel is
\begin{align}\bsp
\cut_{p_2^2}T(p_1^2,p_2^2,p_3^3;m_{12}^2,0,m_{13}^2)
=&2\pi\frac{e^{\gamma_E\epsilon}\Gamma(1-\epsilon)}{\Gamma(2-2\epsilon)}\left(-p_1^2\right)^{-1-\epsilon}\frac{(-z\bz)^{\epsilon}(w_1\bar{w}_1-z\bz)^{1-2\epsilon}}{\bz(w_1-z)(z-\bar{w}_1)}\\
&\,_2F_1\left(1,1-\epsilon,2-2\epsilon,\frac{(z-\zbar)(w_1\bar{w}_1-z\bz)}{\bz(w_1-z)(z-\bar{w}_1)}\right)\\
=&\frac{2\pi}{p_1^2(z-\zbar)}\ln\left(\frac{-\zbar(z-\bar{w}_1)(w_1-z)}{z(w_1-\bz)(\bar{w}_1-\bz)}\right)+\mathcal{O}\left(\epsilon\right) .
\esp\end{align}
The cut in the $p_3^2$ channel is
\begin{align}\bsp
\cut_{p_3^2}T(p_1^2,p_2^2,p_3^3;m_{12}^2,0,m_{13}^2)
=&-2\pi\frac{e^{\gamma_E\epsilon}\Gamma(1-\epsilon)}{\Gamma(2-2\epsilon)}\left(p_1^2\right)^{-1-\epsilon}\frac{u_3^\epsilon(u_3-\mu_{13})^{1-2\epsilon}}{(\zbar-1)(z-\bar{w}_1)(z-w_1)}\\
&\,_2F_1\left(1,1-\epsilon,2-2\epsilon,\frac{(z-\zbar)(u_3-\mu_{13})}{(\zbar-1)(z-\bar{w}_1)(z-w_1)}\right)\\
=&\frac{2\pi}{p_1^2(z-\zbar)}\ln\left(\frac{(z-1)(\zbar-\bar{w}_1)(w_1-\zbar)}{(1-\zbar)(z-\bar{w}_1)(z-w_1)}\right)+\mathcal{O}(\epsilon).
\esp\end{align}

\subsubsection{Double cuts}
The double cut in the $p_i^2$ and $p_j^2$ channels is
\begin{align}\bsp
\cut_{p_i^2,p_j^2}T(p_1^2,p_2^2,p_3^3;m_{12}^2,0,m_{13}^2)
=& 4\pi^2i\Theta_{ij}\frac{e^{\gamma_E\epsilon}}{\Gamma(1-\epsilon)}\left((-1)^a p_1^2\right)^{-1-\epsilon}(z-\zbar)^{-1+2\epsilon}\\
&\left((z-w_1)(\zbar-\bar{w}_1)(z-\bar{w}_1)(\zbar-\bar{w}_1)\right)^{-\epsilon}\\
=&\frac{4\pi^2 i}{p_1^2(z-\zbar)}(-1)^a\Theta_{ij}+\mathcal{O}(\epsilon)\, ,
\esp\end{align}
where $a=0$ for $(i,j)=(1,2)$ or $(1,3)$, and $a=1$ for $(i,j)=(2,3)$. The theta functions are
\begin{align*}
\Theta_{12}&=\theta(z-w_1)\theta(w_1-\zbar)\theta(z-\bar{w}_1)\theta(\zbar-\bar{w}_1)\\
\Theta_{13}&=\theta(w_1-z)\theta(w_1-\zbar)\theta(z-\bar{w}_1)\theta(\bar{w}_1-\zbar)\\
\Theta_{23}&=\theta(z-w_1)\theta(w_1-\zbar)\theta(z-\bar{w}_1)\theta(\bar{w}_1-\zbar)\,.
\end{align*}

\subsection{$T(p_1^2,p_2^2,p_3^3;m_{12}^2,m_{23}^2,m_{13}^2)$}\label{tp1p2p3m12m23m13}

For the triangle of \refF{fig:3mTriangle1} we take the expression from ref.~\cite{Denner:1991kt}, adjusted to match our conventions:
\begin{align}
T(p_1^2,p_2^2,p_3^3;m_{12}^2,m_{23}^2,m_{13}^2)  = \frac{i}{p_1^2(z-\zbar)} \mathcal{T}(p_1^2,p_2^2,p_3^2;m_{12}^2,m_{23}^2,m_{13}^2)\, ,
\label{eq:Denner}
\end{align}
where 
\begin{equation}
\mathcal{T}(p_1^2,p_2^2,p_3^2;m_{12}^2,m_{23}^2,m_{13}^2)= \sum_{i=1}^3 \sum_{\sigma = \pm} \left[ \text{Li}_2 \left( \frac{y_{0i}-1}{y_{i\sigma}} \right) - \text{Li}_2 \left( \frac{y_{0i}}{y_{i\sigma}} \right) \right] ~.
\end{equation}
 The $y_{0i}$ and $y_{i\pm} $ are given by 
\begin{align}\bsp
 y_{0i} & =  \frac{-1}{2u_i\sqrt{\lambda_z} }
\bigl[ u_i( u_{i} -u_{i+1} -u_{i-1} + 2\mu_{i-1,i+1} -\mu_{i,i+1} -\mu_{i-1,i} ) \\[1ex]
 & \quad -(u_{i+1}-u_{i-1})(\mu_{i-1,i} -\mu_{i,i+1} )
- \sqrt{\lambda_z} (u_{i}-\mu_{i-1,i} +\mu_{i,i+1})\bigr], \\[1ex]
    y_{i\pm} & = y_{0i} - \frac{1}{2 u_{i}}
\left[ u_{i}-\mu_{i-1,i}+\mu_{i,i+1} \pm \sqrt{\lambda_{i}} \right].
\esp\end{align}
 Here, the indices $i\pm1$ are defined cyclically.
The variables $u_i$, $\mu_{ij}$ are defined in eq.~\eqref{eq:u_i_def}, and the $\lambda_i$ for $i=z,1,2,3$ are defined in eqs.~\eqref{eq:def_lambda_z}, \eqref{eq:def_lambda_1} and \eqref{eq:lambda3def}. 

To get as close as possible to a rational symbol alphabet, we use the variables $z$, $\bz$, $w_1$, $\bw_1$ and  $\mu_{23}$, which are adapted to the $p_1^2$ channel. Since this triangle is fully symmetric, it is easy to choose variables adapted to any of the other two channels. However, given our choice, square roots of $\lambda_2 \equiv {\lambda(u_2, \mu_{12}, \mu_{23})}$ and $\lambda_3 \equiv {\lambda(u_3,\mu_{13} , \mu_{23})}$ make an unavoidable appearance.  Written in a form where the first entries may be readily identified with the three channel thresholds and the three internal masses, the symbol of the triangle is 
\begin{align}\bsp\label{eq:symbT33masses}
\mathcal{S}\left[\mathcal{T}\left(p_1^2,z,\zbar,w_1,\bar{w}_1,\mu_{23}\right)\right]=&
    w_1 \left(1-\bar{w}_1\right) \otimes
    \frac{T_{1-}}{T_{1+}}
   +\frac{1}{2}
   \left(-z \bar{z}+w_1 \bar{w}_1- \sqrt{\lambda_2} +\mu_{23}\right)\otimes \frac{T_{2-}}{T_{2+}}
\\
&
   +\frac{1}{2} \left(\bar{z} z-z-\bar{z}-w_1 \bar{w}_1+w_1+\bar{w}_1+\sqrt{\lambda_3}-\mu_{23}\right)
   \otimes \frac{T_{3-}}{T_{3+}}
\\
&
   + w_1 \bar{w}_1
   \otimes \frac{T_{2+}}{ (-z) T_{1-}}
   + \left(1-w_1\right) \left(1-\bar{w}_1\right) \otimes 
   \frac{(z-1) T_{1+}}{T_{3-}}
\\
&
 + 4 \mu _{23}  \otimes
   \frac{z T_{3+} }{(1-z) T_{2-} }.
\esp\end{align}
The $T_{i\pm}$ are given by the general formula 
\begin{align}\label{eq:tipmdef}
T_{i\pm} =& -u_i (-u_i +u_{i+1}+u_{i-1}+\mu_{i,i+1}+\mu_{i,i-1}-2\mu_{i+1,i-1}) \nonumber \\
&+(u_{i+1} -u_{i-1}) (\mu_{i,i+1}-\mu_{i,i-1}) \pm \sqrt{\lambda_z \lambda_i}\,.
\end{align}
In particular, we have
\bean
T_{1\pm} &=& -2(z \bz + w_1 \bw_1 - \mu_{23}) + (w_1 + \bw_1)(z+\bz) \pm (w_1 -\bw_1)(z-\bz) \\
T_{2\pm} &=& (z\zbar+w_1\bar{w}_1-\mu_{23}) (z + \bar{z})-2  z \bar{z}(w_1+\bw_1) \pm(z-\bz) \sqrt{\lambda_2} \\
T_{3\pm} &=& z^2(1-\zbar)+\zbar^2(1-z)+(w_1+\bw_1)(2 z\bz-z-\bz) \\
&& +(\mu_{23}-w_1\bw_1)(z+\bz-2) \pm (z-\bz)\sqrt{\lambda_3}\,.
\eean
We note that $T_{1\pm}$ can be written in a simpler form, but where the $\pm$ notation is less clear:
\begin{align}\bsp\label{t1pm}
T_{1+}&=2\left(\mu_{23}-(w_1-\zbar)(\bar{w}_1-z)\right)\\
T_{1-}&=2\left(\mu_{23}-(w_1-z)(\bar{w}_1-\zbar)\right)\, .
\esp\end{align}

Since the triangle depends on the external momenta through the invariants, it depends on $z$ and $\zb$ only through the symmetric combinations $u_2 = z\zb$, $u_3 = (1-z)(1-\zb)$.  Therefore, once we have removed the rational prefactor, the symbol above is antisymmetric under the exchange $z \leftrightarrow \zb$. However, we note that this antisymmetry is not superficially apparent in the last three terms.

\subsubsection{Single cuts}
\label{sec:singlecutstp1p2p3m12m23m13}

As mentioned in section \ref{sec:triangles}, for the triangle with three external and three internal masses we must be very careful with the choice of variables. We now show how it is possible to choose variables such that each of the single cuts has a rational alphabet. However, unlike what happens for all other configurations of masses, for each cut we must choose different variables. For instance, in \refE{eq:symbT33masses} we chose variables that rationalize the symbol of the $p_1^2$ cut (indeed, the $T_{1\pm}$ are rational, as seen in  \refE{t1pm}). In this section, we give the cut results in terms of two slightly different sets of variables: either we normalize invariants by the same invariant associated with the channel being cut, or by a different invariant.  Our notation is that $p_i^2$ is the channel used for normalization, and $p_j^2$ is the cut channel in the case where it is different.

We start with variables where we cut in the same channel we normalize by, namely $p_i^2$. To be more precise, the variables we choose are
\begin{align}\bsp\label{eq:varset1}
z&=\frac{1+u_j-u_k+\sqrt{\lambda(1,u_j,u_k)}}{2}\, ,\quad \bz=\frac{1+u_j-u_k-\sqrt{\lambda(1,u_j,u_k)}}{2}\,,\\
w_i&=\frac{1+\mu_{ij}-\mu_{jk}+\sqrt{\lambda(u_j,\mu_{ij},\mu_{jk})}}{2}\, ,\quad \bar{w}_i=\frac{1+\mu_{ij}-\mu_{jk}-\sqrt{\lambda(u_j,\mu_{ij},\mu_{jk})}}{2}\,,
\esp\end{align}
related to the invariants through
\begin{align}\bsp
&z\bz=u_j=\frac{p_j^2}{p_i^2}\,,\quad (1-z)(1-\bz)=u_k=\frac{p_k^2}{p_i^2} \,,\quad \mu_{jk}=\frac{m_{jk}^2}{p_i^2}\,,\\
&w_i \bar{w}_i=\mu_{ij}= \frac{ m_{ij}^2}{p_i^2}\,,\quad (1-w_i)(1- \bar{w}_i)= \mu_{ik}=\frac{ m_{ik}^2}{p_i^2}\,.
\esp\end{align}
This is a slight abuse of notation, as strictly speaking the $z$ and $\zbar$ variables are different for each $i$. For $i=1$, these are the variables defined in \ref{eq:z_def} and \ref{eq:zw1_variables_def} and the ones used for \refE{eq:symbT33masses}.

In terms of these variables, the single cut in the $p_i^2$ channel is
\begin{align}\bsp
\cut_{p_i^2}T(p_1^2,p_2^2,p_3^3;m_{12}^2,m_{23}^2,m_{13}^2)=&-2\pi\frac{e^{\gamma_E}\Gamma(1-\epsilon)}{\Gamma(2-2\epsilon)}(p_i^2)^{-1-\epsilon}\frac{(w_i-\bar{w}_i)^{1-2\epsilon}}{(z-w_i)(\bz-\bar{w}_i)-\mu_{jk}}\\
&\hypgeo{1}{1-\epsilon}{2-2\epsilon}{\frac{(z-\bz)(w_i-\bar{w}_i)}{(z-w_i)(\bz-\bar{w}_i)-\mu_{jk}}}\\
=&\frac{2\pi}{p_i^2(z-\zbar)}\ln\left(\frac{(w_i-\zbar)(\bar{w}_i-z)-\mu_{jk}}{(w_i-z)(\bar{w}_i-\zbar)-\mu_{jk}}\right)+\mathcal{O}(\epsilon)\,.
\esp\end{align}
Setting $(i,j,k)=(1,2,3)$ and comparing with \refE{eq:symbT33masses}, we see that the expected relation between cuts and coproduct entries holds.\\

Requiring that we normalize invariants by the channel being cut might be too restrictive. We now show how to define variables that do not have this requirement, but in terms of which the symbol alphabet is still rational. We define
\begin{align}\bsp\label{eq:varset2}
z&=\frac{1+u_j-u_k+\sqrt{\lambda(1,u_j,u_k)}}{2} ,\quad \bz=\frac{1+u_j-u_k-\sqrt{\lambda(1,u_j,u_k)}}{2}\,,\\
w_j&=\frac{u_j+\mu_{ij}-\mu_{jk}+\sqrt{\lambda(u_j,\mu_{ij},\mu_{jk})}}{2} ,\quad \bar{w}_j=\frac{u_j+\mu_{ij}-\mu_{jk}-\sqrt{\lambda(u_j,\mu_{ij},\mu_{jk})}}{2}\,,
\esp\end{align}
related to the invariants through slightly more complicated relations,
\begin{align}\bsp
&z\bz=u_j=\frac{p_j^2}{p_i^2}\,,\quad (1-z)(1-\bz)=u_k=\frac{p_k^2}{p_i^2} \,,\quad \mu_{ik}=\frac{m_{ik}^2}{p_i^2}\,,\\
&w_j \bar{w}_j=u_j\mu_{ij}= u_j\frac{ m_{ij}^2}{p_i^2}\,,\quad (u_j-w_j)(u_j- \bar{w}_j)= u_j\mu_{jk}=u_j\frac{ m_{jk}^2}{p_i^2}\,.
\esp\end{align}
As above, there is a slight abuse of notation in the definition of the $z$ and $\zbar$ variables.

In terms of these variables, the single cut in the $p_j^2$ channel is
\begin{align}\bsp
\cut_{p_j^2}T(p_1^2,p_2^2,p_3^3;m_{12}^2,m_{23}^2,m_{13}^2)=&-2\pi\frac{e^{\gamma_E}\Gamma(1-\epsilon)}{\Gamma(2-2\epsilon)}(-p_i^2)^{-1-\epsilon}\frac{(-z\bz)^{\epsilon}(w_j-\bar{w}_j)^{1-2\epsilon}}{(z-w_j)(\bz-\bar{w}_j)-z\bz\mu_{ik}}\\
&\hypgeo{1}{1-\epsilon}{2-2\epsilon}{\frac{(z-\bz)(w_j-\bar{w}_j)}{(z-w_j)(\bz-\bar{w}_j)-z\bz\mu_{ik}}}\\
=&\frac{2\pi}{p_i^2(z-\bz)}\ln\left(\frac{z\bz\mu_{ik}-(z-w_j)(\bz-\bar{w}_j)}{z\bz\mu_{ik}-(z-\bar{w}_j)(\bz-w_j)}\right)+\mathcal{O}(\epsilon)\,.
\esp\end{align}
As promised, the symbol letters are rational.

\subsubsection{Double cuts}

We now give the results for the double cuts in terms of the two sets of variables. For the variables in \refE{eq:varset1}, we compute the double cut in channels $p_i^2$ and $p_j^2$. It is given by
\begin{align}\bsp
\cut_{p_i^2,p_j^2}T(p_1^2,p_2^2,p_3^3;m_{12}^2,m_{23}^2,m_{13}^2)=& \frac{4\pi^2 ie^{\gamma_E}}{\Gamma(1-\epsilon)}\frac{(p_i^2)^{-1-\epsilon}}{(z-\bz)^{1-2\epsilon}}\left(\mu_{jk}-(z-w_i)(\bz-\bar{w}_i)\right)^{-\epsilon}\\
&\left((z-\bar{w}_i)(\bz-w_i)-\mu_{jk}\right)^{-\epsilon}\Theta_{ij} ,
\esp\end{align}
where
\begin{equation*}
\Theta_{ij}=\theta\left(\mu_{jk}-(z-w_i)(\bz-\bar{w}_i)\right)\theta\left((z-\bar{w}_i)(\bz-w_i)-\mu_{jk}\right)\, .
\end{equation*}\\

For the variables of \refE{eq:varset2}, we compute the double cut in channels $p_j^2$ and $p_k^2$. It is given by
\begin{align}\bsp
\cut_{p_j^2,p_k^2}T(p_1^2,p_2^2,p_3^3;m_{12}^2,m_{23}^2,m_{13}^2)=& \frac{4\pi^2 ie^{\gamma_E}}{\Gamma(1-\epsilon)}\frac{(-p_i^2)^{-1-\epsilon}}{(-z\bz)^{-\epsilon}(z-\bz)^{1-2\epsilon}}\Theta_{jk}\\
&\frac{\left(z\bz\mu_{ik}-(z-w_j)(\bz-\bar{w}_j)\right)^{-\epsilon}}{ \left((z-\bar{w}_j)(\bz-w_j)-z\bz\mu_{ik}\right)^{\epsilon}},
\esp\end{align}
where
\begin{equation*}
\Theta_{jk}=\theta\left((z-\bar{w}_j)(\bz-w_j)-z\bz\mu_{ik}\right)\theta\left(z\bz\mu_{ik}-(z-w_j)(\bz-\bar{w}_j)\right)\, .
\end{equation*}